\documentclass[iop]{emulateapj}
\usepackage{natbib}
\usepackage{color}
\usepackage{ulem}

\shorttitle{}
\shortauthors{Hsieh et al.}

\begin{document}

\title{The Connecting Molecular Ridge in the Galactic Center}

\author{
        Hsieh, Pei-Ying\altaffilmark{1,2},
        Ho, Paul T. P.\altaffilmark{1,3,4},         
        Hwang, Chorng-Yuan\altaffilmark{2}
\\pyhsieh@asiaa.sinica.edu.tw}

\affil{$^1$ Academia Sinica Institute of Astronomy and Astrophysics, P.O. Box 23-141, Taipei 10617, Taiwan, R.O.C.}
\affil{$^2$ Institute of Astrophysics, National Central University, No.300, Jhongda Rd., Jhongli City, Taoyuan County 32001, Taiwan, R.O.C.}

\affil{$^3$ East Asian Observatory, 660 N. AÕohoku Place, University Park, Hilo, Hawaii 96720, U.S.A.}

\affil{$^4$ Harvard-Smithsonian Center for Astrophysics, 60 Garden Street, Cambridge, MA 02138, USA}

\begin{abstract}
We report new observations of multiple transitions of the CS molecular lines in the SgrA region of the Galactic center, at an angular resolution of 40$\arcsec$ (=1.5 pc). The objective of this paper is to study the polar arc, which is a molecular ridge near the SgrA region, with apparent non-coplanar motions, and a velocity gradient perpendicular to the Galactic plane. With our high resolution dense-gas maps, we search for the base and the origin of the polar arc, which is expected to be embedded in the Galactic disk. We find that the polar arc is connected to a continuous structure from one of the disk ring/arm in both the spatial and velocity domains. This structure near SgrA* has high CS(J = 4--3)/CS(J = 2--1) ratios $\ge1$. That this structure has eluded detection in previous observations, is likely due to the combination of high excitation and low surface brightness temperature. We call this new structure the connecting ridge.  We discuss the possible mechanism to form this structure and to lift the gas above the Galactic plane.
\end{abstract}

\submitted{}
\accepted{August 19, 2015}


\section{INTRODUCTION}\label{sect-intro}


Since our Galactic center (GC) is the nearest nucleus of a galaxy ($d$=8.5 kpc) \citep{reid93,ghez08,reid14}, it offers an excellent opportunity to study the detailed structures and dynamics in the circumnuclear environment. Our focus is on how the complex activities in the nuclear region, might be related to the large scale structures. Both inflow into and outflow of material from the nuclear regions can be related to the large scale structures.    

On the possibility of inflows, we examine first the dynamical models.  Previous observations show that the GC exhibits two prominent molecular rings: the 200-pc twisted ring (TR) in the central molecular zone (CMZ) \citep[e.g.][] {sco72,bania77,bally88,morris96,dahmen,tsuboi99,oka05,martin04,molinari,martins12,oka12,jones12,jones13,longmore} and the 2-pc circumnuclear disk (CND) \citep[e.g.][]{guesten87,jackson93,serabyn89,amo11,harris,mezger,etx,lau,wright,maria,herrnstein,chris,great,mills13a}.
These two molecular rings are important because of the possibility that they are the gas reservoirs to fuel the central supermassive black hole (SMBH) SgrA* \citep{reid04,ghez08,gill09}, as well as the episodes of massive star formation within the CMZ (e.g., the HII regions in the 50 km s$^{-1}$ cloud, the three young star clusters, as well as the SgrB complex at 120 pc from SgrA*)(Figure~\ref{fig-fig1}a). 
The 200 pc TR could be a nuclear spiral produced by the non-axisymmetric potential \citep{sofue95,binney91,sawada04}. This is supported by the recent work of \citet{krui15}, who have constructed an orbital model for the gas streamers observed within the CMZ \citep{molinari}, by integrating orbits in the empirically constrained gravitational potential. 
Within the TR, \citet{sofue95} had further proposed the existence of the inner nuclear spirals (arm III, arm VI) with an extent of 70 pc, and which are headed towards SgrA. Arm III is also called the polar arc (PA) by Bally et al (1988). We show the CS(J = 1--0) line emission \citep{tsuboi99} of the PA in Figure~\ref{fig-fig1}b. The PA extends north of the SgrA complex at a 40$\degr$ angle away from the Galactic plane, and shows a large velocity gradient from $(l,b,V_{\rm sys}) = (0\degr,0\fdg05,70~\rm km~s^{-1}$) to $(l,b,V_{\rm sys}) = 0\fdg2,0\fdg25,140~\rm km~s^{-1}$.  Below $V_{\rm sys}$ of 70 km s$^{-1}$, the PA lies close to the Galactic plane and becomes confused with the molecular clouds in the SgrA region.
The SgrA region consists the radio continuum emission of SgrA East, SgrA West, and the radio halo (bipolar lobe). Concentrations of the molecular clouds  are located on the boundaries of the continuum sources. The clouds are known as 20 km s$^{-1}$ cloud (M-0.02-0.07) and 50 km s$^{-1}$ cloud (M-0.13-0.08).
Thus, from the dynamical models, gas could be brought from the 200 pc scale into the nuclear region via the PA structure.

Then, there is the possibility of outflows.
As shown in Figure~\ref{fig-fig1}a \citep{yusef04}, the 20 cm radio continuum emission shows several features related to the life-cycle of massive star formation: (1) the radio arc and the nonthermal radio filaments (NTF) \citep[e.g.][]{serabyn87,lang99,yusef84}, the arches cluster and the quintuplet cluster ($\sim10^{4}$M$_{\rm \odot}$) \citep{figer99a,figer99b,figer02}; these features are related through the collective winds of massive WR and OB stars, as well as the interactions between shocked charged particles and the Galactic magnetic field (Lang, Zadeh, Rosner and Bodo 1996) which can generate and illuminate the magnetized filaments; (2) the supernovae remnants of SgrA East \citep{zhao13,sjo,maeda,yusef87,ekers}, the radio halo \citep{pedlar89}, and the south east (SE) lobe \citep{zhao14}; these various lobe-like features could be due to the expansion, flow, and compression of the ISM by the supernova SgrA East; (3) the north west (NW) lobe; this structure could be produced by activities within 0.5 pc of the GC \citep{zhao14}.

The observational kinematical data which support both the inflow and outflow scenarios are already collected. Many studies have already been made with low-excitation molecular lines and with low angular resolutions \citep[e.g.][]{burton83,bally87,bally88,burton92,tsuboi99}.
High angular resolution studies also exist.  For the CND and the central 10 pc, early studies have focused on the interactions between the nearby SNR with NH$_{3}(3,3)$, CS(J =7--6) lines (SgrA East: \citet{ho85,serabyn92}, G359.02-0.09: \citet{ho85,coil00} and the ambient molecular clouds (the 20 km s$^{-1}$ cloud or M-0.13-0.08, and the 50 km s$^{-1}$ cloud or M-0.02-0.07: \citet{ho85,ho91,serabyn92,tsuboi09}.
\citet{serabyn92} and \citep{ho85} showed that the inner edge of the ambient clouds have high density and curves around the SNRs. This implies that the local SNRs are impacting and compressing the circumnuclear environment.  More recent studies of CS(J = 1--0), NH$_{3}(6,6)$, HCN(J = 4--3) lines have focused on the possibility of inflows which impact upon the CND \citep{minh13,liu12}. Streamers of molecular material appear to flow into the central parsec of the nucleus.     

However, connecting the inner region and the outer regions remain difficult. Previous studies could not resolve the central 70 pc (30$\arcmin$) of the Milky Way, principally because of contamination by the emission and absorption of the foreground cold gas.
Hence, the relation and the physical connection between the CMZ, the incoming streamers, the PA, and CND remains uncertain \citep{wardle08,liu12}.  


In this paper, we would like to focus on the inner 10 pc, and seek the connection between the PA and the CND. SMA HCN(J = 4--3) line map of the CND \citep{maria} demonstrated that the high-excitation transitions are less affected by the foreground cold gas along the line of sight.
These previous SMA studies were limited to the CND, and the extended emission was filtered out because of the lack of short spacing information. Here, we use the Caltech Submillimeter Observatory (CSO) 10 m single dish telescope to study the nuclear structure and its dynamics with the CS(J = 5--4) and the CS(J = 4--3) lines. This allows us to sample the large scale structures with adequate spatial resolution, while utilizing the higher excitation lines to suppress foreground confusion.  We compare this data with the archival CS(J = 1--0) line data \citep{tsuboi99} and the CS(J = 2--1) line data \citep[][(Paper II)]{hsieh15} taken with the Nobeyama 45m telescope, in order to study the gas excitation.   

\section{OBSERVATIONS}\label{sect-obs}

We observed the GC with the CSO 10 m telescope using the CS(J = 4--3) (195.9542 GHz),and CS(J = 5--4) (244.9356 GHz) lines during 2013 June.
The 230 GHz sidecab receiver was used with a backend of the Fast Fourier Transform Spectrometer (FFTS1) for all the observations. We used the wide bandwidth mode with the bandwidth of 1000 MHz, where the default resolution was 122 kHz. This yielded an intrinsic velocity resolution of 0.19 km s$^{-1}$ and 0.14 km s$^{-1}$ for the CS(J = 4--3) and CS(J = 5--4) lines, respectively.

The average system temperatures ranged from 190 K to 700 K during the observing runs.
Pointing was checked every hour using Saturn and IRAS 16293. The pointing accuracy was $\sim6\arcsec$. The standard chopper calibration method was used to measure the $T_{\rm A}^{*}$.
We used the on-the-fly (OTF) mapping technique with 20$\arcsec$ and 10$\arcsec$ grids for CS(J = 4--3) and CS(J = 5--4) lines, respectively. We used CLASS to remove the spectral baseline with polynominal functions of orders from 1 to 3. The spectra are gridded onto each pixel and are convolved with a Gaussian function. The final angular resolution and rms noise per channel are $38\arcsec$ and 0.07 K for the CS(J = 4--3) line, and $30\arcsec$ and 0.2 K for the CS(J = 5--4) line.
The temperature scale presented in this paper is $T{\rm _A^*}$, which is the calibrated and atmosphere-corrected antenna temperature. Therefore the main-beam brightness temperature $T_{\rm mb}$ is $T{\rm _A^*}$/$\eta_{\rm mb}$, where the $\eta_{\rm mb}$ is the main-beam efficiency.
We determined the main beam efficiency ($\eta_{\rm mb}$)  by observing the continuum flux of Saturn \citep{mangum}. We adopted the value $\eta_{\rm mb}$=0.4 and 0.3 for the CS(J = 5--4) and CS(J = 4--3) lines, respectively.

\section{RESULTS}\label{sect-reults}

\subsection{Integrated Intensity Maps}\label{sect-mom0}

Figure~\ref{fig-cs43mom0} shows the integrated intensity (moment 0) maps of the CS(J = 1--0) \citep{tsuboi99}, CS(J = 2--1) (Paper II), CS(J = 4--3), and CS(J = 5--4) lines integrated within $\pm160$ km s$^{-1}$. These maps were smoothed to the same resolution of 40$\arcsec$ for comparisons. Overall, the higher-J CS(J = 4--3) and CS(J = 5--4) lines are less sensitive to the diffuse emission as shown in the lower-J CS(J = 1--0), HCN(J = 1--0), HCO$^{+}$(J = 1--0) \citep{tsuboi99,tsuboi11}, and CS(J = 2--1) line (Paper II). The CS(J = 4--3) and CS(J = 5--4) lines are dominated instead by the more compact ridges and dense clouds. The major features of the CND, the 20 km s$^{-1}$ cloud, the 50 km s$^{-1}$ cloud, and the high velocity compact cloud (HVCC; CO 0.02-0.02) \citep{oka99} are in general consistent in all the lines. 

In contrast to these well known structures, we found there is a ridge-like structure shown in the CS(J = 4--3) line map. At the position of this ridge-like structure, compact clouds can be seen in the CS(J = 1--0) and CS(J = 2--1) lines, immersed in the background diffuse emission. 
In Figure~\ref{fig-polar}, we show the larger scale map of the polar arc (PA), as measured in the CS(J = 1--0) line, and the central region as measured in the  CS(J = 4--3) line. We also overlay the 20 cm continuum emission on these maps. 
We find that the CO 0.02-0.02 cloud and the ridge-like structure may form a complete ``northern molecular loop'' which we have highlighted with the yellow dashed-ellipse. The northern molecular loop structure may enclose the northern radio halo. A similar structure can also be seen in the low-J CS(J = 2--1) and CS(J = 1--0) lines, but the CS(J = 4--3) line exhibits clearer features.

\subsection{Channel Maps}

In Figure~\ref{fig-cs43chan1}, we show the CS(J = 4--3) line channel maps at the velocity resolution of 10 km s$^{-1}$. The purpose of showing the channel maps is to demonstrate the kinematics of the individual cloud structures. The coherent kinematics often help to define distinct structures.   The CND appears as a high velocity ($-110$ km s$^{-1}$ to 110 kms$^{-1}$) feature (the green ellipse) centered on  SgrA* (green cross).
The well studied 20 km s$^{-1}$ and 50 km s$^{-1}$ clouds show a velocity gradient from -30 km s$^{-1}$ to 80 km s$^{-1}$, which is rotating with respect to SgrA*
\citep[e.g.][]{martins12,liu12}. The CO 0.02-0.02 cloud \citep{oka99,oka08} is seen from 80 to 110 km s$^{-1}$.
The ridge-like structure is detected roughly from 30 km s$^{-1}$ to 90 km s$^{-1}$ (labeled as green stars). Consistent with Figure~\ref{fig-fig1}, this structure was only detected with individual clumps in the CS(J = 1--0), the CS(J = 2--1) and the CS(J = 5--4) images. Here we call this ridge-like structure the ``connecting ridge'' (CR). The width of the CR is $\sim2\arcmin$ (4.6 pc).
The CR seems not to have been detected before. We searched through the literature.
The large scale survey of $^{13}$CO \citep{bally88}, $^{12}$CO \citep{oka98,oka01}, and NH$_{3}$ \citep{pucell} do not show the CR in their maps. The NH$_{3}$ line map of the entire 200 pc TR made with ATCA \citep{ott14}
suffers from the missing flux, and there is no detection for the CR.
The CR is also absent in the Mopra molecular line maps of the CMZ \citep{jones12,jones13}. This is probably because of the lower sensitivity and resolution for the past observations. It is also possible to be the excitation effect. We will discuss this in the section of discussion.
In the CS(J = 4--3) channel map, the CR seems to extend to the north of the Galactic plane and coincides with the base of the PA (labeled as the green solid-squares). Our CS(J = 4--3) line observations did not cover the PA. Hence we show in Figure~\ref{fig-cs10chan} the channel maps of the CS(J = 1--0) line in the same velocity range of 30 km $^{-1}$ to 110 km $^{-1}$.
At the velocity of 30 km s$^{-1}$, the CR seems to be near the CND and appears to connect to the Galactic plane. We have called this part of the structure in the Galactic plane the ``disk ridge'' (DR) (labeled by the green solid-triangles). 

In Figure~\ref{fig-cs54chan1} we show the channel maps of the CS(J = 5--4) line smoothed to the same angular resolution of 38$\arcsec$ for the CS(J = 4--3) line. The CS(J = 5--4) line is in general consistent with the CS(J = 4--3) line, but traces more compact structures. However, this is likely due to a factor of two lower sensitivity in the CS(J = 5--4) line. For the CR, we only detect some faint clumps from 40 km s$^{-1}$ to 80 km s$^{-1}$.  We also detected the base of the PA similar to the Figure~\ref{fig-cs10chan}.

\subsection{The Position-Velocity Diagram}

In Figure~\ref{fig-pv-mom0}, Figure~\ref{fig-pv-comb} and Figure~\ref{fig-pv-c1} we show the position-velocity diagrams (pv-diagrams) of the DR, the CR, and the PA.  We show these diagrams in order to demonstrate the coherent kinematics within each of the structures, and with respect to each other.  
In Figure~\ref{fig-pv-comb} we merge the pv-diagrams of the DR, the CR, and the PA in order to see the continuity of the ridges.
Our CS(J = 4--3) line data does not cover the PA region, so we also show the pv-diagrams of the CS(J = 1--0) data along the PA for comparison.
As indicated in the CS(J = 4--3) line channel maps (Figure~\ref{fig-cs43chan1}), the CR appears from $\sim$50 km s$^{-1}$ and smoothly connects to the PA at velocity $\ge$80 km s$^{-1}$. Although the CR becomes fainter in the CS(J = 1--0) line, Figure~\ref{fig-pv-comb} suggests that the DR, the CR, and the base of the PA have smooth velocity transitions along the structure. 
We emphasize that this continuous nature of the kinematics is the principal argument in favor of a coherent structure.    

In Figure~\ref{fig-pv-c1} we found that the PA seems to show expanding features. In paper II, we measured the whole PA with expanding velocity of $\sim$74 km s$^{-1}$. The detail discussion and results are in paper II.
In the DR, there are several clumps which show broadened FWZI linewidths of $\sim$90 km s$^{-1}$, which might indicate that they are physically near the CND.
There are also some clumps in the CR which have broad linewidths (FWZI $\sim$50 km s$^{-1}$). The linewidths are about a factor of 2 larger than the values in the rest of the region, and can be seen in the other CS lines.
This is best seen in the velocity dispersion maps presented below.


\subsection{The Ridge From 30 km s$^{-1}$ to 90 km s$^{-1}$}

From the channel maps and the pv-diagrams, we found that the velocity of the CR is roughly from 30 km s$^{-1}$ to 90 km s$^{-1}$.
We made the integrated intensity map and the intensity-weighted mean velocity map of the CS(J = 4--3) line from 30 km s$^{-1}$ to 90 km s$^{-1}$ (Figure~\ref{fig-cs43mom1}). The spectra of the DR, CR, and the base of the PA are shown from Figure~\ref{fig-cs43spec1} to Figure~\ref{fig-cs43spec2}. The spectra of the CR also suggest that the FWZI velocity range is from 30 km s$^{-1}$ to 90 km s$^{-1}$.
The velocity map (Figure~\ref{fig-cs43mom1}b) shows two kinematic systems: (1) along the 20 km s$^{-1}$ cloud to the 50 km s$^{-1}$ cloud, (2) from the DR to the CR. The velocity patterns are different and this kinematical behavior argues for the DR being a distinct feature within the Galactic plane.

In Figure~\ref{fig-cs43mom2}, we show the intensity-weighted velocity dispersion map of the CS(J = 4--3) line. The velocity dispersions of the DR, the CND, the CR, the CO 0.02-0.02 cloud, and near the PA, are larger than the surrounding material by about a factor of 2 or higher. The gas with high velocity dispersions also surround the 20 cm radio halo, which is similar to the northern molecular loop structure seen in Figure~\ref{fig-cs43mom0}. We notice that the gas of the 20 km s$^{-1}$ cloud and the 50 km s$^{-1}$ cloud also have higher velocity dispersions near the radio halo. By inspecting the spectra and the channel maps of the data, we find that the line widths of the 20 km s$^{-1}$ and 50 km s$^{-1}$ clouds are mainly intrinsically broad, and the northern molecular loop consists of multiple distinct components at the velocities of $\sim-12.5$ km s$^{-1}$ and 47.5 km s$^{-1}$.
We emphasize that it is the enhanced velocity dispersions, together with the continuous kinematics, which argue that the CR, DR, and PA, could be a single coherent feature.

\subsection{The CS Line Ratios}

In Figure~\ref{fig-cs-ratio} we show the CS(J = 4--3)/CS(J = 2--1), CS(J = 4--3)/CS(J = 1--0), CS(J = 5--4)/CS(J = 4--3), and CS(J = 5--4)/CS(J = 1--0) intensity ratio maps. We integrated the maps from $-160$ km s$^{-1}$ to $+160$ km s$^{-1}$ with a convolved beam of 40$\arcsec$. We utilized data with significance $\ge3\sigma$. Figure~\ref{fig-cs-ratio} represents the average intensity ratios because multiple components have been integrated along the line of light. The distributions of the CS(J = 4--3)/CS(J = 2--1) and CS(J = 4--3)/CS(J = 1--0) ratios are similar. The CR and the base of the PA have higher ratios than the ambient gas by a factor of $\sim$2 in the CS(J = 4--3)/CS(J = 2--1) ratio map, and the CS(J = 4--3)/CS(J = 1--0) ratio map shows higher contrasts than the CS(J = 4--3)/CS(J = 2--1) ratio map. 
The CND has higher ratios than the 50 km s$^{-1}$ and 20 km s$^{-1}$ clouds, and is comparable to the CR in the CS(J = 4--3)/CS(J = 1--0), CS(J = 4--3)/CS(J = 2--1), and CS(J = 5--4)/CS(J = 1--0) ratio maps. 
We emphasize that the enhanced line ratios is another argument for the CR, DR, and PA, to be a coherent structure.
The high-J line ratios of CS(J = 5--4)/CS(J = 4--3) are higher in the 50 km s$^{-1}$ and 20 km s$^{-1}$ clouds $\ge1.5$ than in the CND $\sim1.2$. We will discuss the difference of the ratio maps in the section of discussion.

\subsection{Derived Parameters}

\subsubsection{Rotational Diagrams}


In Figure~\ref{fig-rot-diag}, we show the rotational diagrams of the major components mentioned above for the CS(J = 5--4), CS(J = 4--3), CS(J = 2--1), and CS(J = 1--0) lines. The usefulness of such a diagram depends on the assumption of thermal equilibrium. We convolved all of the CS line maps to the same angular resolution of 40$\arcsec$ and velocity resolution of 5 km s$^{-1}$. For the CR, we measured the rotational temperature of the broadline clump shown in Figure~\ref{fig-cs43spec2}. We also present the rotational diagrams of the DR and the base of the PA. For these 3 regions, the fluxes used for the rotational diagrams are averaged from the spectra shown in Figure~\ref{fig-cs43spec1}, Figure~\ref{fig-cs43spec2}, and Figure~\ref{fig-cs43spec3}. For the CR, we measured the flux from 30 km s$^{-1}$ to 90 km s$^{-1}$.
The fluxes used in the rotational diagrams are presented from Table~\ref{t.obspar1} to Table~\ref{t.obspar3}.
We could estimate the rotational temperature ($T_{\rm{rot}}$) and column densities ($N_{\rm{mol}}$) of the CS lines by assuming optically thin and local thermodynamic equilibrium (LTE). 
The column density in the upper level ($N_u$) is expressed as
\begin{equation}
N_u = \frac{8\pi k_{\rm{B}} \nu^2 \int{T_{\rm{b}}}dv}{hc^3 A_{ul}},
\end{equation}
where $k_{\rm{B}}$ and $h$ are the Boltzmann and Planck constants, respectively. $A_{ul}$ is an Einstein A coefficient from upper ($u$) to lower state ($l$), $\nu$ is the frequency of the line, $c$ is the speed of light, and $T_{\rm b}$ is the brightness temperature of the line.
In the LTE condition, $N_u$ is written as 
\begin{equation}
N_u = \frac{N_{\rm{total}}}{Q(T_{\rm ex})} g_u \exp \Biggl(-\frac{E_u(\rm erg)}{k_{\rm B}T_{\rm ex}}\Biggr),
\end{equation}
where $N_{\rm{total}}$ is the total column density of a given molecule, $Q(T)$ is a partition function, $E_u$ is an energy at level $u$ above the ground state, and $g_{\rm u}$ is the statistical weight. Hence we could rewrite the equation as
\begin{equation}\label{slope}
ln\Biggl(\frac{N_u}{N_{\rm{total}}}\Biggl)=-\frac{1}{T_{\rm rot}}E_{\rm u}(\rm K)+ln \Biggl(\frac{g_{\rm u}}{Q(T_{\rm rot})}\Biggl).
\end{equation}
Therefore, we could determine the total column density and the excitation temperature by fitting a straight line for the data, plotted in logarithmic form, with Equation~\ref{slope}. The $E_{\rm u}$(K) is $E_{\rm u}$(erg)/$k_{B}$ for multiple lines with different $E_{\rm u}$.
This excitation temperature is called the ``rotational temperature'' ($T_{\rm{rot}}$) because we measured the rotational transitions of CS molecules.  The derived $T_{\rm{rot}}$ is also a lower limit to the kinetic temperature ($T_{\rm{kin}}$), since the gas may not be thermalized. The detailed methods of the rotational diagram were described in \citet{gold99}. 
In Figure~\ref{fig-rot-diag}, we found that the rotational diagrams of all the features could not be described by a single component. The CS(J = 5--4) and CS(J = 4--3) lines define a flatter slope than the CS(J = 1--0) and CS(J = 2--1) lines. This suggests that the molecular gas is a mixture of ``high-temperature'' and ``low-temperature'' components along the line of sight. If we assume the high-J lines (J$\ge$4) represent warmer components, then it indicates the $T_{\rm rot}$ of the warm gas components are higher than the cold components by a factor more than three. Our results on the presence of multiple temperature components are consistent with the NH$_{3}$ line observations \citep{nh3_05}.
The fitted results are presented from Table~\ref{t.obspar1} to Table~\ref{t.obspar3}. The $T_{\rm rot}$ are similar for the DR, the CR and the base of the PA. However, the base of the PA is very close to the CR and the $T_{\rm rot}$ differences might not be discernible. Also the $T_{\rm rot}$ (High) of the DR is almost twice higher than the CR and the PA. This indicates that the DR is closer to the CND. 
We note that the fitted rotational temperatures for the CS lines are low relative to the expectations from the NH$_{3}$ observations.  This is an immediate hint that the assumption of LTE is probably not correct for the CS lines.  This means that the density may not be high enough and will have to be taken into account.  We discuss this further below.  However, regardless of the actual density and temperatures of the gas, we emphasize that the different CS line ratios help us to isolate different cloud components.

\subsubsection{Statistical Equilibrium}

In order to derive the physical properties of molecular gas instead of the optically thin assumption,
we used the 1D radiative transfer code called Radex \citep{radex}
to perform the statistical equilibrium calculations.
Radex uses the escape probability formalism \citep{goldreich}
to solve the statistical equilibrium equations to model the observed
line intensities for given physical conditions.
The statistical equilibrium calculations account for the opacity
effects as well as subthermal excitation, while the rotation diagram analysis relies on optically thin approximation and LTE conditions.

The Rayleigh-Jeans approximated brightness temperature ($T_{b}$ (K) or $\int^{}_{}T_b\,dv $ (K km s$^{-1}$) are calculated for each molecule with varying molecular column density, H$_{2}$ density of the main collision partner,
and kinetic temperature. The observe main beam brightness temperature ($T_{mb}$) is diluted by the beam filling factor $f$,
\begin{equation}
f = \frac{\theta_{s}}{\theta_{s}+\theta_{b}} = \frac{T_b}{T_{mb}},
\end{equation}
where $\theta_{s}$ is the source size and $\theta_{b}$ is the beam size. The brightness temperature of the source will be diluted if the source size is smaller than the beam size.
We adopted line ratios to constrain the statistical
equilibrium while canceling beam filling factors.
Comparison of the models and the observations was performed with the
$\chi^2$-statistics for each set of parameters. We find the best fit
models by minimizing $\chi^2$ value.
In this paper, we use the uniform sphere geometry, single collision partner (H$_{2}$), and background temperature of 2.73 K.
The collisional rate and spectroscopic data of CS molecule were taken from Leiden Atomic and Molecular Database (LAMDA) \citep{lamda,lique}.

We construct a 50$\times$50$\times$50 grid of parameter space for kinetic temperature ($T_{\rm K}$), molecular hydrogen density ($n_{\rm H_2}$), and column density of CS molecule ($N_{\rm CS}$).
\begin{enumerate}
\item $T_{\rm K}$ is from 10 K to 650 K.
\item $n_{\rm H_2}$ is from 10$^{4}$ cm$^{-3}$ to 5$\times$10$^{7}$ cm$^{-3}$.
\item $N_{\rm CS}$ is from 10$^{12}$ cm$^{-2}$ to 10$^{16}$ cm$^{-2}$.
\end{enumerate}
We then estimate the best fitted $T_{\rm K}$, $n_{\rm H_2}$ , and $N_{\rm CS}$ by comparing with the ratios and intensities.

The input CS line fluxes are the same as the values used for rotational diagrams method.
From Figure~\ref{fig-radex1} to Figure~\ref{fig-radex3} we show the $\chi^2$ contours versus $T_{\rm K}$ and $n_{\rm H_2}$. From Table~\ref{t.obspar1} to Table~\ref{t.obspar3} we show the derived parameters with best fitted $\chi^2$.
The $T_{\rm rot}$ and the $N_{\rm CS}$ derived in the rotational diagram method are lower than the values derived in the statistical equilibrium method. 
We also present the excitation temperature and the opacity calculated from
the Radex code. The excitation temperature of the four CS lines are different.
The opacity is optically thin in the CS(J = 1--0) line, and $\ge$1 for the CS(J = 2--1), CS(J = 4--3), and CS(J = 5--4) lines.
These might suggest that the CS lines are likely to be subthermal since the excitation temperature is lower than the kinetic temperature.
From Figure~\ref{fig-radex1} to Figure~\ref{fig-radex3}  we show the CS J-ladder diagram of the data and the model. Our results show that the beam filling factors of this region are from 0.14 to 0.2, which suggest sources are much smaller than the beam size. This might also explain the lower $N_{\rm CS}$ derived in the rotational diagrams, since the brightness temperature used for the rotational diagrams are diluted by the beam.

\section{Discussions}
\subsection{Why the CR is Bright in the CS(J = 4--3) Line?}

We detected the CR in the new CS(J = 4--3) line map. This component also exists in the CS(J = 1--0), CS(J = 2--1), and CS(J = 5--4) lines, but is much more visible in the CS(J = 4--3) line map.  The CS(J = 1--0), CS(J = 2--1), and CS(J = 5--4) lines only detected several compact clouds of the CR. The extended emission was only marginally detected in the CS(J = 1--0) and CS(J = 2--1), but not detected in the CS(J = 5--4) lines. For the CS(J = 5--4) line, insufficient sensitivity is likely the reason that the extended emission of the CR was not detected. The CS(J = 4--3) line intensity of the extended emission of the CR is $\sim$0.4 K, and the noise level of the CS(J = 5--4) line is 0.17 K ($T_{\rm mb}$). Hence the noise will dominate in the CS(J = 5--4) line map if the CS(J = 5--4) and CS(J = 4--3) intensities are comparable. Indeed for the compact clouds detected in both of these lines, the intensities are comparable.

As in previous discussions, it is distinctly possible that the various transitions are sampling different regions with different excitation.  We can then set some limits on the possible ranges of conditions by considering the available transitions.
We check the excitation conditions under which the CS(J = 4--3) line is most prominent.  If the gas is thermalized, i.e. if the density is sufficiently high so that all the energy levels are in thermal equilibrium, the relative intensity of the different transitions is just a function of temperature.  This is the basis of the analysis of equation (3), and the utility of the rotational diagrams.  If the gas is not thermalized, then the relative intensity ratios of the various transitions will be a function of excitation temperature, filling factor, frequency, and opacity. 
The excitation temperature is of course driven by the density \citep[cf.][]{radex}. The CS line ratios are sensitive to the density of the molecular gas because of its large dipole moment, and small frequency spacing between the transitions. Hence the CS molecule is a useful density tracer \citep{radex}.
Our statistical equilibrium calculations show that the physical conditions ($n_{\rm H_2}$, $N_{\rm CS}$, and $T_{\rm K}$) of the DR, CR and the base of the PA are similar. The CS(J = 1--0) line is optically thin, and the other lines are optically thick. The CS(J = 4--3) line has highest opacity as compared to the other lines. 
For the Rayleigh Jeans approximation, the brightness temperature is proportional to $f(T_{\rm ex}-T_{\rm bg})(1-e^{-\tau})$, where $T_{\rm ex}$, $T_{\rm bg}$, $\tau$, $f$ are the excitation temperature, background temperature, opacity, frequency, and filling factor of a transition.
Considering the modeled parameters and the Rayleigh Jeans approximation, the CS(J = 4--3) line is intrinsically brighter than the CS(J = 1--0) and CS(J = 5--4) lines. The brightness temperature of the CS(J = 2--1) line is comparable to the CS(J = 4--3) line.
Our statistical equilibrium results suggest that for $n_{\rm H_2}=10^{5.7-5.9}$ cm$^{-2}$, $T_{\rm K}=18$K, and $N_{\rm CS}=(2-4)\times10^{15}$ cm$^{-2}$, the CS(J = 4--3) will be the brightest line of the CR.

Our resolved ratio maps (Figure~\ref{fig-cs-ratio}) might suggest the variation of excitation conditions. As shown in the CS(J = 4--3)/CS(J = 2--1) ratio map (and similarly with CS(J = 4--3)/CS(J = 1--0)), the connecting ridge has higher line ratios for lines with larger energy differences. This indicates higher temperatures, an is as expected as large energy gaps will be sensitive to higher temperatures. These higher ratios also suggest higher opacity in the connecting ridge than the ambient gas.

It is interesting to ask why the CS(J = 4--3)/CS(J = 2--1) and CS(J = 4--3)/CS(J = 1--0) ratios of the 20 km s$^{-1}$ and 50 km s$^{-1}$ clouds are not as significant as in the CS(J = 5--4) ratios. The CS(J = 4--3)/CS(J = 2--1) and CS(J = 4--3)/CS(J = 1--0) ratios show different distribution from the CS(J = 5--4)/CS(J = 1--0), and CS(J = 5--4)/CS(J = 4--3) ratio maps, where the 20 and 50 km s$^{-1}$ clouds show high ratios. We note that in Figure~\ref{fig-rot-diag}, the dense gas in the Galactic center has structures of multiple-temperature components, where the CS(J = 1--0) and CS(J = 2--1) lines represent cooler component, and CS(J = 4--3), CS(J = 5--4) lines represent warmer component.
If this implies that the gas is a mixture of different excitation conditions (temperature, density), then in the CS(J = 4--3)/CS(J = 2--1) ratios map, we are sampling the cooler part of the 20 and 50 km s$^{-1}$, and the CS(J = 5--4) line traces better the warmer components in these clouds. Therefore, the variations of the ratios from high-J to low-J lines indicate the mixture of multiple phases of the molecular gas, and its sensitivity to different tracers. Our results are consistent with the previous results \citep[e.g.,][]{guesten83,gues85,serabyn89,serabyn92,huett,nh3_05}.

\subsection{The Possible Origins of the CR}

In Figure~\ref{fig-cs43mom1} we found that there are two kinematic systems: (1) the disk rotation of the 20 km s$^{-1}$ and 50 km s$^{-1}$ clouds, (2) the velocity-gradient along the DR, CR, and the base of the PA (Figure~\ref{fig-pv-comb}). The velocity gradient along the DR, CR and the base of the PA seem to suggest a coherent extraplanar structure, and the gradient is not aligned with the Galactic plane.
\citet{sofue96} studied the $^{13}$CO(J = 1--0) line data from the Bell Laboratory Survey (BLT)\citep{bally87}(beam=100$\arcsec$).  They pointed out a pair of molecular spurs are associated with the galactic center lobe (GCL)\citep{sofue84,bland03}. The eastern- and western-spurs are receding and approaching at V$_{lsr}$ of  $\sim$100 km s$^{-1}$  and $\sim-150$ km s$^{-1}$, respectively. The size scale of the molecular spurs is $\sim$30 pc.
In this paper we confirm that the location of the eastern-molecular spur is coincident with the PA with consistent velocity.

In Figure~\ref{fig-pv-c1}, we found that the PA shows expanding motion. We estimate the energy of this feature. The kinetic energy ($E_{\rm k}$) of the PA is,
\begin{equation}
E_{\rm k} = \frac{1}{2}M_{\rm ridge}\Delta V^{2},
\end{equation}
where the $M_{\rm ridge}$ is the molecular hydrogen mass of the ridge, and the $\Delta V$ is expanding velocity of 74 km s$^{-1}$ (paper II).
We derive the gas mass ($M_{\rm H_2}$) of the entire PA from the CS(J = 1--0) map. We adopted the same equation from \citet{tsuboi99}, with $T_{\rm ex}$ of 20 K, and the CS abundance of $10^{-8}$ \citep{irvine}. The corresponding $M_{\rm H_2}$ is 1.5$\times10^{5}$ M$_{\odot}$ and the kinetic energy of the PA is from 8$\times10^{51}$ erg (paper II).
One of the possible origin of the energy input is from multiple supernova explosions, which is able to release the energy into the ISM with efficiency up to 20\% \citep{weaver77,mccray87}. Hence we need around 40 supernova explosions to push the gas (PA) away, or 8 supernova explosions for 100\% efficiency.

In Figure~\ref{fig-10ghz}, the 10 GHz continuum emission of the GCL \citep{handa87} is shown as a dome-shaped, limb-brightened bipolar structure in the Galactic center. It has a vertical extension more than 200 pc. The GCL was detected in the radio continuum emission \citep{sofue84,law08a,law08b}, the dust emission at 8.3$\micron$ \citep{bland03}, and the radio recombination lines \citep{law09}. The GCL was proposed to be the entrainment of dust by the large-scale bipolar wind powered by the central starburst, which had taken place within the last 7 Myr with an energy ejection zone of 100 pc \citep{bland03}. In Figure~\ref{fig-10ghz}, the dust emission of the GCL shows molecular counterparts of the eastern (positive velocity) and western spurs (negative velocity) detected by \citep{sofue96}. The eastern spur is 
the PA shown in the CS(J = 1--0) line \citep{tsuboi99} (Figure~\ref{fig-10ghz}).
The PA shows vertical extension of $\sim$43 pc from the central disk.
The preliminary idea for this paper is that the molecular gas originally in the disk was pushed out by the supernova explosions.
Here are some of the indirect evidences for this idea: (1) the continuity of the velocities of the DR, CR, and PA. The CR, DR has velocity gradient orthogonal to the Galactic disk, (2) the expanding motion of the PA, (3) similar physical conditions of the DR, CR, and PA, (4) the high dispersions surrounding the 20 cm radio halo.
However, since we only have morphological association of the GCL and the CR/PA, it is unclear whether the kinetic energy within the CR/PA was converted from the kinetic energy of stellar feedback. A large molecular line map of the entire GCL will provide better constraints. The detail studies of the expanding features in the Galactic center is presented in paper II.
The other possibility is that the DR, CR, ad PA are formed by the magnetic buoyancy caused by the Parker instability (field strength of $\sim$150 $\mu$G)
\citep{fukui06,machida09,takahashi09}. We will discuss this the next paper (paper II).

\subsection{From Small Scale (pc) to Large Scale (kpc)}

The ubiquitous presence of nuclear winds and outflows are recently shown from nearby star-formaing galaxies to the high-z galaxies \citep{wind05,ho14,rubin14,chen10,murray11}.
In the case of the IC 342, which is one of the nearest galaxy at a distance of 3.5 Mpc \citep{wu14}. \citet{ic342} suggested that the giant molecular associations (GMAs) (gas mass: $10^7-10^8$M$_{\odot}$) in its nuclear region are pushed out by the violent nuclear starburst. The mechanical energy of the stellar wind is able to push out the GMAs from 12 pc to 50 pc relative to the galactic center. On the large scale of kilo-pc, the intensive studies of NGC 253 \citep{ngc253} and M82 \citep{ohyama02} also showed that the central starburst activity is able to push large amounts of molecular gas away from disk. This disruption may prevent the gas from forming stars in the immediate future. The Galactic wind can play the role of suppression of star formation in galaxies.

\section{SUMMARY}

Our high resolution CS line maps provide an alternative way to study the physical conditions and the dynamics of the extraplanar structures which can not be done with the continuum observation.
We detected a new component (CR) in the CS(J = 4--3) line. Our statistical equilibrium results suggest that for $n_{\rm H_2}=10^{5.7-5.9}$ cm$^{-2}$, $T_{\rm K}=18$K, and $N_{\rm CS}=(2-4)\times10^{15}$ cm$^{-2}$, the CS(J = 4--3) will be the brightest line of the CR.
 The CR is spatially and dynamically associated with the extraplanar PA. The kinetic and the spatial continuation of the disk ridge, the CR, and the PA suggest the molecular gas originally in the disk might be lifted, possibly by the energy of 8-80 supernova explosions. The PA is spatially and kinematically associated with the eastern molecular spur of the GCL. A large and sensitive molecular line map is awaited to study the kinematics of the entire GCL.

\acknowledgements
We would like to thank the referee for a constructive report that
improved the paper. We thank Simon Radford for the assistance of the CSO observations. We thank Dr. Yusef-Zadeh, F. to provide us the VLA 20 cm continuum map. This material is based upon work at the Caltech Submillimeter Observatory, which is operated by the California Institute of Technology. These observations are part of the East Asia Core Observatories Association (EACOA) project on CSO. Pei-Ying Hsieh is supported by the National Science Council (NSC) and the Ministry of Science and Technology (MoST) of Taiwan, NSC 100-2112-M-001-006-MY3, NSC 97-2112-M-001-021-MY3, and MoST 103-2112-M-001-032-MY3.

\clearpage


\bibliography{cso.ref}

@ARTICLE{reid93,
   author = {{Reid}, M.~J.},
    title = "{The distance to the center of the Galaxy}",
  journal = {\araa},
 keywords = {Astrometry, Distance, Galactic Nuclei, Milky Way Galaxy, Astronomical Models, Globular Clusters, Water Masers},
     year = 1993,
   volume = 31,
    pages = {345-372},
      doi = {10.1146/annurev.aa.31.090193.002021},
   adsurl = {http://adsabs.harvard.edu/abs/1993ARA
  adsnote = {Provided by the SAO/NASA Astrophysics Data System}
}

@ARTICLE{reid14,
   author = {{Reid}, M.~J. and {Menten}, K.~M. and {Brunthaler}, A. and {Zheng}, X.~W. and 
	{Dame}, T.~M. and {Xu}, Y. and {Wu}, Y. and {Zhang}, B. and 
	{Sanna}, A. and {Sato}, M. and {Hachisuka}, K. and {Choi}, Y.~K. and 
	{Immer}, K. and {Moscadelli}, L. and {Rygl}, K.~L.~J. and {Bartkiewicz}, A.
	},
    title = "{Trigonometric Parallaxes of High Mass Star Forming Regions: The Structure and Kinematics of the Milky Way}",
  journal = {\apj},
archivePrefix = "arXiv",
   eprint = {1401.5377},
 primaryClass = "astro-ph.GA",
 keywords = {Galaxy: fundamental parameters, Galaxy: kinematics and dynamics, Galaxy: structure, gravitational waves, parallaxes, stars: formation},
     year = 2014,
    month = mar,
   volume = 783,
      eid = {130},
    pages = {130},
      doi = {10.1088/0004-637X/783/2/130},
   adsurl = {http://adsabs.harvard.edu/abs/2014ApJ...783..130R},
  adsnote = {Provided by the SAO/NASA Astrophysics Data System}
}


@ARTICLE{oka99,
   author = {{Oka}, T. and {White}, G.~J. and {Hasegawa}, T. and {Sato}, F. and 
	{Tsuboi}, M. and {Miyazaki}, A.},
    title = "{A High-Velocity Molecular Cloud near the Center of the Galaxy}",
  journal = {\apj},
   eprint = {arXiv:astro-ph/9810434},
 keywords = {GALAXY: CENTER, ISM: CLOUDS, ISM: MOLECULES},
     year = 1999,
    month = apr,
   volume = 515,
    pages = {249-255},
      doi = {10.1086/307029},
   adsurl = {http://adsabs.harvard.edu/abs/1999ApJ...515..249O},
  adsnote = {Provided by the SAO/NASA Astrophysics Data System}
}
  
@ARTICLE{oka08,
   author = {{Oka}, T. and {Hasegawa}, T. and {White}, G.~J. and {Sato}, F. and 
	{Tsuboi}, M. and {Miyazaki}, A.},
    title = "{Aperture Synthesis Imaging of a High-Velocity Compact Cloud near the Galactic Center}",
  journal = {\pasj},
 keywords = {galaxies: nuclei, Galaxy: center, ISM: clouds, ISM: molecules},
     year = 2008,
    month = jun,
   volume = 60,
    pages = {429-},
   adsurl = {http://adsabs.harvard.edu/abs/2008PASJ...60..429O},
  adsnote = {Provided by the SAO/NASA Astrophysics Data System}
}

@ARTICLE{bally88,
   author = {{Bally}, J. and {Stark}, A.~A. and {Wilson}, R.~W. and {Henkel}, C.
	},
    title = "{Galactic center molecular clouds. II - Distribution and kinematics}",
  journal = {\apj},
 keywords = {GALACTIC NUCLEI, KINEMATICS, MASS DISTRIBUTION, MILKY WAY GALAXY, MOLECULAR CLOUDS, ASTRONOMICAL MAPS, ASTRONOMICAL SPECTROSCOPY, GAS DYNAMICS, LINE SPECTRA, MILLIMETER WAVES},
     year = 1988,
    month = jan,
   volume = 324,
    pages = {223-247},
      doi = {10.1086/165891},
   adsurl = {http://adsabs.harvard.edu/abs/1988ApJ...324..223B},
  adsnote = {Provided by the SAO/NASA Astrophysics Data System}
}

@ARTICLE{minh13,
   author = {{Minh}, Y.~C. and {Liu}, H.~B. and {Ho}, P.~T.~P. and {Hsieh}, P.-Y. and 
	{Su}, Y.-N. and {Kim}, S.~S. and {Wright}, M.},
    title = "{Green Bank Telescope Observations of the NH$_{3}$ (3, 3) and (6, 6) Transitions toward Sagittarius a Molecular Clouds}",
  journal = {\apj},
 keywords = {Galaxy: center, ISM: individual objects: Sgr A, ISM: molecules, radio lines: ISM},
     year = 2013,
    month = aug,
   volume = 773,
      eid = {31},
    pages = {31},
      doi = {10.1088/0004-637X/773/1/31},
   adsurl = {http://adsabs.harvard.edu/abs/2013ApJ...773...31M},
  adsnote = {Provided by the SAO/NASA Astrophysics Data System}
}

@ARTICLE{tsuboi11,
   author = {{Tsuboi}, M. and {Tadaki}, K.-I. and {Miyazaki}, A. and {Handa}, T.
	},
    title = "{Sagittarius A Molecular Cloud Complex in H$^{13}$CO$^{+}$ and Thermal SiO Emission Lines}",
  journal = {\pasj},
 keywords = {Galaxy: center, ISM: clouds, ISM: magnetic fileds, ISM: supernova remnants},
     year = 2011,
    month = aug,
   volume = 63,
    pages = {763-},
      doi = {10.1093/pasj/63.4.763},
   adsurl = {http://adsabs.harvard.edu/abs/2011PASJ...63..763T},
  adsnote = {Provided by the SAO/NASA Astrophysics Data System}
}

@ARTICLE{tsuboi99,
   author = {{Tsuboi}, M. and {Handa}, T. and {Ukita}, N.},
    title = "{Dense Molecular Clouds in the Galactic Center Region. I. Observations and Data}",
  journal = {\apjs},
 keywords = {ATLASES, GALAXY: CENTER, ISM: CLOUDS, ISM: STRUCTURE, Atlases, Galaxy: Center, ISM: Clouds, ISM: Structure},
     year = 1999,
    month = jan,
   volume = 120,
    pages = {1-39},
      doi = {10.1086/313165},
   adsurl = {http://adsabs.harvard.edu/abs/1999ApJS..120....1T},
  adsnote = {Provided by the SAO/NASA Astrophysics Data System}
}

@ARTICLE{sofue95,
   author = {{Sofue}, Y.},
    title = "{Galactic-Center Molecular Arms, Ring, and Expanding Shell. I. Kinematical Structures in Longitude--Velocity Diagrams}",
  journal = {\pasj},
   eprint = {astro-ph/9508110},
 keywords = {GALAXY: CENTER, GALAXY: KINEMATICS AND DYNAMICS, GALAXY: STRUCTURE, ISM: CLOUDS, ISM: MOLECULES},
     year = 1995,
    month = oct,
   volume = 47,
    pages = {527-549},
   adsurl = {http://adsabs.harvard.edu/abs/1995PASJ...47..527S},
  adsnote = {Provided by the SAO/NASA Astrophysics Data System}
}

@ARTICLE{koyama96,
   author = {{Koyama}, K. and {Maeda}, Y. and {Sonobe}, T. and {Takeshima}, T. and 
	{Tanaka}, Y. and {Yamauchi}, S.},
    title = "{ASCA View of Our Galactic Center: Remains of Past Activities in X-Rays?}",
  journal = {\pasj},
 keywords = {GALAXIES: MILKY WAY, INTERSTELLAR: CLOUDS, X-RAYS: SOURCES, X-RAYS: SPECTRA},
     year = 1996,
    month = apr,
   volume = 48,
    pages = {249-255},
      doi = {10.1093/pasj/48.2.249},
   adsurl = {http://adsabs.harvard.edu/abs/1996PASJ...48..249K},
  adsnote = {Provided by the SAO/NASA Astrophysics Data System}
}

@ARTICLE{morris96,
   author = {{Morris}, M. and {Serabyn}, E.},
    title = "{The Galactic Center Environment}",
  journal = {\araa},
     year = 1996,
   volume = 34,
    pages = {645-702},
      doi = {10.1146/annurev.astro.34.1.645},
   adsurl = {http://adsabs.harvard.edu/abs/1996ARA
  adsnote = {Provided by the SAO/NASA Astrophysics Data System}
}

@ARTICLE{sofue96,
   author = {{Sofue}, Y.},
    title = "{High-Velocity Molecular Gas in the Galactic Center Radio Lobe}",
  journal = {\apjl},
   eprint = {astro-ph/9512162},
 keywords = {GALAXY: CENTER, ISM: MOLECULES, GALAXY: KINEMATICS AND DYNAMICS, ISM: JETS AND OUTFLOWS},
     year = 1996,
    month = mar,
   volume = 459,
    pages = {L69},
      doi = {10.1086/309946},
   adsurl = {http://adsabs.harvard.edu/abs/1996ApJ...459L..69S},
  adsnote = {Provided by the SAO/NASA Astrophysics Data System}
}

@ARTICLE{mori09,
   author = {{Mori}, H. and {Hyodo}, Y. and {Tsuru}, T.~G. and {Nobukawa}, M. and 
	{Koyama}, K.},
    title = "{A Super Bubble Candidate in the Galactic Center and a Local Enhancement G359.77-0.09}",
  journal = {\pasj},
archivePrefix = "arXiv",
   eprint = {0905.2725},
 primaryClass = "astro-ph.HE",
 keywords = {Galaxy: center, ISM: individual (G359.77-0.09), ISM: supernova remnants, X-rays: ISM},
     year = 2009,
    month = aug,
   volume = 61,
    pages = {687-},
   adsurl = {http://adsabs.harvard.edu/abs/2009PASJ...61..687M},
  adsnote = {Provided by the SAO/NASA Astrophysics Data System}
}

@ARTICLE{gold99,
   author = {{Goldsmith}, P.~F. and {Langer}, W.~D.},
    title = "{Population Diagram Analysis of Molecular Line Emission}",
  journal = {\apj},
 keywords = {ISM: CLOUDS, ISM: MOLECULES, RADIATIVE TRANSFER, ISM: Clouds, ISM: Molecules, Radiative Transfer},
     year = 1999,
    month = may,
   volume = 517,
    pages = {209-225},
      doi = {10.1086/307195},
   adsurl = {http://adsabs.harvard.edu/abs/1999ApJ...517..209G},
  adsnote = {Provided by the SAO/NASA Astrophysics Data System}
}
@ARTICLE{shirley,
   author = {{Shirley}, Y.~L. and {Evans}, II, N.~J. and {Young}, K.~E. and 
	{Knez}, C. and {Jaffe}, D.~T.},
    title = "{A CS J=5--{\gt}4 Mapping Survey Toward High-Mass Star-forming Cores Associated with Water Masers}",
  journal = {\apjs},
   eprint = {astro-ph/0308310},
 keywords = {ISM: Dust, Extinction, ISM: Clouds, ISM: Molecules, Stars: Formation},
     year = 2003,
    month = dec,
   volume = 149,
    pages = {375-403},
      doi = {10.1086/379147},
   adsurl = {http://adsabs.harvard.edu/abs/2003ApJS..149..375S},
  adsnote = {Provided by the SAO/NASA Astrophysics Data System}
}

@ARTICLE{lamda,
   author = {{Sch{\"o}ier}, F.~L. and {van der Tak}, F.~F.~S. and {van Dishoeck}, E.~F. and 
	{Black}, J.~H.},
    title = "{An atomic and molecular database for analysis of submillimetre line observations}",
  journal = {\aap},
   eprint = {astro-ph/0411110},
 keywords = {astronomical data bases: miscellaneous, atomic data, molecular data, radiative transfer, ISM: atoms, ISM: molecules},
     year = 2005,
    month = mar,
   volume = 432,
    pages = {369-379},
      doi = {10.1051/0004-6361:20041729},
   adsurl = {http://esoads.eso.org/abs/2005A
  adsnote = {Provided by the SAO/NASA Astrophysics Data System}
}

@ARTICLE{lique,
   author = {{Lique}, F. and {Spielfiedel}, A. and {Cernicharo}, J.},
    title = "{Rotational excitation of carbon monosulfide by collisions with helium}",
  journal = {\aap},
 keywords = {molecular processes, molecular data},
     year = 2006,
    month = jun,
   volume = 451,
    pages = {1125-1132},
      doi = {10.1051/0004-6361:20054363},
   adsurl = {http://esoads.eso.org/abs/2006A
  adsnote = {Provided by the SAO/NASA Astrophysics Data System}
}

@ARTICLE{radex,
   author = {{van der Tak}, F.~F.~S. and {Black}, J.~H. and {Sch{\"o}ier}, F.~L. and 
	{Jansen}, D.~J. and {van Dishoeck}, E.~F.},
    title = "{A computer program for fast non-LTE analysis of interstellar line spectra. With diagnostic plots to interpret observed line intensity ratios}",
  journal = {\aap},
archivePrefix = "arXiv",
   eprint = {0704.0155},
 keywords = {radiative transfer, methods: numerical, radio lines: ISM, infrared: ISM, submillimeter},
     year = 2007,
    month = jun,
   volume = 468,
    pages = {627-635},
      doi = {10.1051/0004-6361:20066820},
   adsurl = {http://adsabs.harvard.edu/abs/2007A
  adsnote = {Provided by the SAO/NASA Astrophysics Data System}
}

@ARTICLE{goldreich,
   author = {{Goldreich}, P. and {Kwan}, J.},
    title = "{Molecular Clouds}",
  journal = {\apj},
     year = 1974,
    month = may,
   volume = 189,
    pages = {441-454},
      doi = {10.1086/152821},
   adsurl = {http://adsabs.harvard.edu/abs/1974ApJ...189..441G},
  adsnote = {Provided by the SAO/NASA Astrophysics Data System}
}


@ARTICLE{ic342,
   author = {{Schinnerer}, E. and {B{\"o}ker}, T. and {Meier}, D.~S. and 
	{Calzetti}, D.},
    title = "{Self-Regulated Fueling of Galaxy Centers: Evidence for Star Formation Feedback in IC 342's Nucleus}",
  journal = {\apjl},
archivePrefix = "arXiv",
   eprint = {0808.0793},
 keywords = {Galaxies: Individual: Alphanumeric: IC 342, Galaxies: ISM, Galaxies: Kinematics and Dynamics, Galaxies: Nuclei},
     year = 2008,
    month = sep,
   volume = 684,
    pages = {L21-L24},
      doi = {10.1086/592109},
   adsurl = {http://adsabs.harvard.edu/abs/2008ApJ...684L..21S},
  adsnote = {Provided by the SAO/NASA Astrophysics Data System}
}

@ARTICLE{ho14,
   author = {{Ho}, I.-T. and {Kewley}, L.~J. and {Dopita}, M.~A. and {Medling}, A.~M. and 
	{Allen}, J.~T. and {Bland-Hawthorn}, J. and {Bloom}, J.~V. and 
	{Bryant}, J.~J. and {Croom}, S.~M. and {Fogarty}, L.~M.~R. and 
	{Goodwin}, M. and {Green}, A.~W. and {Konstantopoulos}, I.~S. and 
	{Lawrence}, J.~S. and {L{\'o}pez-S{\'a}nchez}, {\'A}.~R. and 
	{Owers}, M.~S. and {Richards}, S. and {Sharp}, R.},
    title = "{The SAMI Galaxy Survey: shocks and outflows in a normal star-forming galaxy}",
  journal = {\mnras},
archivePrefix = "arXiv",
   eprint = {1407.2411},
 keywords = {ISM: jets and outflows, galaxies: evolution, galaxies: kinematics and dynamics, galaxies: starburst},
     year = 2014,
    month = nov,
   volume = 444,
    pages = {3894-3910},
      doi = {10.1093/mnras/stu1653},
   adsurl = {http://adsabs.harvard.edu/abs/2014MNRAS.444.3894H},
  adsnote = {Provided by the SAO/NASA Astrophysics Data System}
}

@ARTICLE{rubin14,
   author = {{Rubin}, K.~H.~R. and {Prochaska}, J.~X. and {Koo}, D.~C. and 
	{Phillips}, A.~C. and {Martin}, C.~L. and {Winstrom}, L.~O.},
    title = "{Evidence for Ubiquitous Collimated Galactic-scale Outflows along the Star-forming Sequence at z \~{} 0.5}",
  journal = {\apj},
archivePrefix = "arXiv",
   eprint = {1307.1476},
 primaryClass = "astro-ph.CO",
 keywords = {galaxies: evolution, galaxies: halos, galaxies: ISM, ultraviolet: ISM },
     year = 2014,
    month = oct,
   volume = 794,
      eid = {156},
    pages = {156},
      doi = {10.1088/0004-637X/794/2/156},
   adsurl = {http://adsabs.harvard.edu/abs/2014ApJ...794..156R},
  adsnote = {Provided by the SAO/NASA Astrophysics Data System}
}

@ARTICLE{chen10,
   author = {{Chen}, Y.-M. and {Tremonti}, C.~A. and {Heckman}, T.~M. and 
	{Kauffmann}, G. and {Weiner}, B.~J. and {Brinchmann}, J. and 
	{Wang}, J.},
    title = "{Absorption-line Probes of the Prevalence and Properties of Outflows in Present-day Star-forming Galaxies}",
  journal = {\aj},
archivePrefix = "arXiv",
   eprint = {1003.5425},
 primaryClass = "astro-ph.GA",
 keywords = {galaxies: evolution, galaxies: star formation},
     year = 2010,
    month = aug,
   volume = 140,
    pages = {445-461},
      doi = {10.1088/0004-6256/140/2/445},
   adsurl = {http://adsabs.harvard.edu/abs/2010AJ....140..445C},
  adsnote = {Provided by the SAO/NASA Astrophysics Data System}
}

@ARTICLE{murray11,
   author = {{Murray}, N. and {M{\'e}nard}, B. and {Thompson}, T.~A.},
    title = "{Radiation Pressure from Massive Star Clusters as a Launching Mechanism for Super-galactic Winds}",
  journal = {\apj},
archivePrefix = "arXiv",
   eprint = {1005.4419},
 primaryClass = "astro-ph.CO",
 keywords = {dust, extinction, galaxies: star clusters: general, intergalactic medium, ISM: structure, quasars: absorption lines},
     year = 2011,
    month = jul,
   volume = 735,
      eid = {66},
    pages = {66},
      doi = {10.1088/0004-637X/735/1/66},
   adsurl = {http://adsabs.harvard.edu/abs/2011ApJ...735...66M},
  adsnote = {Provided by the SAO/NASA Astrophysics Data System}
}

@ARTICLE{sharp10,
   author = {{Sharp}, R.~G. and {Bland-Hawthorn}, J.},
    title = "{Three-Dimensional Integral Field Observations of 10 Galactic Winds. I. Extended Phase (gsim10 Myr) of Mass/Energy Injection Before the Wind Blows}",
  journal = {\apj},
archivePrefix = "arXiv",
   eprint = {1001.4315},
 primaryClass = "astro-ph.CO",
 keywords = {galaxies: individual: NGC 253 NGC 1365 NGC 1482 NGC 1808 NGC 3628 NGC 5128 Circinus NGC 6240 NGC 6810 IC 5063},
     year = 2010,
    month = mar,
   volume = 711,
    pages = {818-852},
      doi = {10.1088/0004-637X/711/2/818},
   adsurl = {http://adsabs.harvard.edu/abs/2010ApJ...711..818S},
  adsnote = {Provided by the SAO/NASA Astrophysics Data System}
}

@ARTICLE{bland03,
   author = {{Bland-Hawthorn}, J. and {Cohen}, M.},
    title = "{The Large-Scale Bipolar Wind in the Galactic Center}",
  journal = {\apj},
   eprint = {astro-ph/0208553},
 keywords = {Galaxy: Center, Galaxy: Halo, ISM: Jets and Outflows},
     year = 2003,
    month = jan,
   volume = 582,
    pages = {246-256},
      doi = {10.1086/344573},
   adsurl = {http://adsabs.harvard.edu/abs/2003ApJ...582..246B},
  adsnote = {Provided by the SAO/NASA Astrophysics Data System}
}

@ARTICLE{keeney06,
   author = {{Keeney}, B.~A. and {Danforth}, C.~W. and {Stocke}, J.~T. and 
	{Penton}, S.~V. and {Shull}, J.~M. and {Sembach}, K.~R.},
    title = "{Does the Milky Way Produce a Nuclear Galactic Wind?}",
  journal = {\apj},
   eprint = {astro-ph/0604323},
 keywords = {Galaxy: Center, Galaxies: Intergalactic Medium, ISM: Clouds, ISM: Jets and Outflows, Galaxies: Quasars: Absorption Lines},
     year = 2006,
    month = aug,
   volume = 646,
    pages = {951-964},
      doi = {10.1086/505128},
   adsurl = {http://adsabs.harvard.edu/abs/2006ApJ...646..951K},
  adsnote = {Provided by the SAO/NASA Astrophysics Data System}
}

@ARTICLE{wind05,
   author = {{Veilleux}, S. and {Cecil}, G. and {Bland-Hawthorn}, J.},
    title = "{Galactic Winds}",
  journal = {\araa},
   eprint = {astro-ph/0504435},
     year = 2005,
    month = sep,
   volume = 43,
    pages = {769-826},
      doi = {10.1146/annurev.astro.43.072103.150610},
   adsurl = {http://adsabs.harvard.edu/abs/2005ARA
  adsnote = {Provided by the SAO/NASA Astrophysics Data System}
}

@ARTICLE{nh3_05,
   author = {{Herrnstein}, R.~M. and {Ho}, P.~T.~P.},
    title = "{The Nature of the Molecular Environment within 5 Parsecs of the Galactic Center}",
  journal = {\apj},
   eprint = {astro-ph/0409271},
 keywords = {Galaxy: Center, ISM: Clouds, ISM: Molecules, Radio Lines: ISM},
     year = 2005,
    month = feb,
   volume = 620,
    pages = {287-307},
      doi = {10.1086/426047},
   adsurl = {http://adsabs.harvard.edu/abs/2005ApJ...620..287H},
  adsnote = {Provided by the SAO/NASA Astrophysics Data System}
}

@ARTICLE{sofue84,
   author = {{Sofue}, Y. and {Handa}, T.},
    title = "{A radio lobe over the galactic centre}",
  journal = {\nat},
 keywords = {Galactic Nuclei, Galactic Radio Waves, Galactic Structure, Milky Way Galaxy, Radio Sources (Astronomy), Astronomical Maps, Brightness Distribution, Continuous Spectra, Hydrogen Clouds, Nebulae},
     year = 1984,
    month = aug,
   volume = 310,
    pages = {568},
      doi = {10.1038/310568a0},
   adsurl = {http://adsabs.harvard.edu/abs/1984Natur.310..568S},
  adsnote = {Provided by the SAO/NASA Astrophysics Data System}
}

@ARTICLE{wu14,
   author = {{Wu}, P.-F. and {Tully}, R.~B. and {Rizzi}, L. and {Dolphin}, A.~E. and 
	{Jacobs}, B.~A. and {Karachentsev}, I.~D.},
    title = "{Infrared Tip of the Red Giant Branch and Distances to the Maffei/IC 342 Group}",
  journal = {\aj},
archivePrefix = "arXiv",
   eprint = {1404.2987},
 keywords = {galaxies: distances and redshifts, galaxies: stellar content, stars: Population II},
     year = 2014,
    month = jul,
   volume = 148,
      eid = {7},
    pages = {7},
      doi = {10.1088/0004-6256/148/1/7},
   adsurl = {http://adsabs.harvard.edu/abs/2014AJ....148....7W},
  adsnote = {Provided by the SAO/NASA Astrophysics Data System}
}

@ARTICLE{ngc253,
   author = {{Bolatto}, A.~D. and {Warren}, S.~R. and {Leroy}, A.~K. and 
	{Walter}, F. and {Veilleux}, S. and {Ostriker}, E.~C. and {Ott}, J. and 
	{Zwaan}, M. and {Fisher}, D.~B. and {Weiss}, A. and {Rosolowsky}, E. and 
	{Hodge}, J.},
    title = "{Suppression of star formation in the galaxy NGC 253 by a starburst-driven molecular wind}",
  journal = {\nat},
     year = 2013,
    month = jul,
   volume = 499,
    pages = {450-453},
      doi = {10.1038/nature12351},
   adsurl = {http://adsabs.harvard.edu/abs/2013Natur.499..450B},
  adsnote = {Provided by the SAO/NASA Astrophysics Data System}
}

@ARTICLE{ohyama02,
   author = {{Ohyama}, Y. and {Taniguchi}, Y. and {Iye}, M. and {Yoshida}, M. and 
	{Sekiguchi}, K. and {Takata}, T. and {Saito}, Y. and {Kawabata}, K.~S. and 
	{Kashikawa}, N. and {Aoki}, K. and {Sasaki}, T. and {Kosugi}, G. and 
	{Okita}, K. and {Shimizu}, Y. and {Inata}, M. and {Ebizuka}, N. and 
	{Ozawa}, T. and {Yadoumaru}, Y. and {Taguchi}, H. and {Asai}, R.
	},
    title = "{Decomposition of the Superwind in M 82}",
  journal = {\pasj},
   eprint = {astro-ph/0209442},
 keywords = {galaxies: individual (M 82), galaxies: intergalactic medium, galaxies: starburst, medium, shock waves},
     year = 2002,
    month = dec,
   volume = 54,
    pages = {891-898},
      doi = {10.1093/pasj/54.6.891},
   adsurl = {http://adsabs.harvard.edu/abs/2002PASJ...54..891O},
  adsnote = {Provided by the SAO/NASA Astrophysics Data System}
}

@INPROCEEDINGS{simpson99,
   author = {{Simpson}, J.~P. and {Witteborn}, F.~C. and {Cohen}, M. and 
	{Price}, S.~D.},
    title = "{The Mid-Infrared Spectrum of the Galactic Center, a Starburst Nucleus}",
booktitle = {The Central Parsecs of the Galaxy},
     year = 1999,
   series = {Astronomical Society of the Pacific Conference Series},
   volume = 186,
   editor = {{Falcke}, H. and {Cotera}, A. and {Duschl}, W.~J. and {Melia}, F. and 
	{Rieke}, M.~J.},
    month = jun,
    pages = {527},
   adsurl = {http://adsabs.harvard.edu/abs/1999ASPC..186..527S},
  adsnote = {Provided by the SAO/NASA Astrophysics Data System}
}

@ARTICLE{sakamoto06,
   author = {{Sakamoto}, K. and {Ho}, P.~T.~P. and {Iono}, D. and {Keto}, E.~R. and 
	{Mao}, R.-Q. and {Matsushita}, S. and {Peck}, A.~B. and {Wiedner}, M.~C. and 
	{Wilner}, D.~J. and {Zhao}, J.-H.},
    title = "{Molecular Superbubbles in the Starburst Galaxy NGC 253}",
  journal = {\apj},
   eprint = {astro-ph/0509430},
 keywords = {Galaxies: Individual: NGC Number: NGC 253, Galaxies: ISM, Galaxies: Starburst, ISM: Bubbles},
     year = 2006,
    month = jan,
   volume = 636,
    pages = {685-697},
      doi = {10.1086/498075},
   adsurl = {http://adsabs.harvard.edu/abs/2006ApJ...636..685S},
  adsnote = {Provided by the SAO/NASA Astrophysics Data System}
}

@ARTICLE{sakamoto11,
   author = {{Sakamoto}, K. and {Mao}, R.-Q. and {Matsushita}, S. and {Peck}, A.~B. and 
	{Sawada}, T. and {Wiedner}, M.~C.},
    title = "{Star-forming Cloud Complexes in the Central Molecular Zone of NGC 253}",
  journal = {\apj},
archivePrefix = "arXiv",
   eprint = {1104.2388},
 primaryClass = "astro-ph.CO",
 keywords = {galaxies: individual: NGC 253, galaxies: ISM, galaxies: starburst},
     year = 2011,
    month = jul,
   volume = 735,
      eid = {19},
    pages = {19},
      doi = {10.1088/0004-637X/735/1/19},
   adsurl = {http://adsabs.harvard.edu/abs/2011ApJ...735...19S},
  adsnote = {Provided by the SAO/NASA Astrophysics Data System}
}

@ARTICLE{mccray87,
   author = {{McCray}, R. and {Kafatos}, M.},
    title = "{Supershells and propagating star formation}",
  journal = {\apj},
 keywords = {Interstellar Matter, Star Formation, Stellar Envelopes, Stellar Coronas, Stellar Winds, Supernovae},
     year = 1987,
    month = jun,
   volume = 317,
    pages = {190-196},
      doi = {10.1086/165267},
   adsurl = {http://adsabs.harvard.edu/abs/1987ApJ...317..190M},
  adsnote = {Provided by the SAO/NASA Astrophysics Data System}
}

@ARTICLE{hsieh15,
   author = {{Hsieh}, P.~Y. and {Ho}, P.~T.~P. and {Hwang}, C.~Y.},
  journal = {Paper II, in prep.},
     year = 2015
}

@ARTICLE{yusef04,
   author = {{Yusef-Zadeh}, F. and {Hewitt}, J.~W. and {Cotton}, W.},
    title = "{A 20 Centimeter Survey of the Galactic Center Region. I. Detection of Numerous Linear Filaments}",
  journal = {\apjs},
   eprint = {astro-ph/0409292},
 keywords = {Galaxy: Center, ISM: Clouds, ISM: General, Shock Waves, ISM: Supernova Remnants},
     year = 2004,
    month = dec,
   volume = 155,
    pages = {421-550},
      doi = {10.1086/425257},
   adsurl = {http://adsabs.harvard.edu/abs/2004ApJS..155..421Y},
  adsnote = {Provided by the SAO/NASA Astrophysics Data System}
}

@ARTICLE{pedlar89,
   author = {{Pedlar}, A. and {Anantharamaiah}, K.~R. and {Ekers}, R.~D. and 
	{Goss}, W.~M. and {van Gorkom}, J.~H. and {Schwarz}, U.~J. and 
	{Zhao}, J.-H.},
    title = "{Radio studies of the Galactic center. I - The Sagittarius A complex}",
  journal = {\apj},
 keywords = {Centimeter Waves, Galactic Nuclei, Galactic Structure, Milky Way Galaxy, Radio Astronomy, Active Galactic Nuclei, Sagittarius Constellation, Star Formation, Supernovae, Very Large Array (Vla)},
     year = 1989,
    month = jul,
   volume = 342,
    pages = {769-784},
      doi = {10.1086/167635},
   adsurl = {http://adsabs.harvard.edu/abs/1989ApJ...342..769P},
  adsnote = {Provided by the SAO/NASA Astrophysics Data System}
}

@INPROCEEDINGS{zhao14,
   author = {{Zhao}, J.-H. and {Morris}, M.~R. and {Goss}, W.~M.},
    title = "{A new perspective on the radio active zone at the Galactic center - feedback from nuclear activities}",
 keywords = {Galactic center, ISM, supernova remnants, neutron stars, winds, outflows},
booktitle = {IAU Symposium},
     year = 2014,
   series = {IAU Symposium},
   volume = 303,
archivePrefix = "arXiv",
   eprint = {1311.5512},
 primaryClass = "astro-ph.GA",
   editor = {{Sjouwerman}, L.~O. and {Lang}, C.~C. and {Ott}, J.},
    month = may,
    pages = {364-368},
      doi = {10.1017/S1743921314000921},
   adsurl = {http://adsabs.harvard.edu/abs/2014IAUS..303..364Z},
  adsnote = {Provided by the SAO/NASA Astrophysics Data System}
}

@ARTICLE{weaver77,
   author = {{Weaver}, R. and {McCray}, R. and {Castor}, J. and {Shapiro}, P. and 
	{Moore}, R.},
    title = "{Interstellar bubbles. II - Structure and evolution}",
  journal = {\apj},
 keywords = {Interstellar Gas, Plasma Interactions, Stellar Winds, Adiabatic Flow, Bubbles, Conductive Heat Transfer, Early Stars, Energy Dissipation, Hydrodynamics, Ion Density (Concentration), Radiative Transfer, Stellar Motions},
     year = 1977,
    month = dec,
   volume = 218,
    pages = {377-395},
      doi = {10.1086/155692},
   adsurl = {http://adsabs.harvard.edu/abs/1977ApJ...218..377W},
  adsnote = {Provided by the SAO/NASA Astrophysics Data System}
}

@ARTICLE{yusef09,
   author = {{Yusef-Zadeh}, F. and {Hewitt}, J.~W. and {Arendt}, R.~G. and 
	{Whitney}, B. and {Rieke}, G. and {Wardle}, M. and {Hinz}, J.~L. and 
	{Stolovy}, S. and {Lang}, C.~C. and {Burton}, M.~G. and {Ramirez}, S.
	},
    title = "{Star Formation in the Central 400 pc of the Milky Way: Evidence for a Population of Massive Young Stellar Objects}",
  journal = {\apj},
archivePrefix = "arXiv",
   eprint = {0905.2161},
 primaryClass = "astro-ph.GA",
 keywords = {galaxies: starburst, Galaxy: center, ISM: clouds, masers, stars: formation},
     year = 2009,
    month = sep,
   volume = 702,
    pages = {178-225},
      doi = {10.1088/0004-637X/702/1/178},
   adsurl = {http://adsabs.harvard.edu/abs/2009ApJ...702..178Y},
  adsnote = {Provided by the SAO/NASA Astrophysics Data System}
}

@ARTICLE{condon92,
   author = {{Condon}, J.~J.},
    title = "{Radio emission from normal galaxies}",
  journal = {\araa},
 keywords = {Cosmic Rays, Radio Emission, Radio Sources (Astronomy), Star Formation, Starburst Galaxies, Synchrotron Radiation, Black Holes (Astronomy), H Ii Regions, Infrared Radiation, Relativistic Particles},
     year = 1992,
   volume = 30,
    pages = {575-611},
      doi = {10.1146/annurev.aa.30.090192.003043},
   adsurl = {http://adsabs.harvard.edu/abs/1992ARA
  adsnote = {Provided by the SAO/NASA Astrophysics Data System}
}

@ARTICLE{gill09,
   author = {{Gillessen}, S. and {Eisenhauer}, F. and {Trippe}, S. and {Alexander}, T. and 
	{Genzel}, R. and {Martins}, F. and {Ott}, T.},
    title = "{Monitoring Stellar Orbits Around the Massive Black Hole in the Galactic Center}",
  journal = {\apj},
archivePrefix = "arXiv",
   eprint = {0810.4674},
 keywords = {black hole physics, astrometry, Galaxy: center, infrared: stars},
     year = 2009,
    month = feb,
   volume = 692,
    pages = {1075-1109},
      doi = {10.1088/0004-637X/692/2/1075},
   adsurl = {http://adsabs.harvard.edu/abs/2009ApJ...692.1075G},
  adsnote = {Provided by the SAO/NASA Astrophysics Data System}
}

@ARTICLE{reid04,
   author = {{Reid}, M.~J. and {Brunthaler}, A.},
    title = "{The Proper Motion of Sagittarius A*. II. The Mass of Sagittarius A*}",
  journal = {\apj},
   eprint = {astro-ph/0408107},
 keywords = {Astrometry, Black Hole Physics, Galaxy: Center, Galaxy: Fundamental Parameters, Galaxy: Structure},
     year = 2004,
    month = dec,
   volume = 616,
    pages = {872-884},
      doi = {10.1086/424960},
   adsurl = {http://adsabs.harvard.edu/abs/2004ApJ...616..872R},
  adsnote = {Provided by the SAO/NASA Astrophysics Data System}
}

@ARTICLE{bania77,
   author = {{Bania}, T.~M.},
    title = "{Carbon monoxide in the inner Galaxy}",
  journal = {\apj},
 keywords = {Carbon Monoxide, Galactic Structure, Interstellar Gas, Milky Way Galaxy, Spatial Distribution, Astronomical Maps, Emission Spectra, Galactic Nuclei, Hydrogen, Kinematics, Radio Sources (Astronomy)},
     year = 1977,
    month = sep,
   volume = 216,
    pages = {381-403},
      doi = {10.1086/155478},
   adsurl = {http://adsabs.harvard.edu/abs/1977ApJ...216..381B},
  adsnote = {Provided by the SAO/NASA Astrophysics Data System}
}

@ARTICLE{oka05,
   author = {{Oka}, T. and {Geballe}, T.~R. and {Goto}, M. and {Usuda}, T. and 
	{McCall}, B.~J.},
    title = "{Hot and Diffuse Clouds near the Galactic Center Probed by Metastable H$^{+}$$_{3}$1,}",
  journal = {\apj},
   eprint = {astro-ph/0507463},
 keywords = {Astrochemistry, Galaxy: Center, ISM: Clouds, ISM: Molecules, Molecular Processes, Radiation Mechanisms: Nonthermal},
     year = 2005,
    month = oct,
   volume = 632,
    pages = {882-893},
      doi = {10.1086/432679},
   adsurl = {http://adsabs.harvard.edu/abs/2005ApJ...632..882O},
  adsnote = {Provided by the SAO/NASA Astrophysics Data System}
}

@ARTICLE{martin04,
   author = {{Martin}, C.~L. and {Walsh}, W.~M. and {Xiao}, K. and {Lane}, A.~P. and 
	{Walker}, C.~K. and {Stark}, A.~A.},
    title = "{The AST/RO Survey of the Galactic Center Region. I. The Inner 3 Degrees}",
  journal = {\apjs},
   eprint = {astro-ph/0211025},
 keywords = {Galaxy: Center, Galaxy: Kinematics and Dynamics, ISM: Atoms, ISM: Molecules, Radio Lines: ISM, Surveys},
     year = 2004,
    month = jan,
   volume = 150,
    pages = {239-262},
      doi = {10.1086/379661},
   adsurl = {http://adsabs.harvard.edu/abs/2004ApJS..150..239M},
  adsnote = {Provided by the SAO/NASA Astrophysics Data System}
}

@ARTICLE{oka12,
   author = {{Oka}, T. and {Onodera}, Y. and {Nagai}, M. and {Tanaka}, K. and 
	{Matsumura}, S. and {Kamegai}, K.},
    title = "{ASTE CO J = 3-2 Survey of the Galactic Center}",
  journal = {\apjs},
 keywords = {Galaxy: center, galaxies: nuclei, ISM: clouds},
     year = 2012,
    month = aug,
   volume = 201,
      eid = {14},
    pages = {14},
      doi = {10.1088/0067-0049/201/2/14},
   adsurl = {http://adsabs.harvard.edu/abs/2012ApJS..201...14O},
  adsnote = {Provided by the SAO/NASA Astrophysics Data System}
}


@ARTICLE{guesten87,
   author = {{Guesten}, R. and {Genzel}, R. and {Wright}, M.~C.~H. and {Jaffe}, D.~T. and 
	{Stutzki}, J. and {Harris}, A.~I.},
    title = "{Aperture synthesis observations of the circumnuclear ring in the Galactic center}",
  journal = {\apj},
 keywords = {Astronomical Spectroscopy, Galactic Nuclei, Milky Way Galaxy, Molecular Gases, Neutral Gases, Synthetic Apertures, Absorption Spectra, Carbon Monoxide, Emission Spectra, Hydrocyanic Acid, Interstellar Gas, Radio Astronomy},
     year = 1987,
    month = jul,
   volume = 318,
    pages = {124-138},
      doi = {10.1086/165355},
   adsurl = {http://adsabs.harvard.edu/abs/1987ApJ...318..124G},
  adsnote = {Provided by the SAO/NASA Astrophysics Data System}
}

@ARTICLE{jackson93,
   author = {{Jackson}, J.~M. and {Geis}, N. and {Genzel}, R. and {Harris}, A.~I. and 
	{Madden}, S. and {Poglitsch}, A. and {Stacey}, G.~J. and {Townes}, C.~H.
	},
    title = "{Neutral gas in the central 2 parsecs of the Galaxy}",
  journal = {\apj},
 keywords = {Galactic Nuclei, Infrared Astronomy, Interstellar Gas, Neutral Gases, Gas Dynamics, Mass Flow Rate, Molecular Clouds, Spatial Distribution},
     year = 1993,
    month = jan,
   volume = 402,
    pages = {173-184},
      doi = {10.1086/172120},
   adsurl = {http://adsabs.harvard.edu/abs/1993ApJ...402..173J},
  adsnote = {Provided by the SAO/NASA Astrophysics Data System}
}

@ARTICLE{amo11,
   author = {{Amo-Baladr{\'o}n}, M.~A. and {Mart{\'{\i}}n-Pintado}, J. and 
	{Mart{\'{\i}}n}, S.},
    title = "{Mapping photodissociation and shocks in the vicinity of Sagittarius A*}",
  journal = {\aap},
archivePrefix = "arXiv",
   eprint = {1011.3271},
 primaryClass = "astro-ph.GA",
 keywords = {astrochemistry, ISM: clouds, ISM: molecules, Galaxy: abundances, Galaxy: nucleus, Galaxy: center},
     year = 2011,
    month = feb,
   volume = 526,
      eid = {A54},
    pages = {A54},
      doi = {10.1051/0004-6361/200913545},
   adsurl = {http://adsabs.harvard.edu/abs/2011A
  adsnote = {Provided by the SAO/NASA Astrophysics Data System}
}

@ARTICLE{harris,
   author = {{Harris}, A.~I. and {Jaffe}, D.~T. and {Silber}, M. and {Genzel}, R.
	},
    title = "{CO 7-6 submillimeter emission from the galactic center - Warm molecular gas and the rotation curve in the central 10 parsecs}",
  journal = {\apjl},
 keywords = {Carbon Monoxide, Galactic Nuclei, Interstellar Gas, Molecular Gases, Angular Velocity, Gas Heating, Luminosity, Mass Distribution, Submillimeter Waves},
     year = 1985,
    month = jul,
   volume = 294,
    pages = {L93-L97},
      doi = {10.1086/184516},
   adsurl = {http://adsabs.harvard.edu/abs/1985ApJ...294L..93H},
  adsnote = {Provided by the SAO/NASA Astrophysics Data System}
}

@ARTICLE{mezger,
   author = {{Mezger}, P.~G. and {Zylka}, R. and {Salter}, C.~J. and {Wink}, J.~E. and 
	{Chini}, R. and {Kreysa}, E. and {Tuffs}, R.},
    title = "{Continuum observations of SGR A at mm/submm wavelengths}",
  journal = {\aap},
 keywords = {Astronomical Spectroscopy, Infrared Astronomy, Millimeter Waves, Molecular Clouds, Radio Sources (Astronomy), Submillimeter Waves, Cosmic Dust, Emission Spectra, Optical Thickness, Stellar Mass, Stellar Winds, Supernova Remnants},
     year = 1989,
    month = jan,
   volume = 209,
    pages = {337-348},
   adsurl = {http://adsabs.harvard.edu/abs/1989A
  adsnote = {Provided by the SAO/NASA Astrophysics Data System}
}

@ARTICLE{etx,
   author = {{Etxaluze}, M. and {Smith}, H.~A. and {Tolls}, V. and {Stark}, A.~A. and 
	{Gonz{\'a}lez-Alfonso}, E.},
    title = "{The Galactic Center in the Far-infrared}",
  journal = {\aj},
archivePrefix = "arXiv",
   eprint = {1108.0313},
 primaryClass = "astro-ph.GA",
 keywords = {dust, extinction, Galaxy: center, infrared: ISM, ISM: individual objects: Sgr A*},
     year = 2011,
    month = oct,
   volume = 142,
      eid = {134},
    pages = {134},
      doi = {10.1088/0004-6256/142/4/134},
   adsurl = {http://adsabs.harvard.edu/abs/2011AJ....142..134E},
  adsnote = {Provided by the SAO/NASA Astrophysics Data System}
}

@ARTICLE{molinari,
   author = {{Molinari}, S. and {Bally}, J. and {Noriega-Crespo}, A. and 
	{Compi{\`e}gne}, M. and {Bernard}, J.~P. and {Paradis}, D. and 
	{Martin}, P. and {Testi}, L. and {Barlow}, M. and {Moore}, T. and 
	{Plume}, R. and {Swinyard}, B. and {Zavagno}, A. and {Calzoletti}, L. and 
	{Di Giorgio}, A.~M. and {Elia}, D. and {Faustini}, F. and {Natoli}, P. and 
	{Pestalozzi}, M. and {Pezzuto}, S. and {Piacentini}, F. and 
	{Polenta}, G. and {Polychroni}, D. and {Schisano}, E. and {Traficante}, A. and 
	{Veneziani}, M. and {Battersby}, C. and {Burton}, M. and {Carey}, S. and 
	{Fukui}, Y. and {Li}, J.~Z. and {Lord}, S.~D. and {Morgan}, L. and 
	{Motte}, F. and {Schuller}, F. and {Stringfellow}, G.~S. and 
	{Tan}, J.~C. and {Thompson}, M.~A. and {Ward-Thompson}, D. and 
	{White}, G. and {Umana}, G.},
    title = "{A 100 pc Elliptical and Twisted Ring of Cold and Dense Molecular Clouds Revealed by Herschel Around the Galactic Center}",
  journal = {\apjl},
archivePrefix = "arXiv",
   eprint = {1105.5486},
 primaryClass = "astro-ph.GA",
 keywords = {Galaxy: center, ISM: clouds, stars: formation},
     year = 2011,
    month = jul,
   volume = 735,
      eid = {L33},
    pages = {L33},
      doi = {10.1088/2041-8205/735/2/L33},
   adsurl = {http://adsabs.harvard.edu/abs/2011ApJ...735L..33M},
  adsnote = {Provided by the SAO/NASA Astrophysics Data System}
}

@ARTICLE{mangum,
   author = {{Mangum}, J.~G.},
    title = "{Main-beam efficiency measurements of the Caltech Submillimeter Observatory}",
  journal = {\pasp},
 keywords = {Astronomical Observatories, Efficiency, Gas Giant Planets, Submillimeter Waves, Terrestrial Planets, Brightness Temperature, Calibrating, Emission Spectra, Temperature Measurement},
     year = 1993,
    month = jan,
   volume = 105,
    pages = {117-122},
      doi = {10.1086/133134},
   adsurl = {http://adsabs.harvard.edu/abs/1993PASP..105..117M},
  adsnote = {Provided by the SAO/NASA Astrophysics Data System}
}

@ARTICLE{lau,
   author = {{Lau}, R.~M. and {Herter}, T.~L. and {Morris}, M.~R. and {Becklin}, E.~E. and 
	{Adams}, J.~D.},
    title = "{SOFIA/FORCAST Imaging of the Circumnuclear Ring at the Galactic Center}",
  journal = {\apj},
archivePrefix = "arXiv",
   eprint = {1307.8443},
 primaryClass = "astro-ph.GA",
 keywords = {dust, extinction, Galaxy: center, infrared: ISM, photon-dominated region: PDR},
     year = 2013,
    month = sep,
   volume = 775,
      eid = {37},
    pages = {37},
      doi = {10.1088/0004-637X/775/1/37},
   adsurl = {http://adsabs.harvard.edu/abs/2013ApJ...775...37L},
  adsnote = {Provided by the SAO/NASA Astrophysics Data System}
}

@ARTICLE{wright,
   author = {{Wright}, M.~C.~H. and {Coil}, A.~L. and {McGary}, R.~S. and 
	{Ho}, P.~T.~P. and {Harris}, A.~I.},
    title = "{Molecular Tracers of the Central 12 Parsecs of the Galactic Center}",
  journal = {\apj},
   eprint = {astro-ph/0011331},
 keywords = {Galaxy: Center, ISM: Molecules, Radio Lines: ISM, Techniques: Interferometric},
     year = 2001,
    month = apr,
   volume = 551,
    pages = {254-268},
      doi = {10.1086/320089},
   adsurl = {http://adsabs.harvard.edu/abs/2001ApJ...551..254W},
  adsnote = {Provided by the SAO/NASA Astrophysics Data System}
}



@ARTICLE{maria,
   author = {{Montero-Casta{\~n}o}, M. and {Herrnstein}, R.~M. and {Ho}, P.~T.~P.
	},
    title = "{Gas Infall Toward Sgr A* from the Clumpy Circumnuclear Disk}",
  journal = {\apj},
archivePrefix = "arXiv",
   eprint = {0903.0886},
 primaryClass = "astro-ph.GA",
 keywords = {ISM: clouds, ISM: molecules, Galaxy: center, radio lines: ISM},
     year = 2009,
    month = apr,
   volume = 695,
    pages = {1477-1494},
      doi = {10.1088/0004-637X/695/2/1477},
   adsurl = {http://adsabs.harvard.edu/abs/2009ApJ...695.1477M},
  adsnote = {Provided by the SAO/NASA Astrophysics Data System}
}

@ARTICLE{herrnstein,
   author = {{Herrnstein}, R.~M. and {Ho}, P.~T.~P.},
    title = "{The Nature of the Molecular Environment within 5 Parsecs of the Galactic Center}",
  journal = {\apj},
   eprint = {astro-ph/0409271},
 keywords = {Galaxy: Center, ISM: Clouds, ISM: Molecules, Radio Lines: ISM},
     year = 2005,
    month = feb,
   volume = 620,
    pages = {287-307},
      doi = {10.1086/426047},
   adsurl = {http://adsabs.harvard.edu/abs/2005ApJ...620..287H},
  adsnote = {Provided by the SAO/NASA Astrophysics Data System}
}


@ARTICLE{chris,
   author = {{Christopher}, M.~H. and {Scoville}, N.~Z. and {Stolovy}, S.~R. and 
	{Yun}, M.~S.},
    title = "{HCN and HCO$^{+}$ Observations of the Galactic Circumnuclear Disk}",
  journal = {\apj},
   eprint = {astro-ph/0502532},
 keywords = {Galaxy: Center, ISM: Kinematics and Dynamics, ISM: Molecules, Radio Continuum: ISM, Radio Lines: ISM, Stars: Formation},
     year = 2005,
    month = mar,
   volume = 622,
    pages = {346-365},
      doi = {10.1086/427911},
   adsurl = {http://adsabs.harvard.edu/abs/2005ApJ...622..346C},
  adsnote = {Provided by the SAO/NASA Astrophysics Data System}
}

@ARTICLE{sco72,
   author = {{Scoville}, N.~Z.},
    title = "{Kinematics of Molecular Clouds Near the Galactic Center}",
  journal = {\apjl},
     year = 1972,
    month = aug,
   volume = 175,
    pages = {L127},
      doi = {10.1086/181000},
   adsurl = {http://adsabs.harvard.edu/abs/1972ApJ...175L.127S},
  adsnote = {Provided by the SAO/NASA Astrophysics Data System}
}

@ARTICLE{tsuboi09,
   author = {{Tsuboi}, M. and {Miyazaki}, A. and {Okumura}, S.~K.},
    title = "{A Galactic Center 50-km s$^{-1}$ Molecular Cloud with an Expanding Shell}",
  journal = {\pasj},
 keywords = {Galaxy, molecular cloud, supernova remnant},
     year = 2009,
    month = feb,
   volume = 61,
    pages = {29-},
      doi = {10.1093/pasj/61.1.29},
   adsurl = {http://adsabs.harvard.edu/abs/2009PASJ...61...29T},
  adsnote = {Provided by the SAO/NASA Astrophysics Data System}
}

@ARTICLE{ho91,
   author = {{Ho}, P.~T.~P. and {Ho}, L.~C. and {Szczepanski}, J.~C. and 
	{Jackson}, J.~M. and {Armstrong}, J.~T.},
    title = "{A molecular gas streamer feeding the Galactic Centre}",
  journal = {\nat},
 keywords = {Galactic Nuclei, Galactic Structure, Milky Way Galaxy, Molecular Gases, Supernovae, Ammonia, Gravitational Fields, Molecular Clouds, Radio Astronomy, Supernova Remnants},
     year = 1991,
    month = mar,
   volume = 350,
    pages = {309-312},
      doi = {10.1038/350309a0},
   adsurl = {http://adsabs.harvard.edu/abs/1991Natur.350..309H},
  adsnote = {Provided by the SAO/NASA Astrophysics Data System}
}

@ARTICLE{ho85,
   author = {{Ho}, P.~T.~P. and {Jackson}, J.~M. and {Barrett}, A.~H. and 
	{Armstrong}, J.~T.},
    title = "{Interactions between the continuum sources in the galactic center and their immediate molecular environment}",
  journal = {\apj},
 keywords = {Continuous Radiation, Galactic Nuclei, Interstellar Matter, Milky Way Galaxy, Molecular Clouds, Radiation Sources, Astronomical Maps, Neutral Gases, Radio Astronomy, Stellar Evolution, Supernovae},
     year = 1985,
    month = jan,
   volume = 288,
    pages = {575-579},
      doi = {10.1086/162823},
   adsurl = {http://adsabs.harvard.edu/abs/1985ApJ...288..575H},
  adsnote = {Provided by the SAO/NASA Astrophysics Data System}
}

@ARTICLE{great,
   author = {{Requena-Torres}, M.~A. and {G{\"u}sten}, R. and {Wei{\ss}}, A. and 
	{Harris}, A.~I. and {Mart{\'{\i}}n-Pintado}, J. and {Stutzki}, J. and 
	{Klein}, B. and {Heyminck}, S. and {Risacher}, C.},
    title = "{GREAT confirms transient nature of the circum-nuclear disk}",
  journal = {\aap},
archivePrefix = "arXiv",
   eprint = {1203.6687},
 primaryClass = "astro-ph.GA",
 keywords = {ISM: clouds, ISM: kinematics and dynamics, ISM: molecules, Galaxy: center, radio lines: ISM},
     year = 2012,
    month = jun,
   volume = 542,
      eid = {L21},
    pages = {L21},
      doi = {10.1051/0004-6361/201219068},
   adsurl = {http://adsabs.harvard.edu/abs/2012A
  adsnote = {Provided by the SAO/NASA Astrophysics Data System}
}

@ARTICLE{figer02,
   author = {{Figer}, D.~F. and {Najarro}, F. and {Gilmore}, D. and {Morris}, M. and 
	{Kim}, S.~S. and {Serabyn}, E. and {McLean}, I.~S. and {Gilbert}, A.~M. and 
	{Graham}, J.~R. and {Larkin}, J.~E. and {Levenson}, N.~A. and 
	{Teplitz}, H.~I.},
    title = "{Massive Stars in the Arches Cluster}",
  journal = {\apj},
   eprint = {astro-ph/0208145},
 keywords = {Galaxy: Center, Infrared: Stars- Galaxy: Open Clusters and Associations: Individual: Name: Arches, Stars: Early-Type- Stars: Formation, Techniques: Spectroscopic},
     year = 2002,
    month = dec,
   volume = 581,
    pages = {258-275},
      doi = {10.1086/344154},
   adsurl = {http://adsabs.harvard.edu/abs/2002ApJ...581..258F},
  adsnote = {Provided by the SAO/NASA Astrophysics Data System}
}

@ARTICLE{figer99b,
   author = {{Figer}, D.~F. and {Kim}, S.~S. and {Morris}, M. and {Serabyn}, E. and 
	{Rich}, R.~M. and {McLean}, I.~S.},
    title = "{Hubble Space Telescope/NICMOS Observations of Massive Stellar Clusters near the Galactic Center}",
  journal = {\apj},
   eprint = {astro-ph/9906299},
 keywords = {GALAXY: CENTER, GALAXY: STELLAR CONTENT, GALAXY: OPEN CLUSTERS AND ASSOCIATIONS: GENERAL, STARS: EVOLUTION, STARS: FORMATION, STARS: LUMINOSITY FUNCTION, MASS FUNCTION, Galaxy: Center, Galaxy: Stellar Content, Galaxy: Open Clusters and Associations: General, Stars: Evolution, Stars: Formation, Stars: Luminosity Function, Mass Function},
     year = 1999,
    month = nov,
   volume = 525,
    pages = {750-758},
      doi = {10.1086/307937},
   adsurl = {http://adsabs.harvard.edu/abs/1999ApJ...525..750F},
  adsnote = {Provided by the SAO/NASA Astrophysics Data System}
}

@ARTICLE{figer99a,
   author = {{Figer}, D.~F. and {McLean}, I.~S. and {Morris}, M.},
    title = "{Massive Stars in the Quintuplet Cluster}",
  journal = {\apj},
   eprint = {astro-ph/9903281},
 keywords = {GALAXY: CENTER, ISM: H II REGIONS, GALAXY: OPEN CLUSTERS AND ASSOCIATIONS: INDIVIDUAL: NAME: QUINTUPLET CLUSTER, STARS: WOLF-RAYET, Galaxy: Center, ISM: H II Regions, Galaxy: Open Clusters and Associations: Individual: Name: Quintuplet cluster, Stars: Wolf-Rayet},
     year = 1999,
    month = mar,
   volume = 514,
    pages = {202-220},
      doi = {10.1086/306931},
   adsurl = {http://adsabs.harvard.edu/abs/1999ApJ...514..202F},
  adsnote = {Provided by the SAO/NASA Astrophysics Data System}
}

@ARTICLE{baganoff,
   author = {{Baganoff}, F.~K. and {Maeda}, Y. and {Morris}, M. and {Bautz}, M.~W. and 
	{Brandt}, W.~N. and {Cui}, W. and {Doty}, J.~P. and {Feigelson}, E.~D. and 
	{Garmire}, G.~P. and {Pravdo}, S.~H. and {Ricker}, G.~R. and 
	{Townsley}, L.~K.},
    title = "{Chandra X-Ray Spectroscopic Imaging of Sagittarius A* and the Central Parsec of the Galaxy}",
  journal = {\apj},
   eprint = {astro-ph/0102151},
 keywords = {Accretion, Accretion Disks, Black Hole Physics, Galaxies: Active, Galaxy: Center, X-Rays: ISM, X-Rays: Stars},
     year = 2003,
    month = jul,
   volume = 591,
    pages = {891-915},
      doi = {10.1086/375145},
   adsurl = {http://adsabs.harvard.edu/abs/2003ApJ...591..891B},
  adsnote = {Provided by the SAO/NASA Astrophysics Data System}
}

@ARTICLE{mills13a,
   author = {{Mills}, E.~A.~C. and {Morris}, M.~R.},
    title = "{Detection of Widespread Hot Ammonia in the Galactic Center}",
  journal = {\apj},
archivePrefix = "arXiv",
   eprint = {1306.0953},
 primaryClass = "astro-ph.GA",
 keywords = {Galaxy: center, ISM: clouds, ISM: molecules, radio lines: ISM},
     year = 2013,
    month = aug,
   volume = 772,
      eid = {105},
    pages = {105},
      doi = {10.1088/0004-637X/772/2/105},
   adsurl = {http://adsabs.harvard.edu/abs/2013ApJ...772..105M},
  adsnote = {Provided by the SAO/NASA Astrophysics Data System}
}

@ARTICLE{mills13b,
   author = {{Mills}, E.~A.~C. and {G{\"u}sten}, R. and {Requena-Torres}, M.~A. and 
	{Morris}, M.~R.},
    title = "{The Excitation of HCN and HCO$^{+}$ in the Galactic Center Circumnuclear Disk}",
  journal = {\apj},
archivePrefix = "arXiv",
   eprint = {1309.7412},
 primaryClass = "astro-ph.GA",
 keywords = {Galaxy: center, ISM: molecules, radiative transfer, techniques: imaging spectroscopy},
     year = 2013,
    month = dec,
   volume = 779,
      eid = {47},
    pages = {47},
      doi = {10.1088/0004-637X/779/1/47},
   adsurl = {http://adsabs.harvard.edu/abs/2013ApJ...779...47M},
  adsnote = {Provided by the SAO/NASA Astrophysics Data System}
}


@ARTICLE{huett,
   author = {{Huettemeister}, S. and {Wilson}, T.~L. and {Bania}, T.~M. and 
	{Martin-Pintado}, J.},
    title = "{Kinetic temperatures in Galactic Center molecular clouds}",
  journal = {\aap},
 keywords = {Ammonia, Electron Transitions, Emission Spectra, Galactic Bulge, Gas Temperature, Interstellar Matter, Inversions, Line Spectra, Metastable State, Molecular Clouds, Radio Astronomy, Centimeter Waves, Interstellar Magnetic Fields, Rotational States, Superhigh Frequencies},
     year = 1993,
    month = dec,
   volume = 280,
    pages = {255-267},
   adsurl = {http://adsabs.harvard.edu/abs/1993A
  adsnote = {Provided by the SAO/NASA Astrophysics Data System}
}

@ARTICLE{lang99,
   author = {{Lang}, C.~C. and {Anantharamaiah}, K.~R. and {Kassim}, N.~E. and 
	{Lazio}, T.~J.~W.},
    title = "{Discovery of a Nonthermal Galactic Center Filament (G358.85+0.47) Parallel to the Galactic Plane}",
  journal = {\apjl},
   eprint = {astro-ph/9906285},
 keywords = {GALAXY: CENTER, ISM: MAGNETIC FIELDS, RADIO CONTINUUM: ISM, Galaxy: Center, ISM: Magnetic Fields, Radio Continuum: ISM},
     year = 1999,
    month = aug,
   volume = 521,
    pages = {L41-L44},
      doi = {10.1086/312180},
   adsurl = {http://adsabs.harvard.edu/abs/1999ApJ...521L..41L},
  adsnote = {Provided by the SAO/NASA Astrophysics Data System}
}


@ARTICLE{yusef84,
   author = {{Yusef-Zadeh}, F. and {Morris}, M. and {Chance}, D.},
    title = "{Large, highly organized radio structures near the galactic centre}",
  journal = {\nat},
 keywords = {Astronomical Maps, Galactic Nuclei, Galactic Radio Waves, Milky Way Galaxy, Radio Sources (Astronomy), Centimeter Waves, Filaments, Interstellar Gas, Interstellar Magnetic Fields, Morphology, Very Large Array (Vla)},
     year = 1984,
    month = aug,
   volume = 310,
    pages = {557-561},
      doi = {10.1038/310557a0},
   adsurl = {http://adsabs.harvard.edu/abs/1984Natur.310..557Y},
  adsnote = {Provided by the SAO/NASA Astrophysics Data System}
}



@ARTICLE{zhao13,
   author = {{Zhao}, J.-H. and {Morris}, M.~R. and {Goss}, W.~M.},
    title = "{Radio Detection of a Candidate Neutron Star Associated with Galactic Center Supernova Remnant Sagittarius A East}",
  journal = {\apj},
archivePrefix = "arXiv",
   eprint = {1309.7020},
 primaryClass = "astro-ph.HE",
 keywords = {Galaxy: center, ISM: individual objects: Sagittarius A, ISM: supernova remnants, radio continuum: ISM, stars: neutron, stars: winds, outflows},
     year = 2013,
    month = nov,
   volume = 777,
      eid = {146},
    pages = {146},
      doi = {10.1088/0004-637X/777/2/146},
   adsurl = {http://adsabs.harvard.edu/abs/2013ApJ...777..146Z},
  adsnote = {Provided by the SAO/NASA Astrophysics Data System}
}

@ARTICLE{sjo,
   author = {{Sjouwerman}, L.~O. and {Pihlstr{\"o}m}, Y.~M.},
    title = "{Very Large Array Observations of Galactic Center OH 1720 MHz Masers in Sagittarius A East and in the Circumnuclear Disk}",
  journal = {\apj},
archivePrefix = "arXiv",
   eprint = {0804.0445},
 keywords = {Galaxies: Nuclei, Galaxy: Center, ISM: individual (circumnuclear disk), ISM: Individual: Alphanumeric: M-0.02-0.07, ISM: Individual: Name: Sagittarius A East, Masers},
     year = 2008,
    month = jul,
   volume = 681,
    pages = {1287-1295},
      doi = {10.1086/588753},
   adsurl = {http://adsabs.harvard.edu/abs/2008ApJ...681.1287S},
  adsnote = {Provided by the SAO/NASA Astrophysics Data System}
}

@ARTICLE{maeda,
   author = {{Maeda}, Y. and {Baganoff}, F.~K. and {Feigelson}, E.~D. and 
	{Morris}, M. and {Bautz}, M.~W. and {Brandt}, W.~N. and {Burrows}, D.~N. and 
	{Doty}, J.~P. and {Garmire}, G.~P. and {Pravdo}, S.~H. and {Ricker}, G.~R. and 
	{Townsley}, L.~K.},
    title = "{A Chandra Study of Sagittarius A East: A Supernova Remnant Regulating the Activity of Our Galactic Center?}",
  journal = {\apj},
   eprint = {astro-ph/0102183},
 keywords = {Galaxy: Center, ISM: Individual: Name: Sagittarius A East, ISM: Supernova Remnants, X-Rays: ISM},
     year = 2002,
    month = may,
   volume = 570,
    pages = {671-687},
      doi = {10.1086/339773},
   adsurl = {http://adsabs.harvard.edu/abs/2002ApJ...570..671M},
  adsnote = {Provided by the SAO/NASA Astrophysics Data System}
}

@ARTICLE{yusef87,
   author = {{Yusef-Zadeh}, F. and {Morris}, M.},
    title = "{Structural details of the Sagittarius A complex - Evidence for a large-scale poloidal magnetic field in the Galactic center region}",
  journal = {\apj},
 keywords = {Galactic Nuclei, Interstellar Magnetic Fields, Milky Way Galaxy, Poloidal Flux, Radio Sources (Astronomy), Supernova Remnants, Dynamo Theory, H Ii Regions, Linear Polarization, Sagittarius Constellation, Very Large Array (Vla)},
     year = 1987,
    month = sep,
   volume = 320,
    pages = {545-561},
      doi = {10.1086/165572},
   adsurl = {http://adsabs.harvard.edu/abs/1987ApJ...320..545Y},
  adsnote = {Provided by the SAO/NASA Astrophysics Data System}
}

@ARTICLE{ekers,
   author = {{Ekers}, R.~D. and {van Gorkom}, J.~H. and {Schwarz}, U.~J. and 
	{Goss}, W.~M.},
    title = "{The radio structure of SGR A}",
  journal = {\aap},
 keywords = {Astronomical Maps, Galactic Nuclei, Galactic Structure, Radio Sources (Astronomy), Angular Resolution, Antenna Arrays, Hydrogen Clouds, Milky Way Galaxy, Radio Astronomy, Radio Telescopes, Thermal Emission},
     year = 1983,
    month = jun,
   volume = 122,
    pages = {143-150},
   adsurl = {http://adsabs.harvard.edu/abs/1983A
  adsnote = {Provided by the SAO/NASA Astrophysics Data System}
}


@ARTICLE{martins12,
   author = {{Mart{\'{\i}}n}, S. and {Mart{\'{\i}}n-Pintado}, J. and {Montero-Casta{\~n}o}, M. and 
	{Ho}, P.~T.~P. and {Blundell}, R.},
    title = "{Surviving the hole. I. Spatially resolved chemistry around Sagittarius A$^{∗}$}",
  journal = {\aap},
archivePrefix = "arXiv",
   eprint = {1112.0566},
 primaryClass = "astro-ph.GA",
 keywords = {ISM: molecules, ISM: clouds, radio lines: ISM, Galaxy: center, ISM: kinematics and dynamics},
     year = 2012,
    month = mar,
   volume = 539,
      eid = {A29},
    pages = {A29},
      doi = {10.1051/0004-6361/201117268},
   adsurl = {http://adsabs.harvard.edu/abs/2012A
  adsnote = {Provided by the SAO/NASA Astrophysics Data System}
}


@ARTICLE{law09,
   author = {{Law}, C.~J. and {Backer}, D. and {Yusef-Zadeh}, F. and {Maddalena}, R.
	},
    title = "{Radio Recombination Lines Toward the Galactic Center Lobe}",
  journal = {\apj},
archivePrefix = "arXiv",
   eprint = {0901.1480},
 primaryClass = "astro-ph.GA",
 keywords = {Galaxy: center, radio lines: general},
     year = 2009,
    month = apr,
   volume = 695,
    pages = {1070-1081},
      doi = {10.1088/0004-637X/695/2/1070},
   adsurl = {http://adsabs.harvard.edu/abs/2009ApJ...695.1070L},
  adsnote = {Provided by the SAO/NASA Astrophysics Data System}
}

@ARTICLE{law08a,
   author = {{Law}, C.~J. and {Yusef-Zadeh}, F. and {Cotton}, W.~D.},
    title = "{A Wide-Area VLA Continuum Survey near the Galactic Center at 6 and 20 cm Wavelengths}",
  journal = {\apjs},
archivePrefix = "arXiv",
   eprint = {0803.1412},
 keywords = {Galaxy: Center, Radio Continuum: General, Surveys},
     year = 2008,
    month = aug,
   volume = 177,
    pages = {515-545},
      doi = {10.1086/588218},
   adsurl = {http://adsabs.harvard.edu/abs/2008ApJS..177..515L},
  adsnote = {Provided by the SAO/NASA Astrophysics Data System}
}

@ARTICLE{law08b,
   author = {{Law}, C.~J. and {Yusef-Zadeh}, F. and {Cotton}, W.~D. and {Maddalena}, R.~J.
	},
    title = "{Green Bank Telescope Multiwavelength Survey of the Galactic Center Region}",
  journal = {\apjs},
archivePrefix = "arXiv",
   eprint = {0801.4294},
 keywords = {Galaxy: Center, Radio Continuum: General, Surveys},
     year = 2008,
    month = jul,
   volume = 177,
    pages = {255-274},
      doi = {10.1086/533587},
   adsurl = {http://adsabs.harvard.edu/abs/2008ApJS..177..255L},
  adsnote = {Provided by the SAO/NASA Astrophysics Data System}
}

@INPROCEEDINGS{irvine,
   author = {{Irvine}, W.~M. and {Goldsmith}, P.~F. and {Hjalmarson}, A.},
    title = "{Chemical abundances in molecular clouds}",
 keywords = {Abundance, Astronomical Models, Interstellar Chemistry, Molecular Clouds, Chemical Composition, Galactic Structure, Milky Way Galaxy, Temperature Effects},
booktitle = {Interstellar Processes},
     year = 1987,
   series = {Astrophysics and Space Science Library},
   volume = 134,
   editor = {{Hollenbach}, D.~J. and {Thronson}, Jr., H.~A.},
    pages = {561-609},
   adsurl = {http://adsabs.harvard.edu/abs/1987ASSL..134..561I},
  adsnote = {Provided by the SAO/NASA Astrophysics Data System}
}

@INPROCEEDINGS{burton83,
   author = {{Burton}, W.~B. and {Liszt}, H.~S.},
    title = "{A CO structure near the galactic center with strong positional and kinematic gradients}",
 keywords = {Carbon Monoxide, Galactic Nuclei, Interstellar Gas, Radial Velocity, Molecular Spectra, Telescopes, Temperature Profiles},
booktitle = {Surveys of the Southern Galaxy},
     year = 1983,
   series = {Astrophysics and Space Science Library},
   volume = 105,
   editor = {{Burton}, W.~B. and {Israel}, F.~P.},
    pages = {149-157},
   adsurl = {http://adsabs.harvard.edu/abs/1983ASSL..105..149B},
  adsnote = {Provided by the SAO/NASA Astrophysics Data System}
}

@ARTICLE{burton92,
   author = {{Burton}, W.~B. and {Liszt}, H.~S.},
    title = "{The gas distribution in the central region of the Galaxy. V - (C-12)O in the direction of the Sagittarius source complex}",
  journal = {\aaps},
 keywords = {Astronomical Spectroscopy, Carbon Monoxide, Galactic Nuclei, Interstellar Gas, Milky Way Galaxy, Astronomical Models, Emission Spectra, Galactic Structure, Line Spectra, Molecular Spectra, Sagittarius Constellation},
     year = 1992,
    month = oct,
   volume = 95,
    pages = {9-39},
   adsurl = {http://adsabs.harvard.edu/abs/1992A
  adsnote = {Provided by the SAO/NASA Astrophysics Data System}
}


@ARTICLE{bally87,
   author = {{Bally}, J. and {Stark}, A.~A. and {Wilson}, R.~W. and {Henkel}, C.
	},
    title = "{Galactic center molecular clouds. I - Spatial and spatial-velocity maps}",
  journal = {\apjs},
 keywords = {Astronomical Maps, Galactic Nuclei, Interstellar Gas, Milky Way Galaxy, Molecular Clouds, Carbon Monoxide, Spatial Distribution, Velocity Distribution},
     year = 1987,
    month = sep,
   volume = 65,
    pages = {13-82},
      doi = {10.1086/191217},
   adsurl = {http://adsabs.harvard.edu/abs/1987ApJS...65...13B},
  adsnote = {Provided by the SAO/NASA Astrophysics Data System}
}


@ARTICLE{jones13,
   author = {{Jones}, P.~A. and {Burton}, M.~G. and {Cunningham}, M.~R. and 
	{Tothill}, N.~F.~H. and {Walsh}, A.~J.},
    title = "{Spectral imaging of the central molecular zone in multiple 7-mm molecular lines}",
  journal = {\mnras},
archivePrefix = "arXiv",
   eprint = {1304.7076},
 primaryClass = "astro-ph.GA",
 keywords = {ISM: kinematics and dynamics, ISM: molecules, radio lines: ISM},
     year = 2013,
    month = jul,
   volume = 433,
    pages = {221-234},
      doi = {10.1093/mnras/stt717},
   adsurl = {http://adsabs.harvard.edu/abs/2013MNRAS.433..221J},
  adsnote = {Provided by the SAO/NASA Astrophysics Data System}
}

@ARTICLE{longmore,
   author = {{Longmore}, S.~N. and {Bally}, J. and {Testi}, L. and {Purcell}, C.~R. and 
	{Walsh}, A.~J. and {Bressert}, E. and {Pestalozzi}, M. and {Molinari}, S. and 
	{Ott}, J. and {Cortese}, L. and {Battersby}, C. and {Murray}, N. and 
	{Lee}, E. and {Kruijssen}, J.~M.~D. and {Schisano}, E. and {Elia}, D.
	},
    title = "{Variations in the Galactic star formation rate and density thresholds for star formation}",
  journal = {\mnras},
archivePrefix = "arXiv",
   eprint = {1208.4256},
 primaryClass = "astro-ph.GA",
 keywords = {masers, stars: formation, stars: massive, ISM: clouds, ISM: evolution, Galaxy: centre},
     year = 2013,
    month = feb,
   volume = 429,
    pages = {987-1000},
      doi = {10.1093/mnras/sts376},
   adsurl = {http://adsabs.harvard.edu/abs/2013MNRAS.429..987L},
  adsnote = {Provided by the SAO/NASA Astrophysics Data System}
}


@ARTICLE{jones12,
   author = {{Jones}, P.~A. and {Burton}, M.~G. and {Cunningham}, M.~R. and 
	{Requena-Torres}, M.~A. and {Menten}, K.~M. and {Schilke}, P. and 
	{Belloche}, A. and {Leurini}, S. and {Mart{\'{\i}}n-Pintado}, J. and 
	{Ott}, J. and {Walsh}, A.~J.},
    title = "{Spectral imaging of the Central Molecular Zone in multiple 3-mm molecular lines}",
  journal = {\mnras},
archivePrefix = "arXiv",
   eprint = {1110.1421},
 primaryClass = "astro-ph.GA",
 keywords = {ISM: kinematics and dynamics, ISM: molecules, radio lines: ISM},
     year = 2012,
    month = feb,
   volume = 419,
    pages = {2961-2986},
      doi = {10.1111/j.1365-2966.2011.19941.x},
   adsurl = {http://adsabs.harvard.edu/abs/2012MNRAS.419.2961J},
  adsnote = {Provided by the SAO/NASA Astrophysics Data System}
}

@ARTICLE{dahmen,
   author = {{Dahmen}, G. and {Huttemeister}, S. and {Wilson}, T.~L. and 
	{Mauersberger}, R.},
    title = "{Molecular gas in the Galactic center region. II. Gas mass and N\_, = H\_2/I\_\^{}(12)CO conversion based on a C\^{}(18)O(J = 1 -{\gt} 0) survey}",
  journal = {\aap},
   eprint = {astro-ph/9711117},
 keywords = {GALAXY: CENTER, RADIATIVE TRANSFER, ISM: MOLECULES, ISM: STRUCTURE, GALAXIES: NUCLEI, RADIO LINES: ISM},
     year = 1998,
    month = mar,
   volume = 331,
    pages = {959-976},
   adsurl = {http://adsabs.harvard.edu/abs/1998A
  adsnote = {Provided by the SAO/NASA Astrophysics Data System}
}

@ARTICLE{ghez08,
   author = {{Ghez}, A.~M. and {Salim}, S. and {Weinberg}, N.~N. and {Lu}, J.~R. and 
	{Do}, T. and {Dunn}, J.~K. and {Matthews}, K. and {Morris}, M.~R. and 
	{Yelda}, S. and {Becklin}, E.~E. and {Kremenek}, T. and {Milosavljevic}, M. and 
	{Naiman}, J.},
    title = "{Measuring Distance and Properties of the Milky Way's Central Supermassive Black Hole with Stellar Orbits}",
  journal = {\apj},
archivePrefix = "arXiv",
   eprint = {0808.2870},
 keywords = {Black Hole Physics, Galaxy: Center, Galaxy: Kinematics and Dynamics, Infrared: Stars, Techniques: High Angular Resolution},
     year = 2008,
    month = dec,
   volume = 689,
    pages = {1044-1062},
      doi = {10.1086/592738},
   adsurl = {http://adsabs.harvard.edu/abs/2008ApJ...689.1044G},
  adsnote = {Provided by the SAO/NASA Astrophysics Data System}
}


@ARTICLE{wardle08,
   author = {{Wardle}, M. and {Yusef-Zadeh}, F.},
    title = "{On the Formation of Compact Stellar Disks around Sagittarius A*}",
  journal = {\apjl},
archivePrefix = "arXiv",
   eprint = {0805.3274},
 keywords = {Accretion, Accretion Disks, Galaxy: Center, ISM: Clouds, Stars: Formation},
     year = 2008,
    month = aug,
   volume = 683,
    pages = {L37-L40},
      doi = {10.1086/591471},
   adsurl = {http://adsabs.harvard.edu/abs/2008ApJ...683L..37W},
  adsnote = {Provided by the SAO/NASA Astrophysics Data System}
}

@ARTICLE{liu12,
   author = {{Liu}, H.~B. and {Hsieh}, P.-Y. and {Ho}, P.~T.~P. and {Su}, Y.-N. and 
	{Wright}, M. and {Sun}, A.-L. and {Minh}, Y.~C.},
    title = "{Milky Way Supermassive Black Hole: Dynamical Feeding from the Circumnuclear Environment}",
  journal = {\apj},
archivePrefix = "arXiv",
   eprint = {1207.6309},
 primaryClass = "astro-ph.GA",
 keywords = {Galaxy: center, Galaxy: kinematics and dynamics, Galaxy: structure, ISM: clouds},
     year = 2012,
    month = sep,
   volume = 756,
      eid = {195},
    pages = {195},
      doi = {10.1088/0004-637X/756/2/195},
   adsurl = {http://adsabs.harvard.edu/abs/2012ApJ...756..195L},
  adsnote = {Provided by the SAO/NASA Astrophysics Data System}
}

@ARTICLE{coil00,
   author = {{Coil}, A.~L. and {Ho}, P.~T.~P.},
    title = "{The Dynamics of Molecular Material within 15 PARSECS of the Galactic Center}",
  journal = {\apj},
   eprint = {arXiv:astro-ph/9910043},
 keywords = {GALAXY: CENTER, ISM: CLOUDS, ISM: MOLECULES},
     year = 2000,
    month = apr,
   volume = 533,
    pages = {245-259},
      doi = {10.1086/308650},
   adsurl = {http://adsabs.harvard.edu/abs/2000ApJ...533..245C},
  adsnote = {Provided by the SAO/NASA Astrophysics Data System}
}

@ARTICLE{serabyn92,
   author = {{Serabyn}, E. and {Lacy}, J.~H. and {Achtermann}, J.~M.},
    title = "{The compression of the M-0.02-0.07 molecular cloud by the Sagittarius A East shell source}",
  journal = {\apj},
 keywords = {H II REGIONS, INTERSTELLAR MATTER, MOLECULAR CLOUDS, SAGITTARIUS CONSTELLATION, SUPERNOVA REMNANTS, FINE STRUCTURE, MILKY WAY GALAXY, STAR FORMATION},
     year = 1992,
    month = aug,
   volume = 395,
    pages = {166-173},
      doi = {10.1086/171640},
   adsurl = {http://adsabs.harvard.edu/abs/1992ApJ...395..166S},
  adsnote = {Provided by the SAO/NASA Astrophysics Data System}
}

@INPROCEEDINGS{gusten04,
   author = {{G{\"u}sten}, R. and {Philipp}, S.~D.},
    title = "{Galactic Center Molecular Clouds}",
booktitle = {The Dense Interstellar Medium in Galaxies},
     year = 2004,
   eprint = {astro-ph/0402019},
   editor = {{Pfalzner}, S. and {Kramer}, C. and {Staubmeier}, C. and {Heithausen}, A.
	},
    pages = {253},
   adsurl = {http://adsabs.harvard.edu/abs/2004dimg.conf..253G},
  adsnote = {Provided by the SAO/NASA Astrophysics Data System}
}

@ARTICLE{miyazaki00,
   author = {{Miyazaki}, A. and {Tsuboi}, M.},
    title = "{Dense Molecular Clouds in the Galactic Center Region. II. Statistical Properties of the Galactic Center Molecular Clouds}",
  journal = {\apj},
 keywords = {Galaxy: Center, ISM: Clouds, ISM: Molecules},
     year = 2000,
    month = jun,
   volume = 536,
    pages = {357-367},
      doi = {10.1086/308899},
   adsurl = {http://adsabs.harvard.edu/abs/2000ApJ...536..357M},
  adsnote = {Provided by the SAO/NASA Astrophysics Data System}
}

@ARTICLE{uchida85,
   author = {{Uchida}, Y. and {Sofue}, Y. and {Shibata}, K.},
    title = "{Origin of the galactic centre lobes}",
  journal = {\nat},
 keywords = {Galactic Evolution, Galactic Nuclei, Galactic Structure, Radio Galaxies, Accretion Disks, Lobes, Stellar Evolution},
     year = 1985,
    month = oct,
   volume = 317,
    pages = {699-701},
      doi = {10.1038/317699a0},
   adsurl = {http://adsabs.harvard.edu/abs/1985Natur.317..699U},
  adsnote = {Provided by the SAO/NASA Astrophysics Data System}
}

@ARTICLE{enokiya,
   author = {{Enokiya}, R. and {Torii}, K. and {Schultheis}, M. and {Asahina}, Y. and 
	{Matsumoto}, R. and {Furuhashi}, E. and {Nakamura}, K. and {Dobashi}, K. and 
	{Yoshiike}, S. and {Sato}, J. and {Furukawa}, N. and {Moribe}, N. and 
	{Ohama}, A. and {Sano}, H. and {Okamoto}, R. and {Mori}, Y. and 
	{Hanaoka}, N. and {Nishimura}, A. and {Hayakawa}, T. and {Okuda}, T. and 
	{Yamamoto}, H. and {Kawamura}, A. and {Mizuno}, N. and {Onishi}, T. and 
	{Morris}, M.~R. and {Fukui}, Y.},
    title = "{Discovery of Possible Molecular Counterparts to the Infrared Double Helix Nebula in the Galactic Center}",
  journal = {\apj},
archivePrefix = "arXiv",
   eprint = {1310.8229},
 primaryClass = "astro-ph.SR",
 keywords = {ISM: clouds, radio lines: ISM},
     year = 2014,
    month = jan,
   volume = 780,
      eid = {72},
    pages = {72},
      doi = {10.1088/0004-637X/780/1/72},
   adsurl = {http://adsabs.harvard.edu/abs/2014ApJ...780...72E},
  adsnote = {Provided by the SAO/NASA Astrophysics Data System}
}

@ARTICLE{binney91,
   author = {{Binney}, J. and {Gerhard}, O.~E. and {Stark}, A.~A. and {Bally}, J. and 
	{Uchida}, K.~I.},
    title = "{Understanding the kinematics of Galactic centre gas}",
  journal = {\mnras},
 keywords = {Carbon Monoxide, Galactic Nuclei, Galactic Structure, Gas Dynamics, Hydrogen, Milky Way Galaxy, Emission Spectra, Galactic Rotation, Infrared Photometry, Mass To Light Ratios, Molecular Clouds},
     year = 1991,
    month = sep,
   volume = 252,
    pages = {210-218},
   adsurl = {http://adsabs.harvard.edu/abs/1991MNRAS.252..210B},
  adsnote = {Provided by the SAO/NASA Astrophysics Data System}
}

@ARTICLE{sawada04,
   author = {{Sawada}, T. and {Hasegawa}, T. and {Handa}, T. and {Cohen}, R.~J.
	},
    title = "{A molecular face-on view of the Galactic Centre region}",
  journal = {\mnras},
   eprint = {astro-ph/0401286},
 keywords = {ISM: molecules, Galaxy: centre, Galaxy: kinematics and dynamics, radio lines: ISM},
     year = 2004,
    month = apr,
   volume = 349,
    pages = {1167-1178},
      doi = {10.1111/j.1365-2966.2004.07603.x},
   adsurl = {http://adsabs.harvard.edu/abs/2004MNRAS.349.1167S},
  adsnote = {Provided by the SAO/NASA Astrophysics Data System}
}

@ARTICLE{kru15,
   author = {{Kruijssen}, J.~M.~D. and {Dale}, J.~E. and {Longmore}, S.~N.
	},
    title = "{The dynamical evolution of molecular clouds near the Galactic Centre - I. Orbital structure and evolutionary timeline}",
  journal = {\mnras},
archivePrefix = "arXiv",
   eprint = {1412.0664},
 keywords = {stars: formation, ISM: clouds, ISM: kinematics and dynamics, Galaxy: centre, galaxies: ISM},
     year = 2015,
    month = feb,
   volume = 447,
    pages = {1059-1079},
      doi = {10.1093/mnras/stu2526},
   adsurl = {http://adsabs.harvard.edu/abs/2015MNRAS.447.1059K},
  adsnote = {Provided by the SAO/NASA Astrophysics Data System}
}

@ARTICLE{krui15,
   author = {{Kruijssen}, J.~M.~D. and {Dale}, J.~E. and {Longmore}, S.~N.
	},
    title = "{The dynamical evolution of molecular clouds near the Galactic Centre - I. Orbital structure and evolutionary timeline}",
  journal = {\mnras},
archivePrefix = "arXiv",
   eprint = {1412.0664},
 keywords = {stars: formation, ISM: clouds, ISM: kinematics and dynamics, Galaxy: centre, galaxies: ISM},
     year = 2015,
    month = feb,
   volume = 447,
    pages = {1059-1079},
      doi = {10.1093/mnras/stu2526},
   adsurl = {http://adsabs.harvard.edu/abs/2015MNRAS.447.1059K},
  adsnote = {Provided by the SAO/NASA Astrophysics Data System}
}

@ARTICLE{fukui06,
   author = {{Fukui}, Y. and {Yamamoto}, H. and {Fujishita}, M. and {Kudo}, N. and 
	{Torii}, K. and {Nozawa}, S. and {Takahashi}, K. and {Matsumoto}, R. and 
	{Machida}, M. and {Kawamura}, A. and {Yonekura}, Y. and {Mizuno}, N. and 
	{Onishi}, T. and {Mizuno}, A.},
    title = "{Molecular Loops in the Galactic Center: Evidence for Magnetic Flotation}",
  journal = {Science},
     year = 2006,
    month = oct,
   volume = 314,
    pages = {106-109},
      doi = {10.1126/science.1130425},
   adsurl = {http://adsabs.harvard.edu/abs/2006Sci...314..106F},
  adsnote = {Provided by the SAO/NASA Astrophysics Data System}
}

@ARTICLE{machida09,
   author = {{Machida}, M. and {Matsumoto}, R. and {Nozawak}, S. and {Takahashi}, K. and 
	{Fukui}, Y. and {Kudo}, N. and {Torii}, K. and {Yamamoto}, H. and 
	{Fujishita}, M. and {Tomisaki}, K.},
    title = "{Formation of Galactic Center Magnetic Loops}",
  journal = {\pasj},
archivePrefix = "arXiv",
   eprint = {0812.3711},
 keywords = {Galaxy: disk, magnetic field, ISM: general, magnetic loops, magnetohydrodynamics: MHD},
     year = 2009,
    month = jun,
   volume = 61,
    pages = {411-},
      doi = {10.1093/pasj/61.3.411},
   adsurl = {http://adsabs.harvard.edu/abs/2009PASJ...61..411M},
  adsnote = {Provided by the SAO/NASA Astrophysics Data System}
}

@ARTICLE{takahashi09,
   author = {{Takahashi}, K. and {Nozawa}, S. and {Matsumoto}, R. and {Machida}, M. and 
	{Fukui}, Y. and {Kudo}, N. and {Torii}, K. and {Yamamoto}, H. and 
	{Fujishita}, M.},
    title = "{Similarity between the Molecular Loops in the Galactic Center and the Solar Chromospheric Arch Filaments}",
  journal = {\pasj},
archivePrefix = "arXiv",
   eprint = {0905.4357},
 primaryClass = "astro-ph.SR",
 keywords = {Galaxy: magnetic loops, ISM, Sun: chromosphere, Sun: magnetic fields, magnetohydrodynamics},
     year = 2009,
    month = oct,
   volume = 61,
    pages = {957-},
      doi = {10.1093/pasj/61.5.957},
   adsurl = {http://adsabs.harvard.edu/abs/2009PASJ...61..957T},
  adsnote = {Provided by the SAO/NASA Astrophysics Data System}
}

@ARTICLE{torii,
   author = {{Torii}, K. and {Kudo}, N. and {Fujishita}, M. and {Kawase}, T. and 
	{Yamamoto}, H. and {Kawamura}, A. and {Mizuno}, N. and {Onishi}, T. and 
	{Mizuno}, A. and {Machida}, M. and {Takahashi}, K. and {Nozawa}, S. and 
	{Matsumoto}, R. and {Fukui}, Y.},
    title = "{A Detailed Observational Study of Molecular Loops 1 and 2 in the Galactic Center}",
  journal = {\pasj},
archivePrefix = "arXiv",
   eprint = {0906.2076},
 keywords = {ISM: clouds, ISM: magnetic fields, magnetic loops, radio lines: ISM},
     year = 2010,
    month = oct,
   volume = 62,
    pages = {1307-1332},
      doi = {10.1093/pasj/62.5.1307},
   adsurl = {http://adsabs.harvard.edu/abs/2010PASJ...62.1307T},
  adsnote = {Provided by the SAO/NASA Astrophysics Data System}
}

@ARTICLE{handa87,
   author = {{Handa}, T. and {Sofue}, Y. and {Nakai}, N. and {Hirabayashi}, H. and 
	{Inoue}, M.},
    title = "{A radio continuum survey of the Galactic plane at 10 GHz}",
  journal = {\pasj},
 keywords = {Continuous Radiation, Galactic Radio Waves, Radio Sources (Astronomy), Sky Surveys (Astronomy), Astronomical Catalogs, H Ii Regions, Relativistic Electron Beams, Synchrotron Radiation},
     year = 1987,
   volume = 39,
    pages = {709-753},
   adsurl = {http://adsabs.harvard.edu/abs/1987PASJ...39..709H},
  adsnote = {Provided by the SAO/NASA Astrophysics Data System}
}

@ARTICLE{oka98,
   author = {{Oka}, T. and {Hasegawa}, T. and {Sato}, F. and {Tsuboi}, M. and 
	{Miyazaki}, A.},
    title = "{A Large-Scale CO Survey of the Galactic Center}",
  journal = {\apjs},
 keywords = {GALAXY: CENTER, GALAXY: KINEMATICS AND DYNAMICS, ISM: MOLECULES, SURVEYS, Galaxy: Center, Galaxy: Kinematics and Dynamics, ISM: Molecules, Surveys},
     year = 1998,
    month = oct,
   volume = 118,
    pages = {455-515},
      doi = {10.1086/313138},
   adsurl = {http://adsabs.harvard.edu/abs/1998ApJS..118..455O},
  adsnote = {Provided by the SAO/NASA Astrophysics Data System}
}

@ARTICLE{ott14,
   author = {{Ott}, J. and {Wei{\ss}}, A. and {Staveley-Smith}, L. and {Henkel}, C. and 
	{Meier}, D.~S.},
    title = "{ATCA Survey of Ammonia in the Galactic Center: The Temperatures of Dense Gas Clumps between Sgr A* and Sgr B2}",
  journal = {\apj},
archivePrefix = "arXiv",
   eprint = {1402.4531},
 keywords = {Galaxy: center, ISM: clouds, ISM: kinematics and dynamics, ISM: molecules, ISM: structure, stars: formation},
     year = 2014,
    month = apr,
   volume = 785,
      eid = {55},
    pages = {55},
      doi = {10.1088/0004-637X/785/1/55},
   adsurl = {http://adsabs.harvard.edu/abs/2014ApJ...785...55O},
  adsnote = {Provided by the SAO/NASA Astrophysics Data System}
}

@ARTICLE{oka01,
   author = {{Oka}, T. and {Hasegawa}, T. and {Sato}, F. and {Tsuboi}, M. and 
	{Miyazaki}, A.},
    title = "{A Molecular Cloud and an Expanding Cavity Adjacent to the Nonthermal Filaments of the Galactic Center Radio Arc}",
  journal = {\pasj},
 keywords = {GALAXIES: NUCLEI, GALAXY: CENTER, ISM: CLOUDS, ISM: MOLECULES},
     year = 2001,
    month = oct,
   volume = 53,
    pages = {779-786},
      doi = {10.1093/pasj/53.5.779},
   adsurl = {http://adsabs.harvard.edu/abs/2001PASJ...53..779O},
  adsnote = {Provided by the SAO/NASA Astrophysics Data System}
}

@ARTICLE{pucell,
   author = {{Purcell}, C.~R. and {Longmore}, S.~N. and {Walsh}, A.~J. and 
	{Whiting}, M.~T. and {Breen}, S.~L. and {Britton}, T. and {Brooks}, K.~J. and 
	{Burton}, M.~G. and {Cunningham}, M.~R. and {Green}, J.~A. and 
	{Harvey-Smith}, L. and {Hindson}, L. and {Hoare}, M.~G. and 
	{Indermuehle}, B. and {Jones}, P.~A. and {Lo}, N. and {Lowe}, V. and 
	{Phillips}, C.~J. and {Thompson}, M.~A. and {Urquhart}, J.~S. and 
	{Voronkov}, M.~A. and {White}, G.~L.},
    title = "{The H$_{2}$O Southern Galactic Plane Survey: NH$_{3}$ (1,1) and (2,2) catalogues}",
  journal = {\mnras},
archivePrefix = "arXiv",
   eprint = {1207.6159},
 keywords = {surveys, stars: early-type, stars: formation, ISM: evolution, Galaxy: structure, radio lines: ISM},
     year = 2012,
    month = nov,
   volume = 426,
    pages = {1972-1991},
      doi = {10.1111/j.1365-2966.2012.21800.x},
   adsurl = {http://adsabs.harvard.edu/abs/2012MNRAS.426.1972P},
  adsnote = {Provided by the SAO/NASA Astrophysics Data System}
}


@ARTICLE{serabyn87,
   author = {{Serabyn}, E. and {Guesten}, R.},
    title = "{A molecular counterpart to the galactic center arc}",
  journal = {\aap},
 keywords = {Electron Transitions, Galactic Nuclei, Interstellar Matter, Ionized Gases, Molecular Clouds, Gas Dynamics, Molecular Gases},
     year = 1987,
    month = oct,
   volume = 184,
    pages = {133-143},
   adsurl = {http://adsabs.harvard.edu/abs/1987A
  adsnote = {Provided by the SAO/NASA Astrophysics Data System}
}

@INPROCEEDINGS{serabyn89,
   author = {{Serabyn}, E. and {G{\"u}sten}, R. and {Evans}, II, N.~J.},
    title = "{CS Multitransition Observations of the Circumnuclear Disk}",
booktitle = {The Center of the Galaxy},
     year = 1989,
   series = {IAU Symposium},
   volume = 136,
   editor = {{Morris}, M.},
    pages = {417},
   adsurl = {http://adsabs.harvard.edu/abs/1989IAUS..136..417S},
  adsnote = {Provided by the SAO/NASA Astrophysics Data System}
}

@ARTICLE{ao13,
   author = {{Ao}, Y. and {Henkel}, C. and {Menten}, K.~M. and {Requena-Torres}, M.~A. and 
	{Stanke}, T. and {Mauersberger}, R. and {Aalto}, S. and {M{\"u}hle}, S. and 
	{Mangum}, J.},
    title = "{The thermal state of molecular clouds in the Galactic center: evidence for non-photon-driven heating}",
  journal = {\aap},
archivePrefix = "arXiv",
   eprint = {1211.7142},
 keywords = {Galaxy: center, ISM: clouds, ISM: molecules, radio lines: ISM},
     year = 2013,
    month = feb,
   volume = 550,
      eid = {A135},
    pages = {A135},
      doi = {10.1051/0004-6361/201220096},
   adsurl = {http://adsabs.harvard.edu/abs/2013A
  adsnote = {Provided by the SAO/NASA Astrophysics Data System}
}

@ARTICLE{guesten83,
   author = {{Guesten}, R. and {Henkel}, C.},
    title = "{H2 densities and masses of the molecular clouds close to the galactic center}",
  journal = {\aap},
 keywords = {Astronomical Maps, Density Distribution, Formaldehyde, Galactic Nuclei, Molecular Clouds, Absorption Spectra, Electron Transitions, Interstellar Matter, Line Spectra},
     year = 1983,
    month = aug,
   volume = 125,
    pages = {136-145},
   adsurl = {http://adsabs.harvard.edu/abs/1983A
  adsnote = {Provided by the SAO/NASA Astrophysics Data System}
}

@ARTICLE{gues85,
   author = {{Guesten}, R. and {Walmsley}, C.~M. and {Ungerechts}, H. and 
	{Churchwell}, E.},
    title = "{Temperature determinations in molecular clouds of the galactic center}",
  journal = {\aap},
 keywords = {Astronomical Spectroscopy, Galactic Nuclei, Milky Way Galaxy, Molecular Clouds, Temperature Measurement, Acetonitrile, Ammonia, Plasma Heating, Transition Probabilities},
     year = 1985,
    month = jan,
   volume = 142,
    pages = {381-387},
   adsurl = {http://adsabs.harvard.edu/abs/1985A
  adsnote = {Provided by the SAO/NASA Astrophysics Data System}
}



\begin{thebibliography}{111}
\expandafter\ifx\csname natexlab\endcsname\relax\def\natexlab#1{#1}\fi

\bibitem[{{Amo-Baladr{\'o}n} {et~al.}(2011){Amo-Baladr{\'o}n},
  {Mart{\'{\i}}n-Pintado}, \& {Mart{\'{\i}}n}}]{amo11}
{Amo-Baladr{\'o}n}, M.~A., {Mart{\'{\i}}n-Pintado}, J., \& {Mart{\'{\i}}n}, S.
  2011, \aap, 526, A54

\bibitem[{{Baganoff} {et~al.}(2003){Baganoff}, {Maeda}, {Morris}, {Bautz},
  {Brandt}, {Cui}, {Doty}, {Feigelson}, {Garmire}, {Pravdo}, {Ricker}, \&
  {Townsley}}]{baganoff}
{Baganoff}, F.~K., {et~al.} 2003, \apj, 591, 891

\bibitem[{{Bally} {et~al.}(1987){Bally}, {Stark}, {Wilson}, \&
  {Henkel}}]{bally87}
{Bally}, J., {Stark}, A.~A., {Wilson}, R.~W., \& {Henkel}, C. 1987, \apjs, 65,
  13

\bibitem[{{Bally} {et~al.}(1988){Bally}, {Stark}, {Wilson}, \&
  {Henkel}}]{bally88}
---. 1988, \apj, 324, 223

\bibitem[{{Bania}(1977)}]{bania77}
{Bania}, T.~M. 1977, \apj, 216, 381

\bibitem[{{Binney} {et~al.}(1991){Binney}, {Gerhard}, {Stark}, {Bally}, \&
  {Uchida}}]{binney91}
{Binney}, J., {Gerhard}, O.~E., {Stark}, A.~A., {Bally}, J., \& {Uchida}, K.~I.
  1991, \mnras, 252, 210

\bibitem[{{Bland-Hawthorn} \& {Cohen}(2003)}]{bland03}
{Bland-Hawthorn}, J., \& {Cohen}, M. 2003, \apj, 582, 246

\bibitem[{{Bolatto} {et~al.}(2013){Bolatto}, {Warren}, {Leroy}, {Walter},
  {Veilleux}, {Ostriker}, {Ott}, {Zwaan}, {Fisher}, {Weiss}, {Rosolowsky}, \&
  {Hodge}}]{ngc253}
{Bolatto}, A.~D., {et~al.} 2013, \nat, 499, 450

\bibitem[{{Burton} \& {Liszt}(1983)}]{burton83}
{Burton}, W.~B., \& {Liszt}, H.~S. 1983, in Astrophysics and Space Science
  Library, Vol. 105, Surveys of the Southern Galaxy, ed. W.~B. {Burton} \&
  F.~P. {Israel}, 149--157

\bibitem[{{Burton} \& {Liszt}(1992)}]{burton92}
{Burton}, W.~B., \& {Liszt}, H.~S. 1992, \aaps, 95, 9

\bibitem[{{Chen} {et~al.}(2010){Chen}, {Tremonti}, {Heckman}, {Kauffmann},
  {Weiner}, {Brinchmann}, \& {Wang}}]{chen10}
{Chen}, Y.-M., {Tremonti}, C.~A., {Heckman}, T.~M., {Kauffmann}, G., {Weiner},
  B.~J., {Brinchmann}, J., \& {Wang}, J. 2010, \aj, 140, 445

\bibitem[{{Christopher} {et~al.}(2005){Christopher}, {Scoville}, {Stolovy}, \&
  {Yun}}]{chris}
{Christopher}, M.~H., {Scoville}, N.~Z., {Stolovy}, S.~R., \& {Yun}, M.~S.
  2005, \apj, 622, 346

\bibitem[{{Coil} \& {Ho}(2000)}]{coil00}
{Coil}, A.~L., \& {Ho}, P.~T.~P. 2000, \apj, 533, 245

\bibitem[{{Condon}(1992)}]{condon92}
{Condon}, J.~J. 1992, \araa, 30, 575

\bibitem[{{Dahmen} {et~al.}(1998){Dahmen}, {Huttemeister}, {Wilson}, \&
  {Mauersberger}}]{dahmen}
{Dahmen}, G., {Huttemeister}, S., {Wilson}, T.~L., \& {Mauersberger}, R. 1998,
  \aap, 331, 959

\bibitem[{{Ekers} {et~al.}(1983){Ekers}, {van Gorkom}, {Schwarz}, \&
  {Goss}}]{ekers}
{Ekers}, R.~D., {van Gorkom}, J.~H., {Schwarz}, U.~J., \& {Goss}, W.~M. 1983,
  \aap, 122, 143

\bibitem[{{Enokiya} {et~al.}(2014){Enokiya}, {Torii}, {Schultheis}, {Asahina},
  {Matsumoto}, {Furuhashi}, {Nakamura}, {Dobashi}, {Yoshiike}, {Sato},
  {Furukawa}, {Moribe}, {Ohama}, {Sano}, {Okamoto}, {Mori}, {Hanaoka},
  {Nishimura}, {Hayakawa}, {Okuda}, {Yamamoto}, {Kawamura}, {Mizuno}, {Onishi},
  {Morris}, \& {Fukui}}]{enokiya}
{Enokiya}, R., {et~al.} 2014, \apj, 780, 72

\bibitem[{{Etxaluze} {et~al.}(2011){Etxaluze}, {Smith}, {Tolls}, {Stark}, \&
  {Gonz{\'a}lez-Alfonso}}]{etx}
{Etxaluze}, M., {Smith}, H.~A., {Tolls}, V., {Stark}, A.~A., \&
  {Gonz{\'a}lez-Alfonso}, E. 2011, \aj, 142, 134

\bibitem[{{Figer} {et~al.}(1999{\natexlab{a}}){Figer}, {Kim}, {Morris},
  {Serabyn}, {Rich}, \& {McLean}}]{figer99b}
{Figer}, D.~F., {Kim}, S.~S., {Morris}, M., {Serabyn}, E., {Rich}, R.~M., \&
  {McLean}, I.~S. 1999{\natexlab{a}}, \apj, 525, 750

\bibitem[{{Figer} {et~al.}(1999{\natexlab{b}}){Figer}, {McLean}, \&
  {Morris}}]{figer99a}
{Figer}, D.~F., {McLean}, I.~S., \& {Morris}, M. 1999{\natexlab{b}}, \apj, 514,
  202

\bibitem[{{Figer} {et~al.}(2002){Figer}, {Najarro}, {Gilmore}, {Morris}, {Kim},
  {Serabyn}, {McLean}, {Gilbert}, {Graham}, {Larkin}, {Levenson}, \&
  {Teplitz}}]{figer02}
{Figer}, D.~F., {et~al.} 2002, \apj, 581, 258

\bibitem[{{Fukui} {et~al.}(2006){Fukui}, {Yamamoto}, {Fujishita}, {Kudo},
  {Torii}, {Nozawa}, {Takahashi}, {Matsumoto}, {Machida}, {Kawamura},
  {Yonekura}, {Mizuno}, {Onishi}, \& {Mizuno}}]{fukui06}
{Fukui}, Y., {et~al.} 2006, Science, 314, 106

\bibitem[{{Ghez} {et~al.}(2008){Ghez}, {Salim}, {Weinberg}, {Lu}, {Do}, {Dunn},
  {Matthews}, {Morris}, {Yelda}, {Becklin}, {Kremenek}, {Milosavljevic}, \&
  {Naiman}}]{ghez08}
{Ghez}, A.~M., {et~al.} 2008, \apj, 689, 1044

\bibitem[{{Gillessen} {et~al.}(2009){Gillessen}, {Eisenhauer}, {Trippe},
  {Alexander}, {Genzel}, {Martins}, \& {Ott}}]{gill09}
{Gillessen}, S., {Eisenhauer}, F., {Trippe}, S., {Alexander}, T., {Genzel}, R.,
  {Martins}, F., \& {Ott}, T. 2009, \apj, 692, 1075

\bibitem[{{Goldreich} \& {Kwan}(1974)}]{goldreich}
{Goldreich}, P., \& {Kwan}, J. 1974, \apj, 189, 441

\bibitem[{{Goldsmith} \& {Langer}(1999)}]{gold99}
{Goldsmith}, P.~F., \& {Langer}, W.~D. 1999, \apj, 517, 209

\bibitem[{{Guesten} {et~al.}(1987){Guesten}, {Genzel}, {Wright}, {Jaffe},
  {Stutzki}, \& {Harris}}]{guesten87}
{Guesten}, R., {Genzel}, R., {Wright}, M.~C.~H., {Jaffe}, D.~T., {Stutzki}, J.,
  \& {Harris}, A.~I. 1987, \apj, 318, 124

\bibitem[{{G{\"u}sten} \& {Philipp}(2004)}]{gusten04}
{G{\"u}sten}, R., \& {Philipp}, S.~D. 2004, in The Dense Interstellar Medium in
  Galaxies, ed. S.~{Pfalzner}, C.~{Kramer}, C.~{Staubmeier}, \&
  A.~{Heithausen}, 253

\bibitem[{{Handa} {et~al.}(1987){Handa}, {Sofue}, {Nakai}, {Hirabayashi}, \&
  {Inoue}}]{handa87}
{Handa}, T., {Sofue}, Y., {Nakai}, N., {Hirabayashi}, H., \& {Inoue}, M. 1987,
  \pasj, 39, 709

\bibitem[{{Harris} {et~al.}(1985){Harris}, {Jaffe}, {Silber}, \&
  {Genzel}}]{harris}
{Harris}, A.~I., {Jaffe}, D.~T., {Silber}, M., \& {Genzel}, R. 1985, \apjl,
  294, L93

\bibitem[{{Herrnstein} \& {Ho}(2005{\natexlab{a}})}]{herrnstein}
{Herrnstein}, R.~M., \& {Ho}, P.~T.~P. 2005{\natexlab{a}}, \apj, 620, 287

\bibitem[{{Herrnstein} \& {Ho}(2005{\natexlab{b}})}]{nh3_05}
---. 2005{\natexlab{b}}, \apj, 620, 287

\bibitem[{{Ho} {et~al.}(2014){Ho}, {Kewley}, {Dopita}, {Medling}, {Allen},
  {Bland-Hawthorn}, {Bloom}, {Bryant}, {Croom}, {Fogarty}, {Goodwin}, {Green},
  {Konstantopoulos}, {Lawrence}, {L{\'o}pez-S{\'a}nchez}, {Owers}, {Richards},
  \& {Sharp}}]{ho14}
{Ho}, I.-T., {et~al.} 2014, \mnras, 444, 3894

\bibitem[{{Ho} {et~al.}(1991){Ho}, {Ho}, {Szczepanski}, {Jackson}, \&
  {Armstrong}}]{ho91}
{Ho}, P.~T.~P., {Ho}, L.~C., {Szczepanski}, J.~C., {Jackson}, J.~M., \&
  {Armstrong}, J.~T. 1991, \nat, 350, 309

\bibitem[{{Ho} {et~al.}(1985){Ho}, {Jackson}, {Barrett}, \& {Armstrong}}]{ho85}
{Ho}, P.~T.~P., {Jackson}, J.~M., {Barrett}, A.~H., \& {Armstrong}, J.~T. 1985,
  \apj, 288, 575

\bibitem[{{Hsieh} {et~al.}(2015){Hsieh}, {Ho}, \& {Hwang}}]{hsieh15}
{Hsieh}, P.~Y., {Ho}, P.~T.~P., \& {Hwang}, C.~Y. 2015, Paper II, in prep.

\bibitem[{{Huettemeister} {et~al.}(1993){Huettemeister}, {Wilson}, {Bania}, \&
  {Martin-Pintado}}]{huett}
{Huettemeister}, S., {Wilson}, T.~L., {Bania}, T.~M., \& {Martin-Pintado}, J.
  1993, \aap, 280, 255

\bibitem[{{Irvine} {et~al.}(1987){Irvine}, {Goldsmith}, \&
  {Hjalmarson}}]{irvine}
{Irvine}, W.~M., {Goldsmith}, P.~F., \& {Hjalmarson}, A. 1987, in Astrophysics
  and Space Science Library, Vol. 134, Interstellar Processes, ed. D.~J.
  {Hollenbach} \& H.~A. {Thronson}, Jr., 561--609

\bibitem[{{Jackson} {et~al.}(1993){Jackson}, {Geis}, {Genzel}, {Harris},
  {Madden}, {Poglitsch}, {Stacey}, \& {Townes}}]{jackson93}
{Jackson}, J.~M., {Geis}, N., {Genzel}, R., {Harris}, A.~I., {Madden}, S.,
  {Poglitsch}, A., {Stacey}, G.~J., \& {Townes}, C.~H. 1993, \apj, 402, 173

\bibitem[{{Jones} {et~al.}(2013){Jones}, {Burton}, {Cunningham}, {Tothill}, \&
  {Walsh}}]{jones13}
{Jones}, P.~A., {Burton}, M.~G., {Cunningham}, M.~R., {Tothill}, N.~F.~H., \&
  {Walsh}, A.~J. 2013, \mnras, 433, 221

\bibitem[{{Jones} {et~al.}(2012){Jones}, {Burton}, {Cunningham},
  {Requena-Torres}, {Menten}, {Schilke}, {Belloche}, {Leurini},
  {Mart{\'{\i}}n-Pintado}, {Ott}, \& {Walsh}}]{jones12}
{Jones}, P.~A., {et~al.} 2012, \mnras, 419, 2961

\bibitem[{{Keeney} {et~al.}(2006){Keeney}, {Danforth}, {Stocke}, {Penton},
  {Shull}, \& {Sembach}}]{keeney06}
{Keeney}, B.~A., {Danforth}, C.~W., {Stocke}, J.~T., {Penton}, S.~V., {Shull},
  J.~M., \& {Sembach}, K.~R. 2006, \apj, 646, 951

\bibitem[{{Koyama} {et~al.}(1996){Koyama}, {Maeda}, {Sonobe}, {Takeshima},
  {Tanaka}, \& {Yamauchi}}]{koyama96}
{Koyama}, K., {Maeda}, Y., {Sonobe}, T., {Takeshima}, T., {Tanaka}, Y., \&
  {Yamauchi}, S. 1996, \pasj, 48, 249

\bibitem[{{Kruijssen} {et~al.}(2015{\natexlab{a}}){Kruijssen}, {Dale}, \&
  {Longmore}}]{kru15}
{Kruijssen}, J.~M.~D., {Dale}, J.~E., \& {Longmore}, S.~N. 2015{\natexlab{a}},
  \mnras, 447, 1059

\bibitem[{{Kruijssen} {et~al.}(2015{\natexlab{b}}){Kruijssen}, {Dale}, \&
  {Longmore}}]{krui15}
---. 2015{\natexlab{b}}, \mnras, 447, 1059

\bibitem[{{Lang} {et~al.}(1999){Lang}, {Anantharamaiah}, {Kassim}, \&
  {Lazio}}]{lang99}
{Lang}, C.~C., {Anantharamaiah}, K.~R., {Kassim}, N.~E., \& {Lazio}, T.~J.~W.
  1999, \apjl, 521, L41

\bibitem[{{Lau} {et~al.}(2013){Lau}, {Herter}, {Morris}, {Becklin}, \&
  {Adams}}]{lau}
{Lau}, R.~M., {Herter}, T.~L., {Morris}, M.~R., {Becklin}, E.~E., \& {Adams},
  J.~D. 2013, \apj, 775, 37

\bibitem[{{Law} {et~al.}(2009){Law}, {Backer}, {Yusef-Zadeh}, \&
  {Maddalena}}]{law09}
{Law}, C.~J., {Backer}, D., {Yusef-Zadeh}, F., \& {Maddalena}, R. 2009, \apj,
  695, 1070

\bibitem[{{Law} {et~al.}(2008{\natexlab{a}}){Law}, {Yusef-Zadeh}, \&
  {Cotton}}]{law08a}
{Law}, C.~J., {Yusef-Zadeh}, F., \& {Cotton}, W.~D. 2008{\natexlab{a}}, \apjs,
  177, 515

\bibitem[{{Law} {et~al.}(2008{\natexlab{b}}){Law}, {Yusef-Zadeh}, {Cotton}, \&
  {Maddalena}}]{law08b}
{Law}, C.~J., {Yusef-Zadeh}, F., {Cotton}, W.~D., \& {Maddalena}, R.~J.
  2008{\natexlab{b}}, \apjs, 177, 255

\bibitem[{{Lique} {et~al.}(2006){Lique}, {Spielfiedel}, \&
  {Cernicharo}}]{lique}
{Lique}, F., {Spielfiedel}, A., \& {Cernicharo}, J. 2006, \aap, 451, 1125

\bibitem[{{Liu} {et~al.}(2012){Liu}, {Hsieh}, {Ho}, {Su}, {Wright}, {Sun}, \&
  {Minh}}]{liu12}
{Liu}, H.~B., {Hsieh}, P.-Y., {Ho}, P.~T.~P., {Su}, Y.-N., {Wright}, M., {Sun},
  A.-L., \& {Minh}, Y.~C. 2012, \apj, 756, 195

\bibitem[{{Longmore} {et~al.}(2013){Longmore}, {Bally}, {Testi}, {Purcell},
  {Walsh}, {Bressert}, {Pestalozzi}, {Molinari}, {Ott}, {Cortese}, {Battersby},
  {Murray}, {Lee}, {Kruijssen}, {Schisano}, \& {Elia}}]{longmore}
{Longmore}, S.~N., {et~al.} 2013, \mnras, 429, 987

\bibitem[{{Machida} {et~al.}(2009){Machida}, {Matsumoto}, {Nozawak},
  {Takahashi}, {Fukui}, {Kudo}, {Torii}, {Yamamoto}, {Fujishita}, \&
  {Tomisaki}}]{machida09}
{Machida}, M., {et~al.} 2009, \pasj, 61, 411

\bibitem[{{Maeda} {et~al.}(2002){Maeda}, {Baganoff}, {Feigelson}, {Morris},
  {Bautz}, {Brandt}, {Burrows}, {Doty}, {Garmire}, {Pravdo}, {Ricker}, \&
  {Townsley}}]{maeda}
{Maeda}, Y., {et~al.} 2002, \apj, 570, 671

\bibitem[{{Martin} {et~al.}(2004){Martin}, {Walsh}, {Xiao}, {Lane}, {Walker},
  \& {Stark}}]{martin04}
{Martin}, C.~L., {Walsh}, W.~M., {Xiao}, K., {Lane}, A.~P., {Walker}, C.~K., \&
  {Stark}, A.~A. 2004, \apjs, 150, 239

\bibitem[{{Mart{\'{\i}}n} {et~al.}(2012){Mart{\'{\i}}n},
  {Mart{\'{\i}}n-Pintado}, {Montero-Casta{\~n}o}, {Ho}, \&
  {Blundell}}]{martins12}
{Mart{\'{\i}}n}, S., {Mart{\'{\i}}n-Pintado}, J., {Montero-Casta{\~n}o}, M.,
  {Ho}, P.~T.~P., \& {Blundell}, R. 2012, \aap, 539, A29

\bibitem[{{McCray} \& {Kafatos}(1987)}]{mccray87}
{McCray}, R., \& {Kafatos}, M. 1987, \apj, 317, 190

\bibitem[{{Mezger} {et~al.}(1989){Mezger}, {Zylka}, {Salter}, {Wink}, {Chini},
  {Kreysa}, \& {Tuffs}}]{mezger}
{Mezger}, P.~G., {Zylka}, R., {Salter}, C.~J., {Wink}, J.~E., {Chini}, R.,
  {Kreysa}, E., \& {Tuffs}, R. 1989, \aap, 209, 337

\bibitem[{{Mills} {et~al.}(2013){Mills}, {G{\"u}sten}, {Requena-Torres}, \&
  {Morris}}]{mills13b}
{Mills}, E.~A.~C., {G{\"u}sten}, R., {Requena-Torres}, M.~A., \& {Morris},
  M.~R. 2013, \apj, 779, 47

\bibitem[{{Mills} \& {Morris}(2013)}]{mills13a}
{Mills}, E.~A.~C., \& {Morris}, M.~R. 2013, \apj, 772, 105

\bibitem[{{Minh} {et~al.}(2013){Minh}, {Liu}, {Ho}, {Hsieh}, {Su}, {Kim}, \&
  {Wright}}]{minh13}
{Minh}, Y.~C., {Liu}, H.~B., {Ho}, P.~T.~P., {Hsieh}, P.-Y., {Su}, Y.-N.,
  {Kim}, S.~S., \& {Wright}, M. 2013, \apj, 773, 31

\bibitem[{{Miyazaki} \& {Tsuboi}(2000)}]{miyazaki00}
{Miyazaki}, A., \& {Tsuboi}, M. 2000, \apj, 536, 357

\bibitem[{{Molinari} {et~al.}(2011){Molinari}, {Bally}, {Noriega-Crespo},
  {Compi{\`e}gne}, {Bernard}, {Paradis}, {Martin}, {Testi}, {Barlow}, {Moore},
  {Plume}, {Swinyard}, {Zavagno}, {Calzoletti}, {Di Giorgio}, {Elia},
  {Faustini}, {Natoli}, {Pestalozzi}, {Pezzuto}, {Piacentini}, {Polenta},
  {Polychroni}, {Schisano}, {Traficante}, {Veneziani}, {Battersby}, {Burton},
  {Carey}, {Fukui}, {Li}, {Lord}, {Morgan}, {Motte}, {Schuller},
  {Stringfellow}, {Tan}, {Thompson}, {Ward-Thompson}, {White}, \&
  {Umana}}]{molinari}
{Molinari}, S., {et~al.} 2011, \apjl, 735, L33

\bibitem[{{Montero-Casta{\~n}o} {et~al.}(2009){Montero-Casta{\~n}o},
  {Herrnstein}, \& {Ho}}]{maria}
{Montero-Casta{\~n}o}, M., {Herrnstein}, R.~M., \& {Ho}, P.~T.~P. 2009, \apj,
  695, 1477

\bibitem[{{Mori} {et~al.}(2009){Mori}, {Hyodo}, {Tsuru}, {Nobukawa}, \&
  {Koyama}}]{mori09}
{Mori}, H., {Hyodo}, Y., {Tsuru}, T.~G., {Nobukawa}, M., \& {Koyama}, K. 2009,
  \pasj, 61, 687

\bibitem[{{Morris} \& {Serabyn}(1996)}]{morris96}
{Morris}, M., \& {Serabyn}, E. 1996, \araa, 34, 645

\bibitem[{{Murray} {et~al.}(2011){Murray}, {M{\'e}nard}, \&
  {Thompson}}]{murray11}
{Murray}, N., {M{\'e}nard}, B., \& {Thompson}, T.~A. 2011, \apj, 735, 66

\bibitem[{{Ohyama} {et~al.}(2002){Ohyama}, {Taniguchi}, {Iye}, {Yoshida},
  {Sekiguchi}, {Takata}, {Saito}, {Kawabata}, {Kashikawa}, {Aoki}, {Sasaki},
  {Kosugi}, {Okita}, {Shimizu}, {Inata}, {Ebizuka}, {Ozawa}, {Yadoumaru},
  {Taguchi}, \& {Asai}}]{ohyama02}
{Ohyama}, Y., {et~al.} 2002, \pasj, 54, 891

\bibitem[{{Oka} {et~al.}(2005){Oka}, {Geballe}, {Goto}, {Usuda}, \&
  {McCall}}]{oka05}
{Oka}, T., {Geballe}, T.~R., {Goto}, M., {Usuda}, T., \& {McCall}, B.~J. 2005,
  \apj, 632, 882

\bibitem[{{Oka} {et~al.}(2008){Oka}, {Hasegawa}, {White}, {Sato}, {Tsuboi}, \&
  {Miyazaki}}]{oka08}
{Oka}, T., {Hasegawa}, T., {White}, G.~J., {Sato}, F., {Tsuboi}, M., \&
  {Miyazaki}, A. 2008, \pasj, 60, 429

\bibitem[{{Oka} {et~al.}(2012){Oka}, {Onodera}, {Nagai}, {Tanaka}, {Matsumura},
  \& {Kamegai}}]{oka12}
{Oka}, T., {Onodera}, Y., {Nagai}, M., {Tanaka}, K., {Matsumura}, S., \&
  {Kamegai}, K. 2012, \apjs, 201, 14

\bibitem[{{Oka} {et~al.}(1999){Oka}, {White}, {Hasegawa}, {Sato}, {Tsuboi}, \&
  {Miyazaki}}]{oka99}
{Oka}, T., {White}, G.~J., {Hasegawa}, T., {Sato}, F., {Tsuboi}, M., \&
  {Miyazaki}, A. 1999, \apj, 515, 249

\bibitem[{{Pedlar} {et~al.}(1989){Pedlar}, {Anantharamaiah}, {Ekers}, {Goss},
  {van Gorkom}, {Schwarz}, \& {Zhao}}]{pedlar89}
{Pedlar}, A., {Anantharamaiah}, K.~R., {Ekers}, R.~D., {Goss}, W.~M., {van
  Gorkom}, J.~H., {Schwarz}, U.~J., \& {Zhao}, J.-H. 1989, \apj, 342, 769

\bibitem[{{Reid}(1993)}]{reid93}
{Reid}, M.~J. 1993, \araa, 31, 345

\bibitem[{{Reid} \& {Brunthaler}(2004)}]{reid04}
{Reid}, M.~J., \& {Brunthaler}, A. 2004, \apj, 616, 872

\bibitem[{{Reid} {et~al.}(2014){Reid}, {Menten}, {Brunthaler}, {Zheng}, {Dame},
  {Xu}, {Wu}, {Zhang}, {Sanna}, {Sato}, {Hachisuka}, {Choi}, {Immer},
  {Moscadelli}, {Rygl}, \& {Bartkiewicz}}]{reid14}
{Reid}, M.~J., {et~al.} 2014, \apj, 783, 130

\bibitem[{{Requena-Torres} {et~al.}(2012){Requena-Torres}, {G{\"u}sten},
  {Wei{\ss}}, {Harris}, {Mart{\'{\i}}n-Pintado}, {Stutzki}, {Klein},
  {Heyminck}, \& {Risacher}}]{great}
{Requena-Torres}, M.~A., {et~al.} 2012, \aap, 542, L21

\bibitem[{{Rubin} {et~al.}(2014){Rubin}, {Prochaska}, {Koo}, {Phillips},
  {Martin}, \& {Winstrom}}]{rubin14}
{Rubin}, K.~H.~R., {Prochaska}, J.~X., {Koo}, D.~C., {Phillips}, A.~C.,
  {Martin}, C.~L., \& {Winstrom}, L.~O. 2014, \apj, 794, 156

\bibitem[{{Sakamoto} {et~al.}(2011){Sakamoto}, {Mao}, {Matsushita}, {Peck},
  {Sawada}, \& {Wiedner}}]{sakamoto11}
{Sakamoto}, K., {Mao}, R.-Q., {Matsushita}, S., {Peck}, A.~B., {Sawada}, T., \&
  {Wiedner}, M.~C. 2011, \apj, 735, 19

\bibitem[{{Sakamoto} {et~al.}(2006){Sakamoto}, {Ho}, {Iono}, {Keto}, {Mao},
  {Matsushita}, {Peck}, {Wiedner}, {Wilner}, \& {Zhao}}]{sakamoto06}
{Sakamoto}, K., {et~al.} 2006, \apj, 636, 685

\bibitem[{{Sawada} {et~al.}(2004){Sawada}, {Hasegawa}, {Handa}, \&
  {Cohen}}]{sawada04}
{Sawada}, T., {Hasegawa}, T., {Handa}, T., \& {Cohen}, R.~J. 2004, \mnras, 349,
  1167

\bibitem[{{Schinnerer} {et~al.}(2008){Schinnerer}, {B{\"o}ker}, {Meier}, \&
  {Calzetti}}]{ic342}
{Schinnerer}, E., {B{\"o}ker}, T., {Meier}, D.~S., \& {Calzetti}, D. 2008,
  \apjl, 684, L21

\bibitem[{{Sch{\"o}ier} {et~al.}(2005){Sch{\"o}ier}, {van der Tak}, {van
  Dishoeck}, \& {Black}}]{lamda}
{Sch{\"o}ier}, F.~L., {van der Tak}, F.~F.~S., {van Dishoeck}, E.~F., \&
  {Black}, J.~H. 2005, \aap, 432, 369

\bibitem[{{Scoville}(1972)}]{sco72}
{Scoville}, N.~Z. 1972, \apjl, 175, L127

\bibitem[{{Serabyn} {et~al.}(1992){Serabyn}, {Lacy}, \&
  {Achtermann}}]{serabyn92}
{Serabyn}, E., {Lacy}, J.~H., \& {Achtermann}, J.~M. 1992, \apj, 395, 166

\bibitem[{{Sharp} \& {Bland-Hawthorn}(2010)}]{sharp10}
{Sharp}, R.~G., \& {Bland-Hawthorn}, J. 2010, \apj, 711, 818

\bibitem[{{Shirley} {et~al.}(2003){Shirley}, {Evans}, {Young}, {Knez}, \&
  {Jaffe}}]{shirley}
{Shirley}, Y.~L., {Evans}, II, N.~J., {Young}, K.~E., {Knez}, C., \& {Jaffe},
  D.~T. 2003, \apjs, 149, 375

\bibitem[{{Simpson} {et~al.}(1999){Simpson}, {Witteborn}, {Cohen}, \&
  {Price}}]{simpson99}
{Simpson}, J.~P., {Witteborn}, F.~C., {Cohen}, M., \& {Price}, S.~D. 1999, in
  Astronomical Society of the Pacific Conference Series, Vol. 186, The Central
  Parsecs of the Galaxy, ed. H.~{Falcke}, A.~{Cotera}, W.~J. {Duschl},
  F.~{Melia}, \& M.~J. {Rieke}, 527

\bibitem[{{Sjouwerman} \& {Pihlstr{\"o}m}(2008)}]{sjo}
{Sjouwerman}, L.~O., \& {Pihlstr{\"o}m}, Y.~M. 2008, \apj, 681, 1287

\bibitem[{{Sofue}(1995)}]{sofue95}
{Sofue}, Y. 1995, \pasj, 47, 527

\bibitem[{{Sofue}(1996)}]{sofue96}
---. 1996, \apjl, 459, L69

\bibitem[{{Sofue} \& {Handa}(1984)}]{sofue84}
{Sofue}, Y., \& {Handa}, T. 1984, \nat, 310, 568

\bibitem[{{Takahashi} {et~al.}(2009){Takahashi}, {Nozawa}, {Matsumoto},
  {Machida}, {Fukui}, {Kudo}, {Torii}, {Yamamoto}, \&
  {Fujishita}}]{takahashi09}
{Takahashi}, K., {et~al.} 2009, \pasj, 61, 957

\bibitem[{{Torii} {et~al.}(2010){Torii}, {Kudo}, {Fujishita}, {Kawase},
  {Yamamoto}, {Kawamura}, {Mizuno}, {Onishi}, {Mizuno}, {Machida}, {Takahashi},
  {Nozawa}, {Matsumoto}, \& {Fukui}}]{torii}
{Torii}, K., {et~al.} 2010, \pasj, 62, 1307

\bibitem[{{Tsuboi} {et~al.}(1999){Tsuboi}, {Handa}, \& {Ukita}}]{tsuboi99}
{Tsuboi}, M., {Handa}, T., \& {Ukita}, N. 1999, \apjs, 120, 1

\bibitem[{{Tsuboi} {et~al.}(2009){Tsuboi}, {Miyazaki}, \& {Okumura}}]{tsuboi09}
{Tsuboi}, M., {Miyazaki}, A., \& {Okumura}, S.~K. 2009, \pasj, 61, 29

\bibitem[{{Tsuboi} {et~al.}(2011){Tsuboi}, {Tadaki}, {Miyazaki}, \&
  {Handa}}]{tsuboi11}
{Tsuboi}, M., {Tadaki}, K.-I., {Miyazaki}, A., \& {Handa}, T. 2011, \pasj, 63,
  763

\bibitem[{{Uchida} {et~al.}(1985){Uchida}, {Sofue}, \& {Shibata}}]{uchida85}
{Uchida}, Y., {Sofue}, Y., \& {Shibata}, K. 1985, \nat, 317, 699

\bibitem[{{van der Tak} {et~al.}(2007){van der Tak}, {Black}, {Sch{\"o}ier},
  {Jansen}, \& {van Dishoeck}}]{radex}
{van der Tak}, F.~F.~S., {Black}, J.~H., {Sch{\"o}ier}, F.~L., {Jansen}, D.~J.,
  \& {van Dishoeck}, E.~F. 2007, \aap, 468, 627

\bibitem[{{Veilleux} {et~al.}(2005){Veilleux}, {Cecil}, \&
  {Bland-Hawthorn}}]{wind05}
{Veilleux}, S., {Cecil}, G., \& {Bland-Hawthorn}, J. 2005, \araa, 43, 769

\bibitem[{{Wardle} \& {Yusef-Zadeh}(2008)}]{wardle08}
{Wardle}, M., \& {Yusef-Zadeh}, F. 2008, \apjl, 683, L37

\bibitem[{{Weaver} {et~al.}(1977){Weaver}, {McCray}, {Castor}, {Shapiro}, \&
  {Moore}}]{weaver77}
{Weaver}, R., {McCray}, R., {Castor}, J., {Shapiro}, P., \& {Moore}, R. 1977,
  \apj, 218, 377

\bibitem[{{Wright} {et~al.}(2001){Wright}, {Coil}, {McGary}, {Ho}, \&
  {Harris}}]{wright}
{Wright}, M.~C.~H., {Coil}, A.~L., {McGary}, R.~S., {Ho}, P.~T.~P., \&
  {Harris}, A.~I. 2001, \apj, 551, 254

\bibitem[{{Wu} {et~al.}(2014){Wu}, {Tully}, {Rizzi}, {Dolphin}, {Jacobs}, \&
  {Karachentsev}}]{wu14}
{Wu}, P.-F., {Tully}, R.~B., {Rizzi}, L., {Dolphin}, A.~E., {Jacobs}, B.~A., \&
  {Karachentsev}, I.~D. 2014, \aj, 148, 7

\bibitem[{{Yusef-Zadeh} {et~al.}(2004){Yusef-Zadeh}, {Hewitt}, \&
  {Cotton}}]{yusef04}
{Yusef-Zadeh}, F., {Hewitt}, J.~W., \& {Cotton}, W. 2004, \apjs, 155, 421

\bibitem[{{Yusef-Zadeh} \& {Morris}(1987)}]{yusef87}
{Yusef-Zadeh}, F., \& {Morris}, M. 1987, \apj, 320, 545

\bibitem[{{Yusef-Zadeh} {et~al.}(1984){Yusef-Zadeh}, {Morris}, \&
  {Chance}}]{yusef84}
{Yusef-Zadeh}, F., {Morris}, M., \& {Chance}, D. 1984, \nat, 310, 557

\bibitem[{{Yusef-Zadeh} {et~al.}(2009){Yusef-Zadeh}, {Hewitt}, {Arendt},
  {Whitney}, {Rieke}, {Wardle}, {Hinz}, {Stolovy}, {Lang}, {Burton}, \&
  {Ramirez}}]{yusef09}
{Yusef-Zadeh}, F., {et~al.} 2009, \apj, 702, 178

\bibitem[{{Zhao} {et~al.}(2013){Zhao}, {Morris}, \& {Goss}}]{zhao13}
{Zhao}, J.-H., {Morris}, M.~R., \& {Goss}, W.~M. 2013, \apj, 777, 146

\bibitem[{{Zhao} {et~al.}(2014){Zhao}, {Morris}, \& {Goss}}]{zhao14}
{Zhao}, J.-H., {Morris}, M.~R., \& {Goss}, W.~M. 2014, in IAU Symposium, Vol.
  303, IAU Symposium, ed. L.~O. {Sjouwerman}, C.~C. {Lang}, \& J.~{Ott},
  364--368

\end{thebibliography}

\clearpage

\begin{center}
\begin{deluxetable}{lcccc}
\tablewidth{0pt}
\tabletypesize{\scriptsize}
\setlength{\tabcolsep}{0.02in} 
\tablecaption{The Disk Ridge \label{t.obspar1}}
\tablehead{
}
\startdata
Line & CS(J = 1--0) & CS(J = 2--1) & CS(J = 4--3) & CS(J = 5--4) \\
\hline
$\int{T_{\rm{b}}}dv$\tablenotemark{a} & 69.9$\pm$14.1 & 85.3$\pm$11.8 & 66.1$\pm$11.0 & 52.3$\pm$15.4 \\
FWHM $dv$ (km s$^{-1}$) & 17 & 15 & 18 & 13 \\
\hline
Rotational Diagrams Method &   &  &  &  \\
\hline
$N_{\rm CS}$ (cm$^{-2}$) & \multicolumn{4}{c}{(4.2$\pm$0.6)$\times10^{14}$}\\
$T_{\rm rot}$ (K) & \multicolumn{4}{c}{10.0$\pm0.8$}\\
$T_{\rm rot}$(Low) (K) & \multicolumn{4}{c}{4.0$\pm0.8$}\\
$T_{\rm rot}$(High) (K) & \multicolumn{4}{c}{17.0$\pm8.6$}\\
\hline
Statistical Equilibrium Modeling Results &   &  &  &  \\
\hline
Minimum $\chi^{2}$\tablenotemark{b} & \multicolumn{4}{c}{0.6}\\
$N_{\rm CS}$ (cm$^{-2}$) & \multicolumn{4}{c}{4.0$\times10^{15}$}\\
$T_{\rm K}$ (K) & \multicolumn{4}{c}{18.2}\\
$\log(n_{\rm H_2})$ density (cm$^{-2}$) & \multicolumn{4}{c}{5.9}\\
Opacity & 0.7 & 2.8 & 5.3 & 4.5\\
Excitation temperature (K) & 20.3  & 17.1  &  14.8 & 12.6  \\

\enddata

\tablenotetext{a}{The integrated flux is average over the 9 points shown in Figure~\ref{fig-cs43spec1}, where 1 point is $\sim~40\arcsec$.
}
\tablenotetext{b}{The degree of freedom of our parameters is 3 and the derived $\chi^{2}$ corresponds to probability of 0.9.}

\end{deluxetable}
\end{center}

\begin{center}
\begin{deluxetable}{lcccc}
\tablewidth{0pt}
\tabletypesize{\scriptsize}
\setlength{\tabcolsep}{0.02in} 
\tablecaption{The Connecting Ridge \label{t.obspar2}}
\tablehead{
}
\startdata
Line & CS(J = 1--0) & CS(J = 2--1) & CS(J = 4--3) & CS(J = 5--4) \\
\hline
$\int{T_{\rm{b}}}dv$\tablenotemark{a} & 32.4$\pm$15.6 & 49.6$\pm$19.3 & 41.9$\pm$7.7 & 20.6$\pm$10.9 \\
FWHM $dv$ (km s$^{-1}$) & 17 & 15 & 18 & 13 \\
\hline
Rotational Diagrams Method &   &  &  &  \\
\hline
$N_{\rm CS}$ (cm$^{-2}$) & \multicolumn{4}{c}{(2.4$\pm$0.9)$\times10^{14}$}\\
$T_{\rm rot}$ (K) & \multicolumn{4}{c}{9.0$\pm1.5$}\\
$T_{\rm rot}$(Low) (K) & \multicolumn{4}{c}{5.0$\pm3.2$}\\
$T_{\rm rot}$(High) (K) & \multicolumn{4}{c}{10.0$\pm4.9$}\\
\hline
Statistical Equilibrium Modeling Results &   &  &  &  \\
\hline
Minimum $\chi^{2}$\tablenotemark{b} & \multicolumn{4}{c}{0.39}\\
$N_{\rm CS}$ (cm$^{-2}$) & \multicolumn{4}{c}{2.0$\times10^{15}$}\\
$T_{\rm K}$ (K) & \multicolumn{4}{c}{18.2}\\
$\log(n_{\rm H_2})$ density (cm$^{-2}$) & \multicolumn{4}{c}{5.7}\\
Opacity & 0.32 & 1.56 & 3.0 & 2.17\\
Excitation temperature (K) & 24.0 & 16.0 & 12.0 & 9.4 \\

\enddata

\tablenotetext{a}{The integrated flux is average over the 9 points shown in Figure~\ref{fig-cs43spec2}, where 1 point is $\sim~40\arcsec$.
}
\tablenotetext{b}{The degree of freedom of our parameters is 3 and the derived $\chi^{2}$ corresponds to probability of 0.9.}

\end{deluxetable}
\end{center}

\begin{center}
\begin{deluxetable}{lcccc}
\tablewidth{0pt}
\tabletypesize{\scriptsize}
\setlength{\tabcolsep}{0.02in} 
\tablecaption{The Base of the Polar Arc \label{t.obspar3}}
\tablehead{
}
\startdata
Line & CS(J = 1--0) & CS(J = 2--1) & CS(J = 4--3) & CS(J = 5--4) \\
\hline
$\int{T_{\rm{b}}}dv$\tablenotemark{a} & 52.4$\pm$11.9 & 63.8$\pm$12.4 & 54.2$\pm$9.7 & 25.6$\pm$10.5 \\
\hline
Rotational Diagrams Method &   &  &  &  \\
\hline
$N_{\rm CS}$ (cm$^{-2}$) & \multicolumn{4}{c}{(3.4$\pm$0.6)$\times10^{14}$}\\
$T_{\rm rot}$ (K) & \multicolumn{4}{c}{9.0$\pm0.8$}\\
$T_{\rm rot}$(Low) (K) & \multicolumn{4}{c}{4.0$\pm1.0$}\\
$T_{\rm rot}$(High) (K) & \multicolumn{4}{c}{10.0$\pm3.7$}\\
\hline
Statistical Equilibrium Modeling Results &   &  &  &  \\
\hline
Minimum $\chi^{2}$\tablenotemark{b} & \multicolumn{4}{c}{1.5}\\
$N_{\rm CS}$ (cm$^{-2}$) & \multicolumn{4}{c}{3.2$\times10^{15}$}\\
$T_{\rm K}$ (K) & \multicolumn{4}{c}{18.2}\\
$\log(n_{\rm H_2})$ density (cm$^{-2}$) & \multicolumn{4}{c}{5.6}\\
Opacity & 0.61 & 2.78 & 5.19 & 3.89\\
Excitation temperature (K) & 21.8 & 15.6 & 12.1 & 9.5 \\

\enddata

\tablenotetext{a}{The integrated flux is average over the 9 points shown in Figure~\ref{fig-cs43spec3}, where 1 point is $\sim~40\arcsec$.
}
\tablenotetext{b}{The degree of freedom of our parameters is 3 and the derived $\chi^{2}$ corresponds to probability of 0.7.}

\end{deluxetable}
\end{center}

\begin{figure}
\begin{center}
\epsscale{0.5}
\includegraphics[angle=0,scale=0.3]{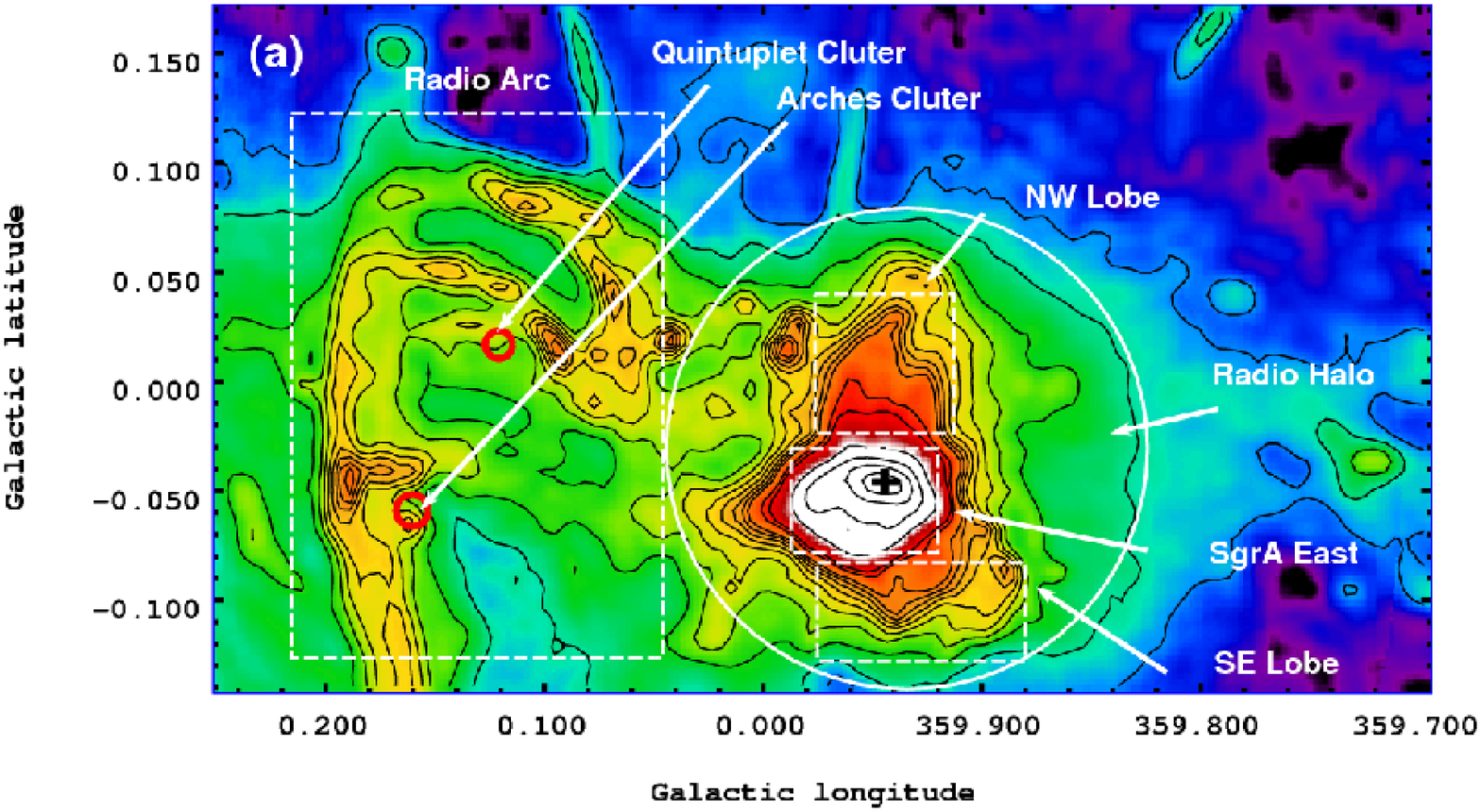}
\includegraphics[angle=0,scale=0.3]{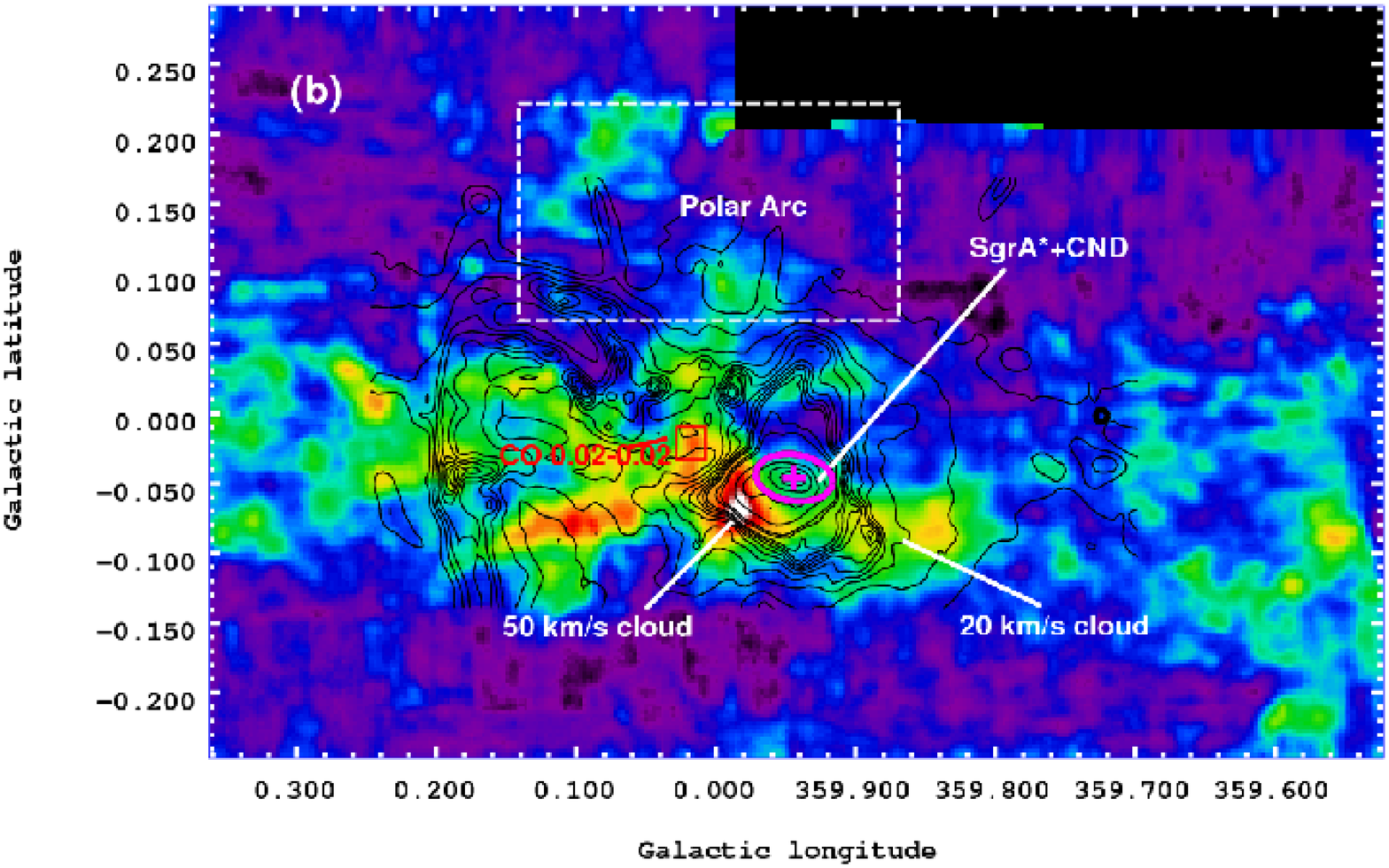}
\caption[]{(a) 20 cm continuum map taken by the VLA \citep{yusef04} (beam$=$30$\arcsec$). Several well studied features related to the massive stars in the Galactic center are labeled. The contour levels are 0.15, 0.2-1 (step of 0.1), 1.3, 1.5, 3-9 (step of 1) Jy beam$^{-1}$. (b) Integrated intensity map of the CS(J = 1--0) line (color) \citep{tsuboi99} with the beam size of 60$\arcsec$ overlaid with the 20 cm radio continuum map (contours, the levels are the same with (a)). The cross and ellipse label the positions of SgA* and the CND, respectively. The major features of the polar arc, the 20 km s$^{-1}$ cloud, 50 km s$^{-1}$ cloud, and the HVCC CO0.02-0.02 \citep{oka99} are labeled. The color bar denotes the unit of K m s$^{-1}$.}
\label{fig-fig1}
\end{center}
\end{figure}

\begin{figure}
\begin{center}
\epsscale{0.5}
\includegraphics[angle=0,scale=0.3]{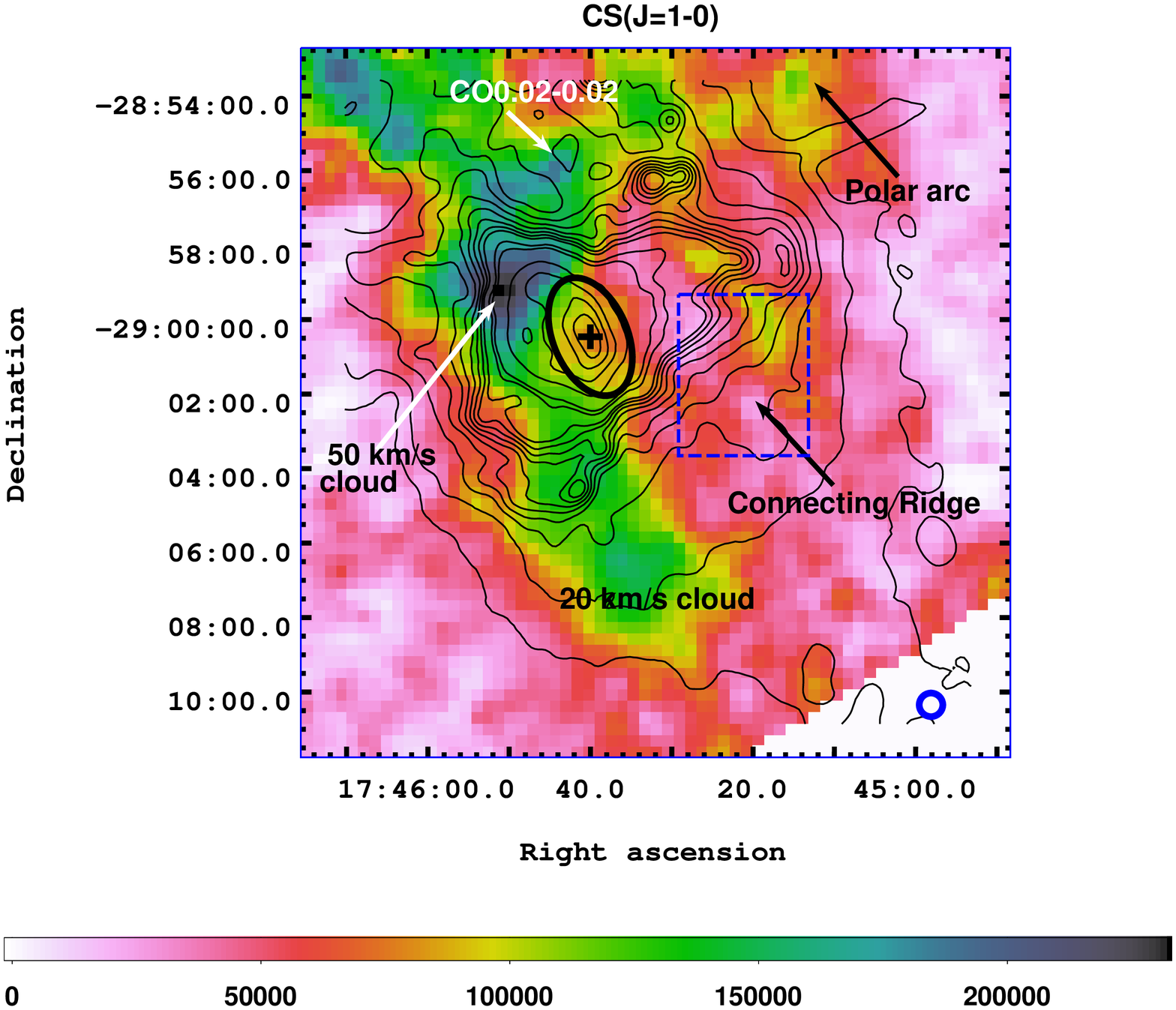}
\includegraphics[angle=0,scale=0.3]{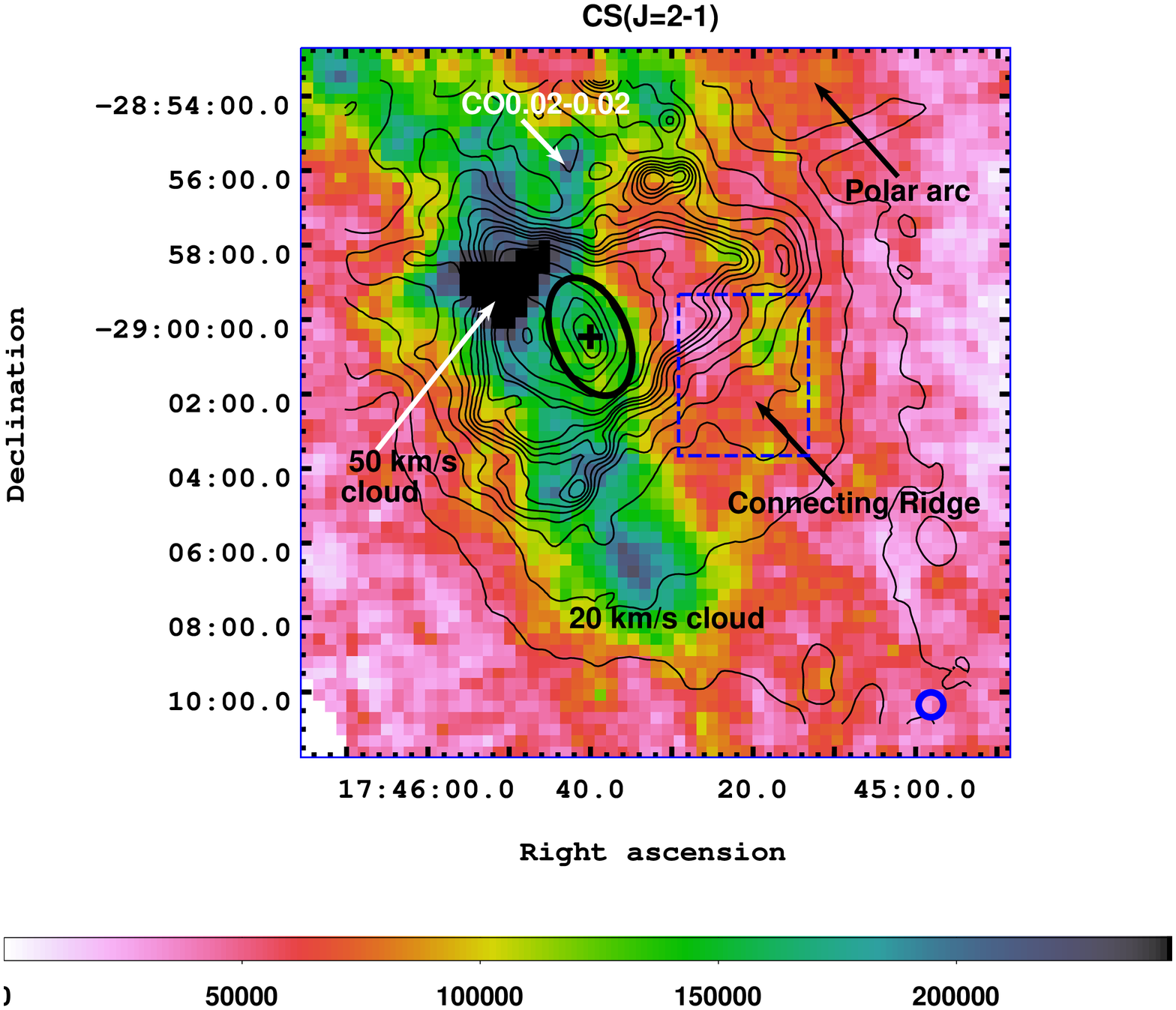}
\includegraphics[angle=0,scale=0.3]{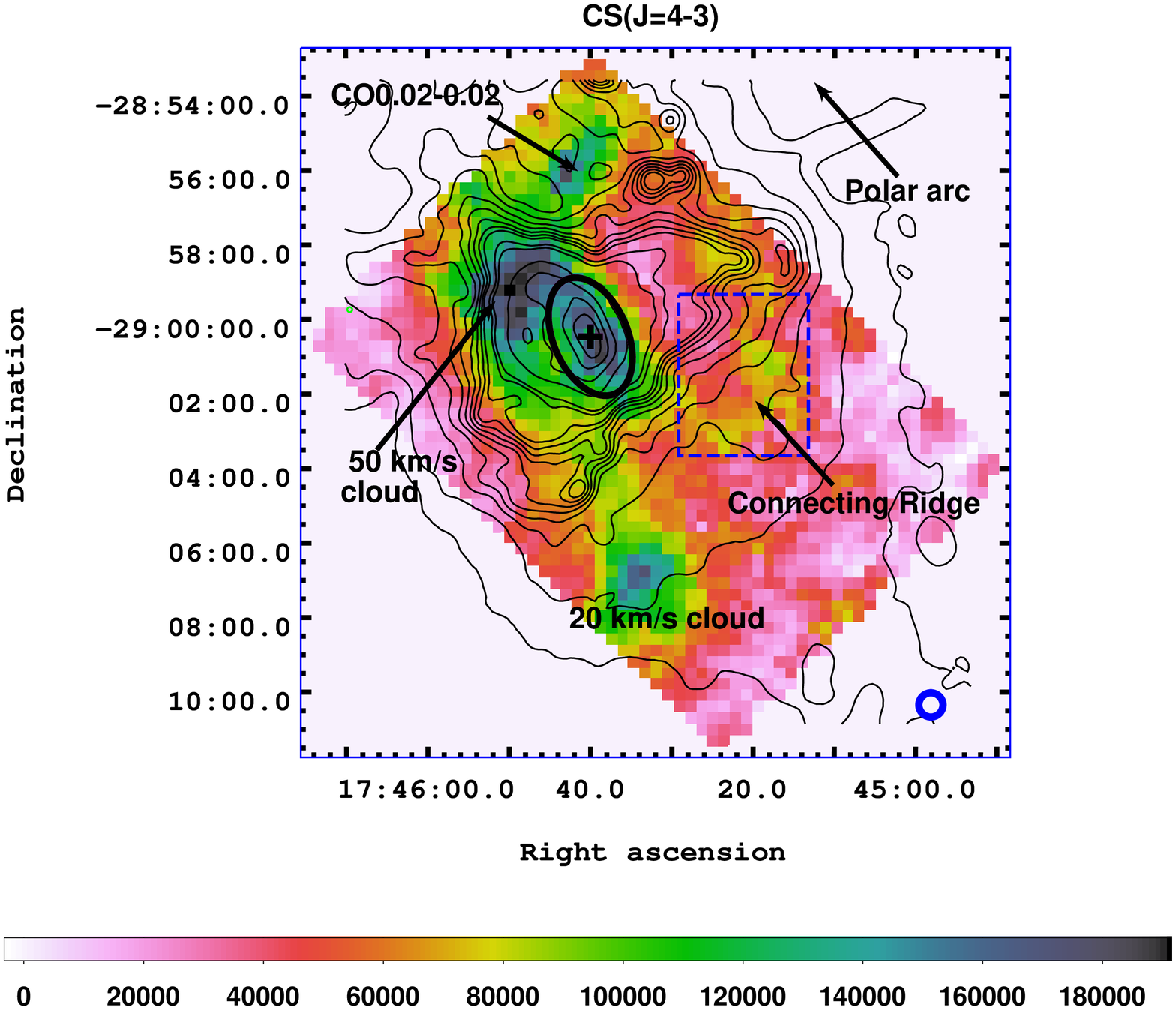}
\includegraphics[angle=0,scale=0.3]{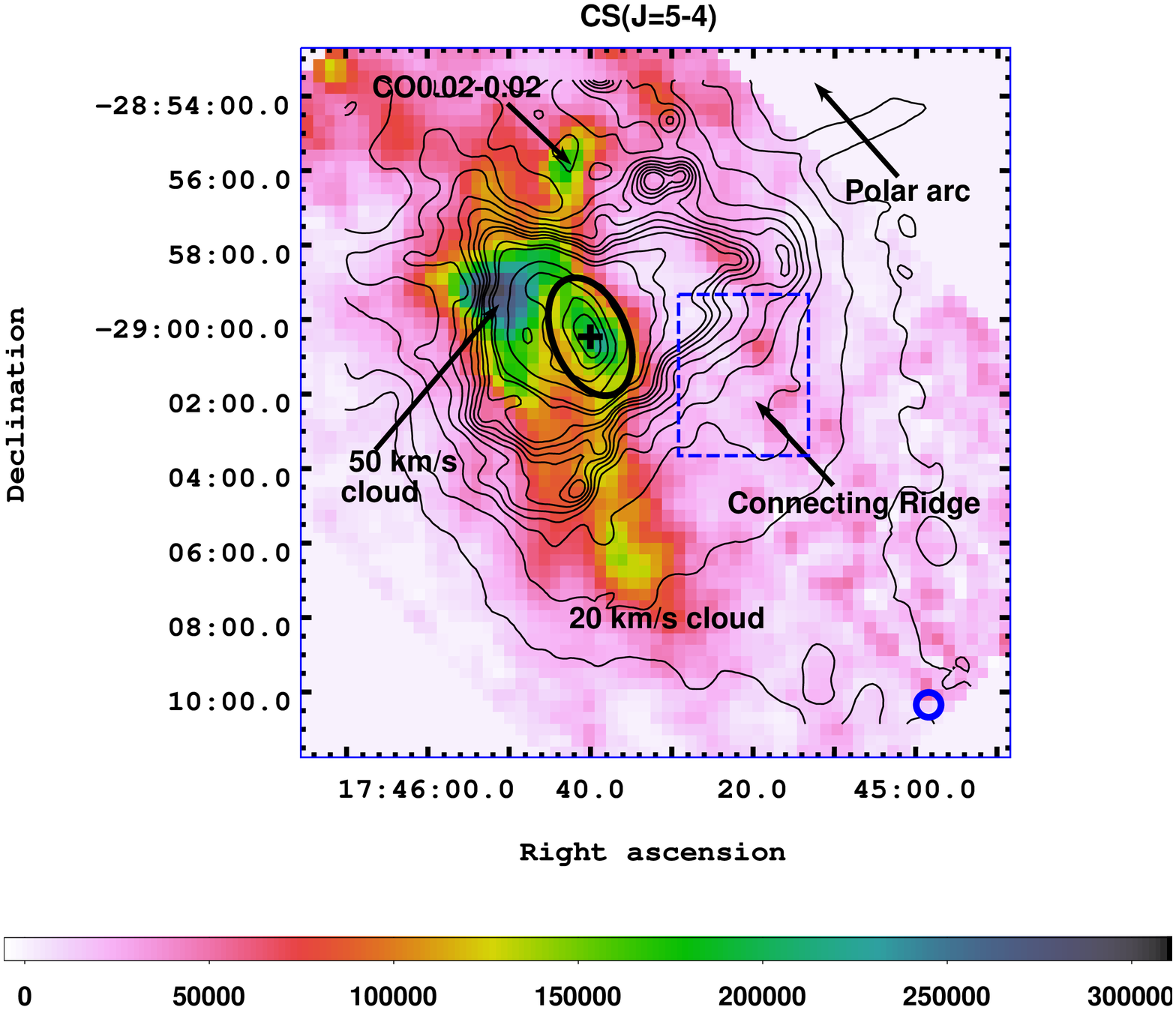}
\caption[]{Integrated intensity maps of the CS(J = 1--0) \citep{tsuboi99}, CS(J = 2--1) (Paper II), CS(J = 4--3), and CS(J = 5--4) lines from $-160$ km s$^{-1}$ to 160 km s$^{-1}$ (color map). The color bar denotes the intensity in unit of K m s$^{-1}$. The beam sizes were smoothed to 40$\arcsec$ (blue circle in the bottom right corner). The cross and ellipse mark the positions of SgA* and the CND, respectively. The contours of the 20 cm continuum emission are overlaid on the color maps. The connecting ridge (CR) found in the CS(J = 4--3) line map was labeled.}
\label{fig-cs43mom0}
\end{center}
\end{figure}

\begin{figure}
\begin{center}
\includegraphics[angle=0,scale=0.37]{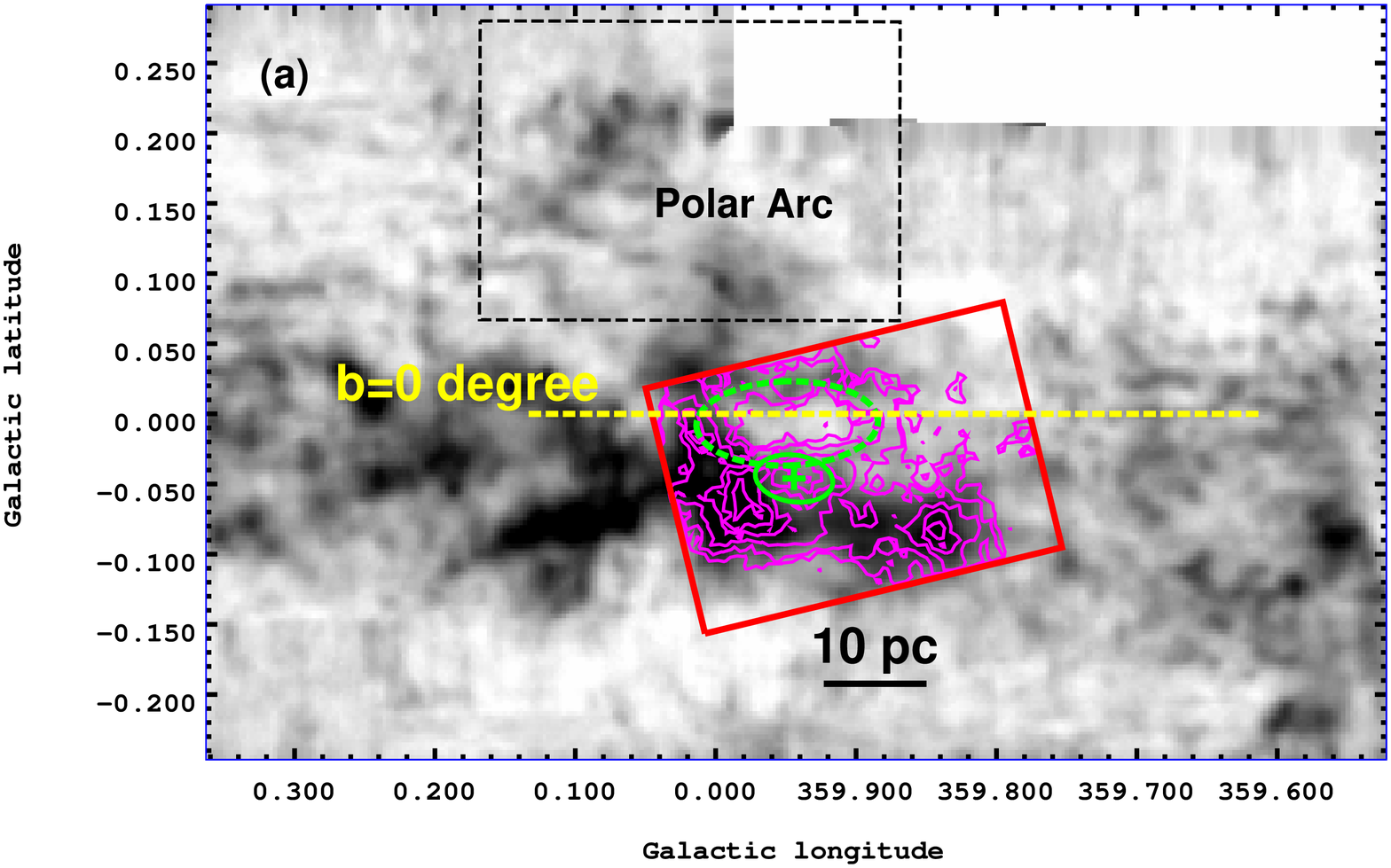}
\includegraphics[angle=0,scale=0.37]{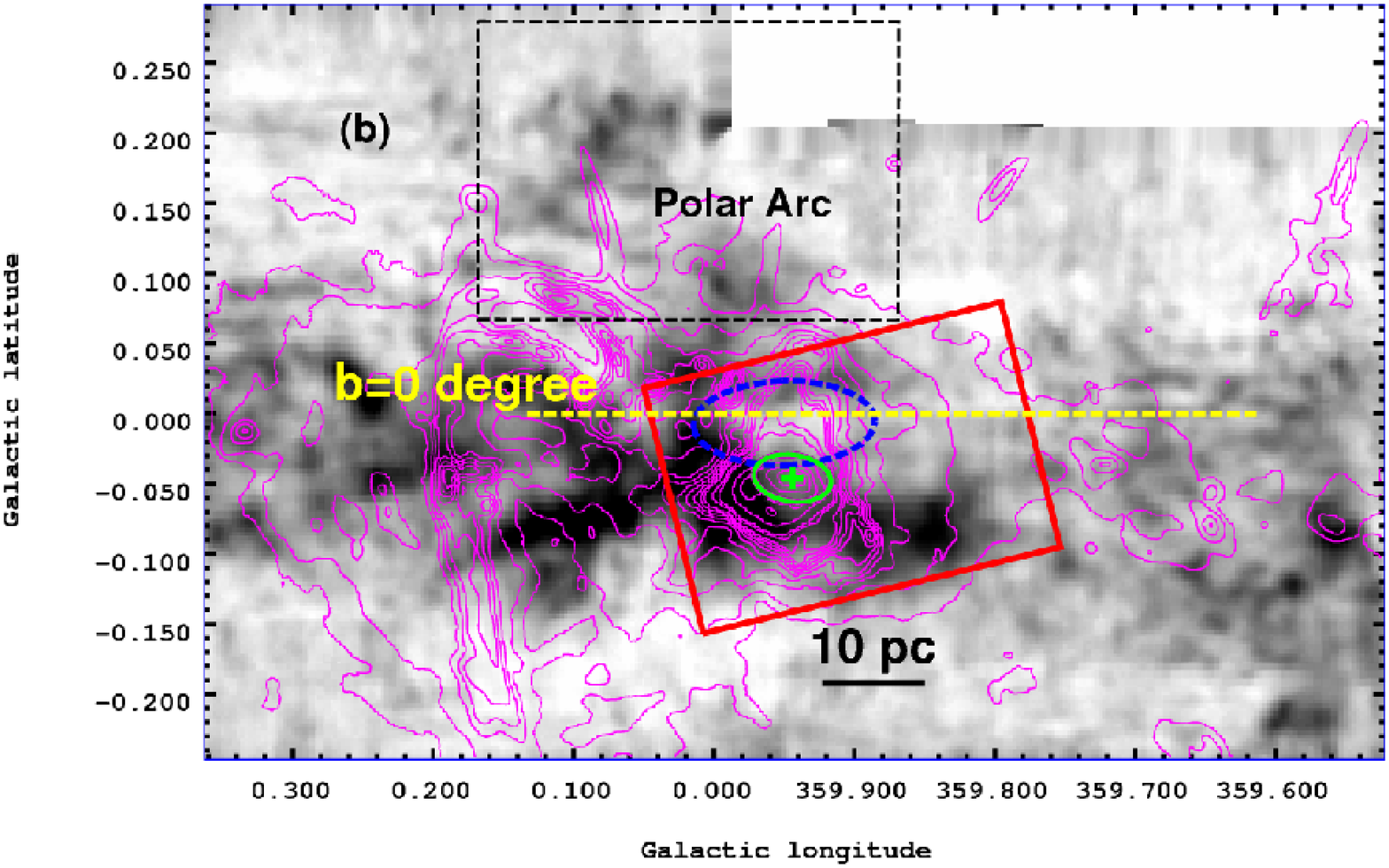}
\caption[]{(a) Integrated intensity maps of the 45m CS(J = 1--0) line (grey), the CSO CS(J = 4--3) line (pink contours). The contour levels are 15, 20, 30, 40, 50, 60 K km s$^{-1}$. The red rectangle shows the map size of the CSO CS(J = 4--3) line observations. The green ellipse marks the CND, and the green cross marks the SgrA*. The Galactic latitude of 0$\degr$ is shown as dashed-line. The green dashed-ellipse indicates the northern molecular loop mentioned in Sect.\ref{sect-mom0}. (b) Same with (a), but the contours are the 20 cm radio continuum (pink contours). The contour levels are 0.15, 0.2-1 (step of 0.1), 1.3, 1.5, 3-9 (step of 1) Jy beam$^{-1}$.}
\label{fig-polar}
\end{center}
\end{figure}

\clearpage

\begin{figure}
\begin{center}
\epsscale{0.5}
\includegraphics[angle=0,scale=0.8]{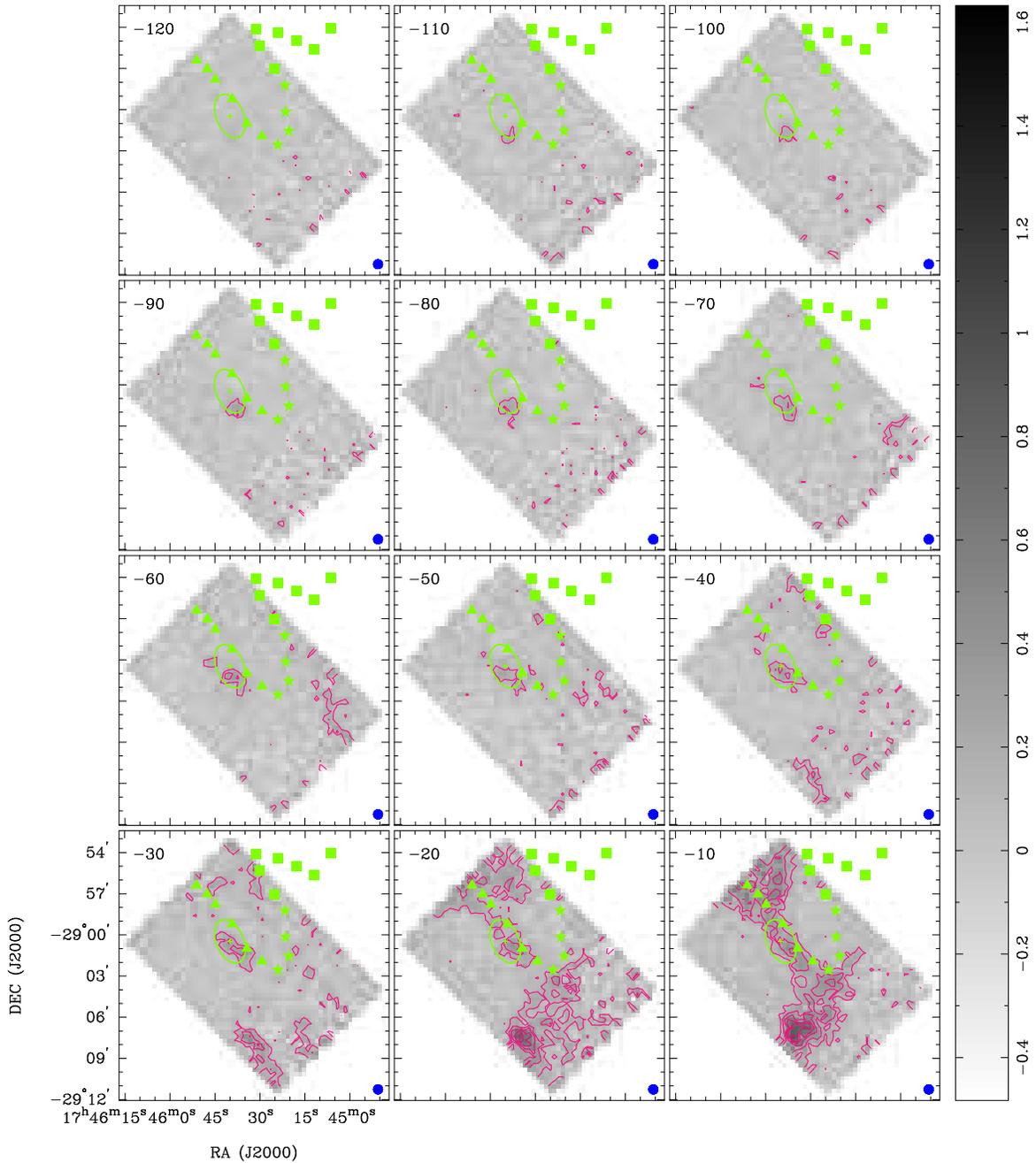}
\caption[]{Channel maps of the CS(J = 4--3) line emission. The beam size is 38$\arcsec$ (the blue circles on the bottom right corner). The systematic velocities are labeled in the top left corner in unit of km s$^{-1}$. The contours are -5, 3, 6, 9, 12, 15, 18, 21, 24$\sigma$, where $\sigma$ is 0.05 K. The triangles show the ``disk ridge''. The stars show the ``connecting ridge'' revealed in our CS(J = 4--3) line map. The squares mark the ``polar arc'' called by \citet{bally88}. Our CS(J = 4--3) line map does not cover the polar arc, and we show the larger CS(J = 1--0) line map \citep{tsuboi99} in Figure~\ref{fig-cs10chan} for comparison. The color bar is in unit of K ($T_{\rm A}$*).}
\label{fig-cs43chan1}
\end{center}
\end{figure}

\begin{figure}
\addtocounter{figure}{-1}
\begin{center}
\epsscale{0.5}
\includegraphics[angle=0,scale=0.8]{f4b.ps}
\caption[Channel maps of the CS(J = 4--3) line emission.]{\small Continue.}
\label{fig-cs43chan2}
\end{center}
\end{figure}

\begin{figure}
\begin{center}
\epsscale{0.7}
\includegraphics[angle=0,scale=0.9]{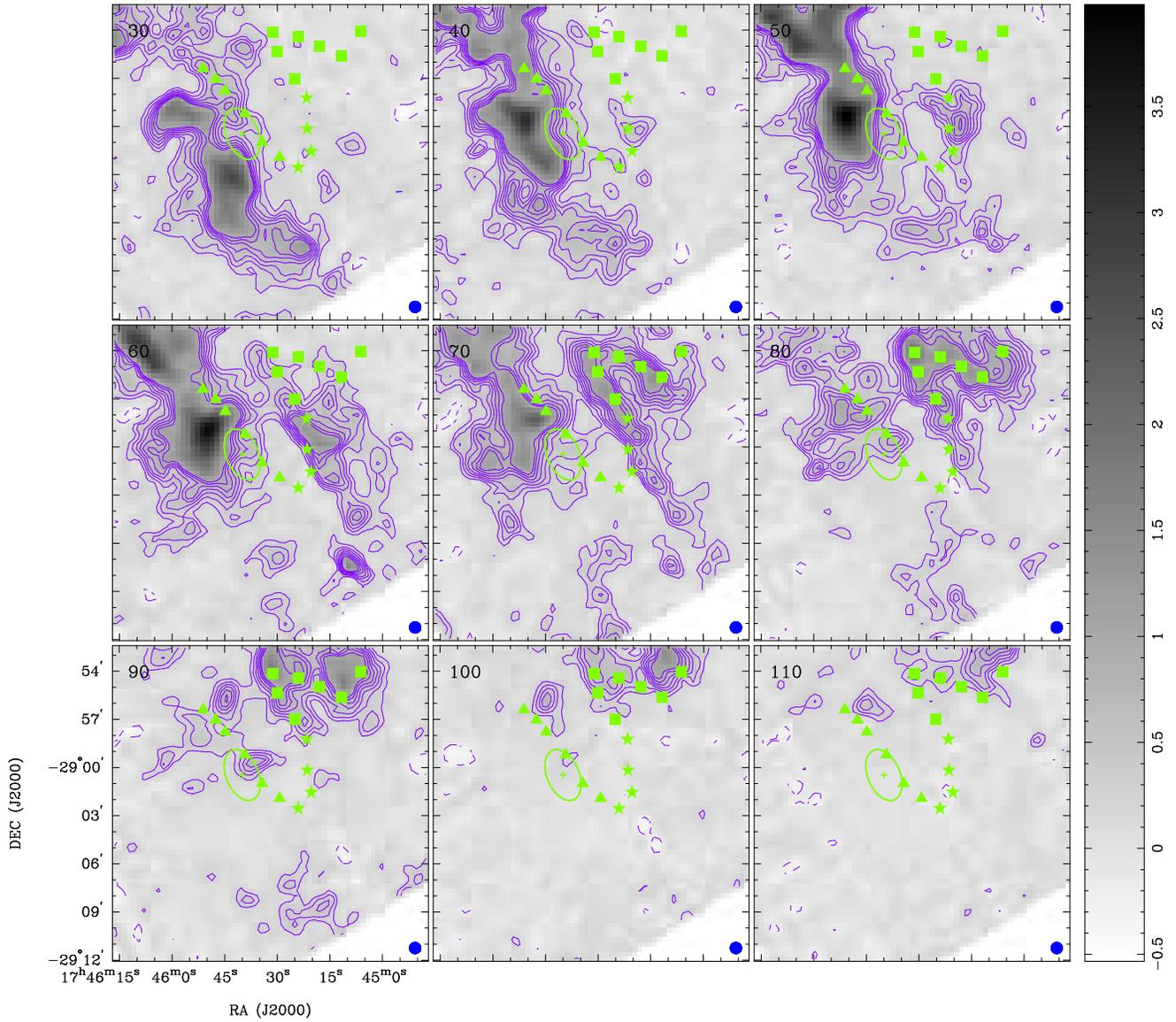}
\caption[]{Channel maps of the CS(J = 1--0) line emission \citep{tsuboi99}. The systematic velocities are shown in the top left corner. The contours are -5, 6, 9, 12, 15, 18, 21, 24$\sigma$, where $\sigma$ is 0.04 K.  We show the CS(J = 1--0) channel maps for comparing with Figure~\ref{fig-cs43chan1} from 30 km s$^{-1}$ to 110 km s$^{-1}$. The labels are the same with Figure~\ref{fig-cs43chan1}. The color bar is in unit of K ($T_{\rm A}$*).}
\label{fig-cs10chan}
\end{center}
\end{figure}

\clearpage

\begin{figure}
\begin{center}
\epsscale{0.5}
\includegraphics[angle=0,scale=0.8]{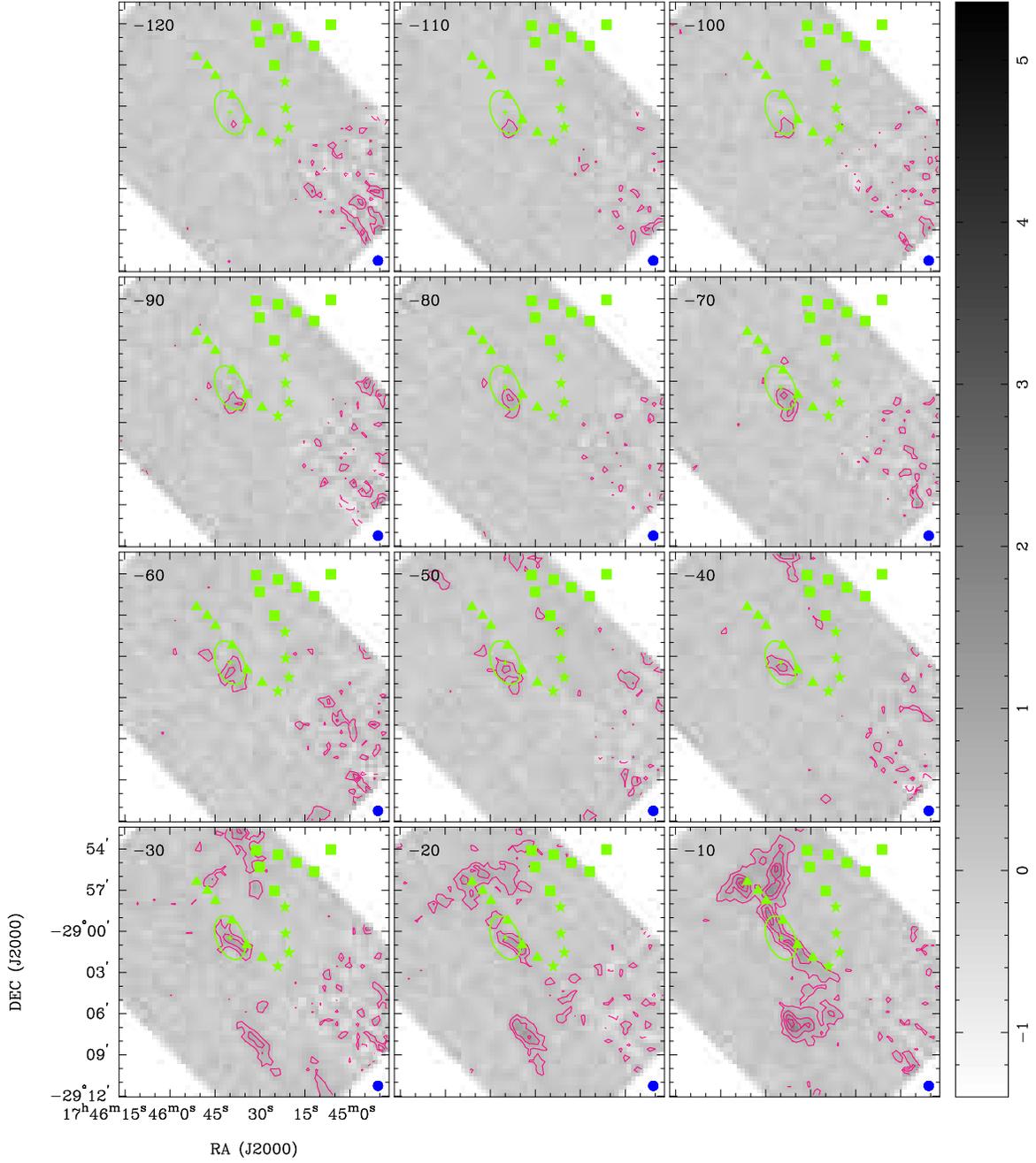}
\caption[]{Channel maps of the CS(J = 5--4) line emission. The beam size was smoothed to 38$\arcsec$ to compare with the CS(J = 4--3) line map. The numbers on the top left corner are the systematic velocity. The contours are, -5, 3, 6, 9, 12, 15, 18, 21, 24$\sigma$, where $\sigma$ is 0.13 K. Our CS(J = 5--4) line maps are slightly larger than the CS(J = 4--3) line map, and cover the base of the PA. The color bar is in unit of K ($T_{\rm A}$*).}
\label{fig-cs54chan1}
\end{center}
\end{figure}

\begin{figure}
\addtocounter{figure}{-1}
\begin{center}
\epsscale{0.5}
\includegraphics[angle=0,scale=0.8]{f6b.ps}
\caption[Channel map of CS(J = 5--4) line emission.]{\small Continue.}
\label{fig-cs54chan2}
\end{center}
\end{figure}

\begin{figure}
\begin{center}
\epsscale{0.2}
\includegraphics[angle=0,scale=0.6]{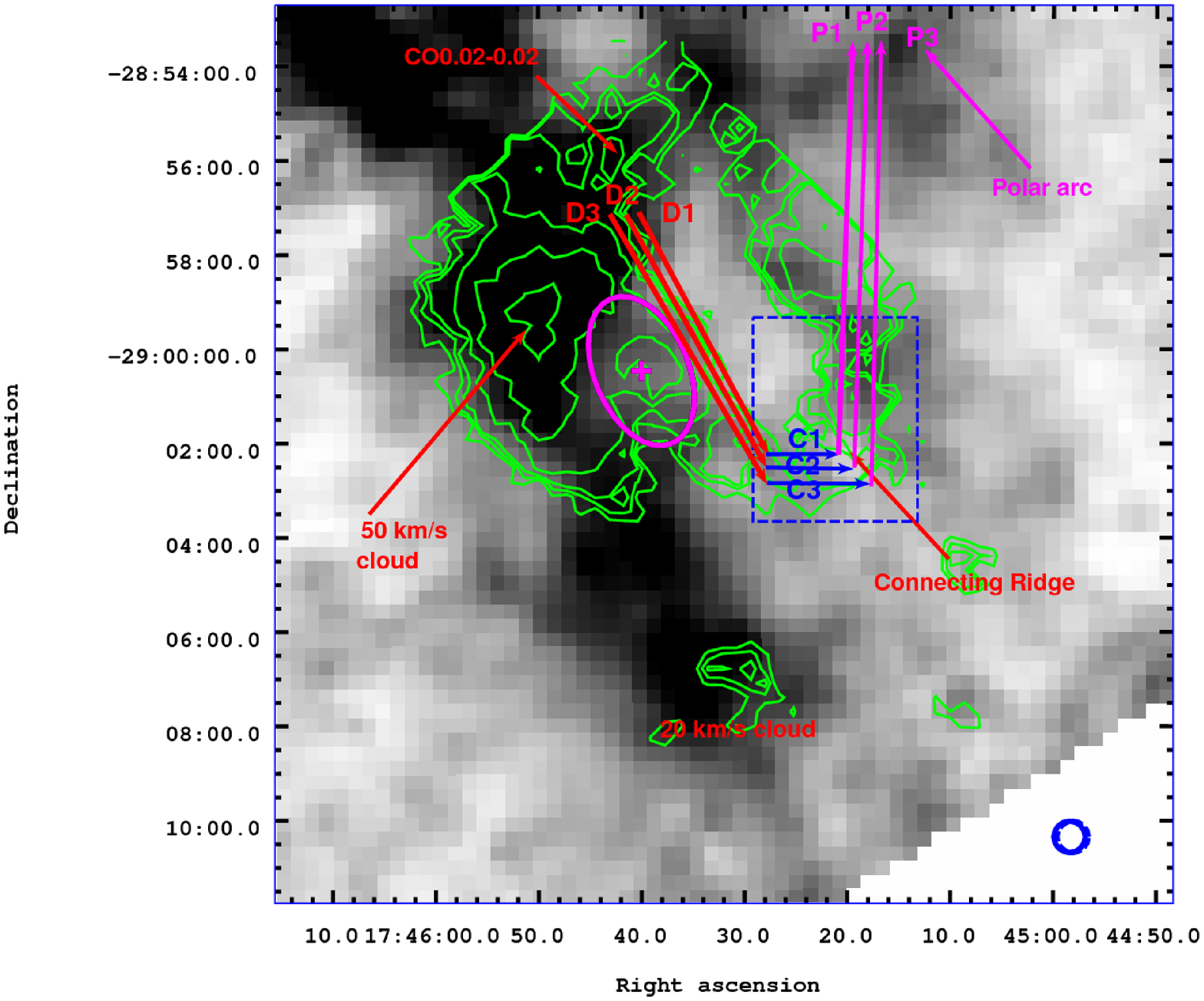}
\caption[]{CS(J = 1--0) integrated intensity map within $\pm160$ km s$^{-1}$ (grey) overlaid on the CSO CS(J = 4--3) line integrated from 30 km s$^{-1}$ to 90 km s$^{-1}$(contours). The contour levels of the CS(J = 4--3) line are, 11, 13, 15, 20, 30, 50 K km s$^{-1}$.
The location of the CND and SgrA* were labeled by the ellipse and cross, respectively. The slices for the pv-diagrams are shown in Figure~\ref{fig-pv-comb} and Figure~\ref{fig-pv-c1}. The directions of the slices are shown as arrow, D1, D2, D3 indicate the pv-diagrams of the disk ridge, C1, C2, and C3 indicate the slices of the CR, and P1, P2, P3 indicate the slices along the base of the PA.}
\label{fig-pv-mom0}
\end{center}
\end{figure}

\begin{figure}
\begin{center}
\epsscale{0.5}
\includegraphics[angle=-90,scale=0.7]{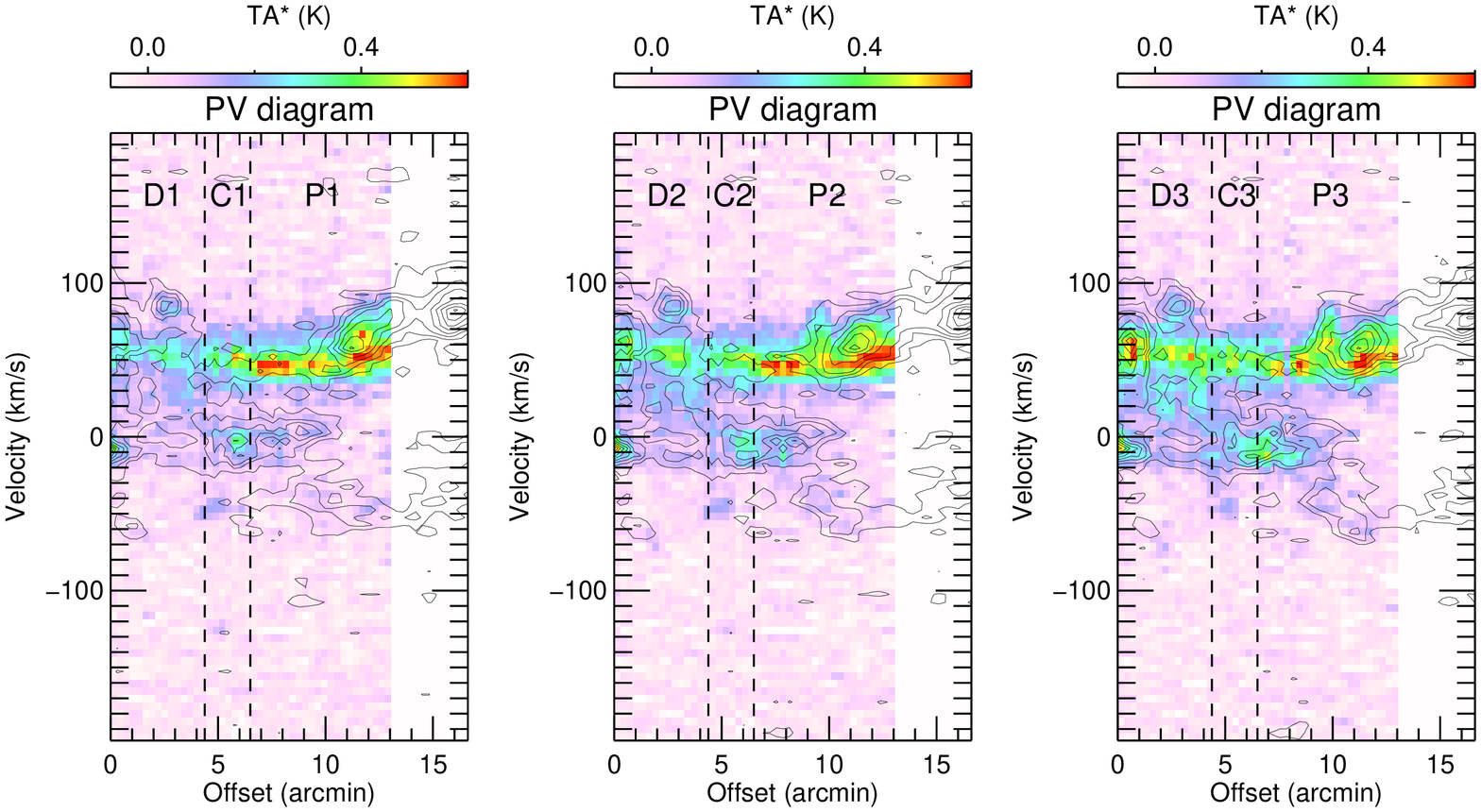}
\caption[]{Merged pv-diagrams of the disk ridge (D1,D2,D3), CR (C1,C2,C3), and the PA (P1,P2,P3) traced by the CS(J = 4--3) (color) and the CS(J = 1--0) line (contours). The first contour is 2.5 K, and in step of 2.5 K. The black dashed lines mark the cuts of the DR, CR, and PA in Figure~\ref{fig-pv-mom0}. 
The origin of the x-axis starts from the top points of D1, D2, and D3. The pv-diagrams of D1-C1-P1, D2-C2-P2, and D3-C3-P3 are merged to show the connections.}
\label{fig-pv-comb}
\end{center}
\end{figure}

\begin{figure}
\begin{center}
\epsscale{0.2}
\includegraphics[angle=-90,scale=0.9]{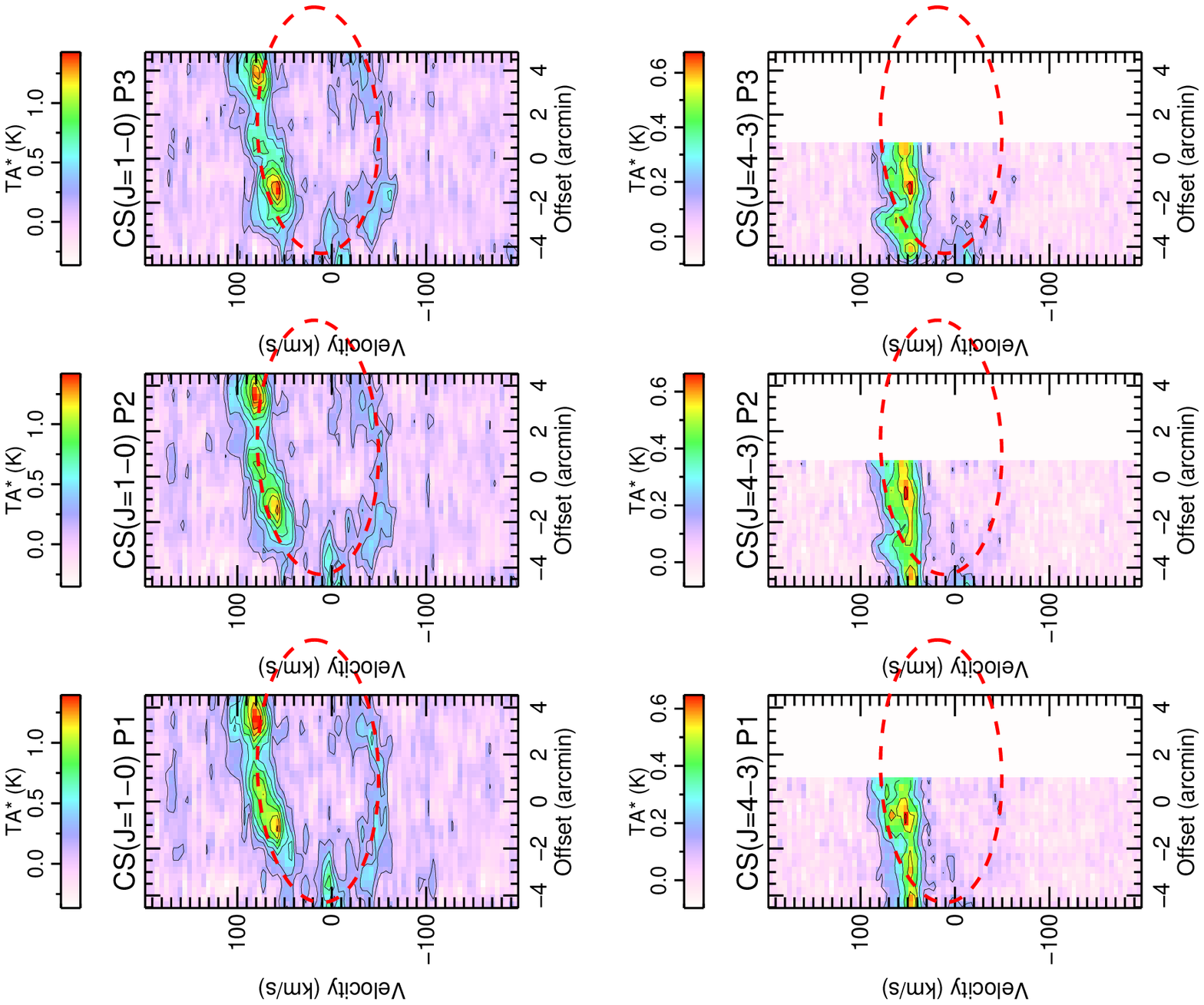}
\caption[]{PV-diagrams of the PA. Top: the CS(J = 1--0) line data sliced with P1, P2, and P3. Bottom: The CS(J = 4--3) line data sliced with P1, P2, and P3. The red ellipse marks the expanding motions of the PA.}
\label{fig-pv-c1}
\end{center}
\end{figure}

\clearpage

\begin{figure}
\begin{center}
\includegraphics[angle=0,scale=0.3]{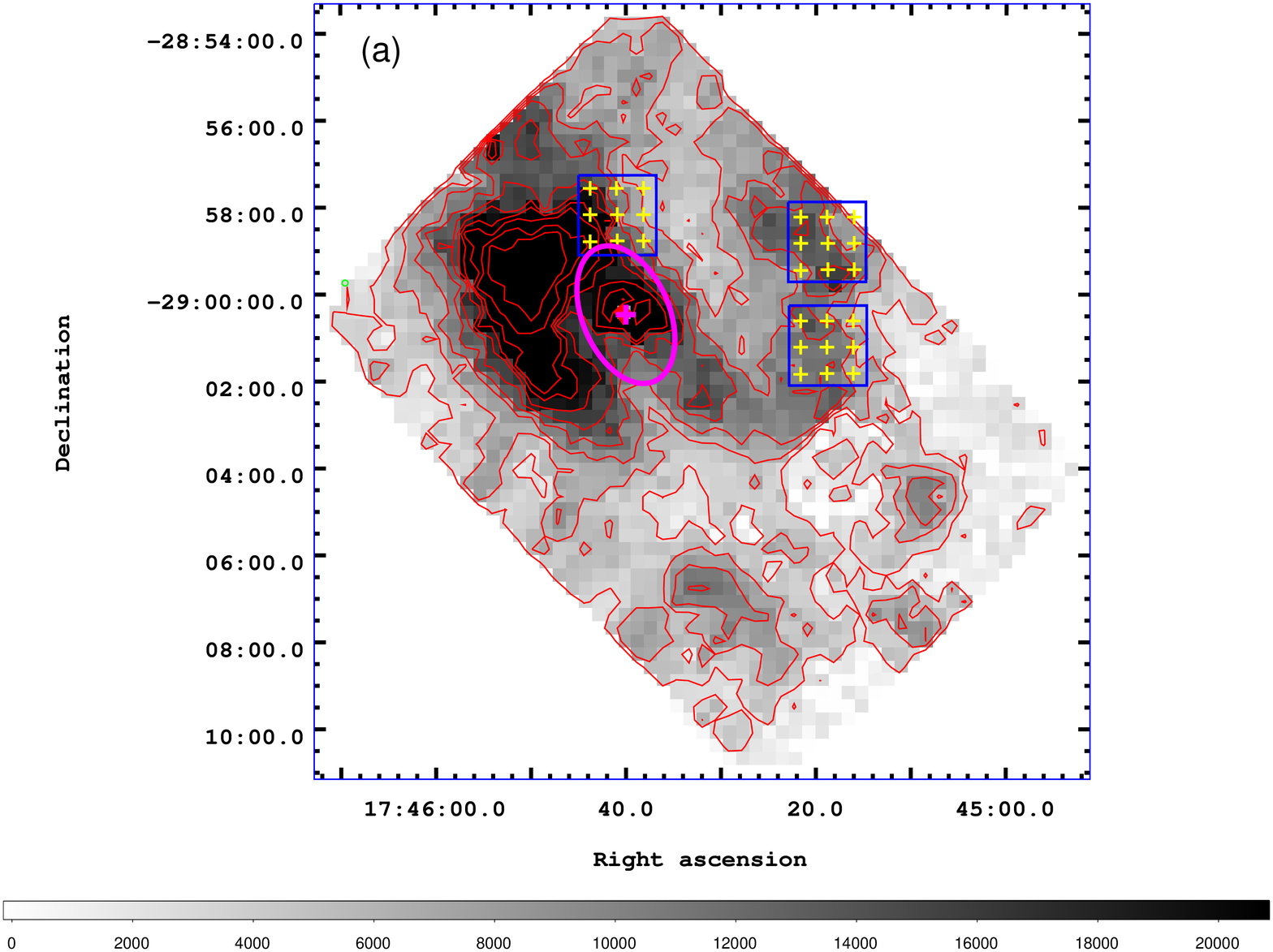}
\includegraphics[angle=0,scale=0.3]{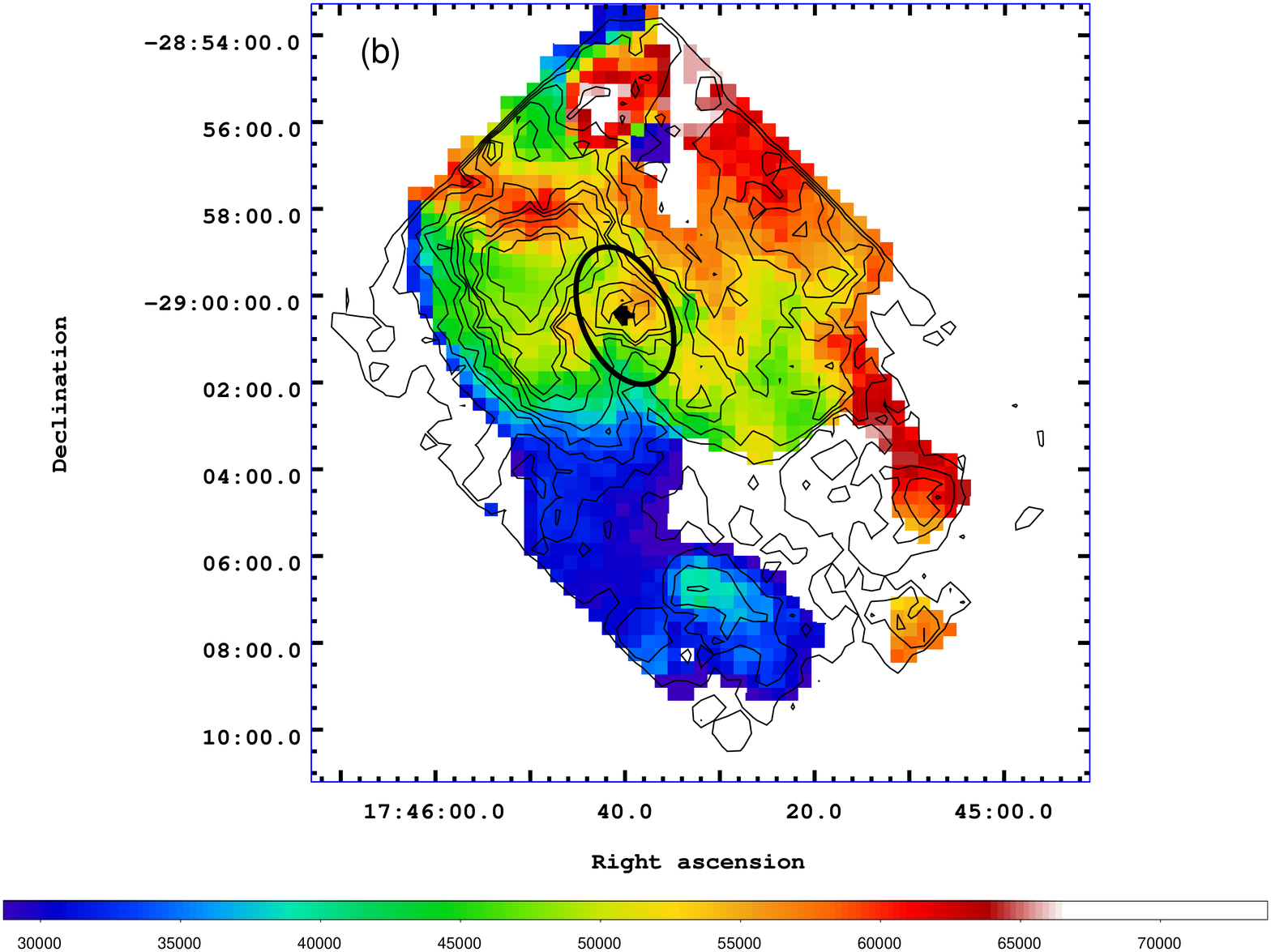}
\caption[]{(a) Integrated intensity map from 30 km s$^{-1}$ to 90 km s$^{-1}$ of the CS(J = 4--3) line. The color bar denotes the unit of the grey map in K m s$^{-1}$ ($T_{\rm A}$*). The contours are 3, 6, 9, 12, 15, 18, 21, 24, 30, 35 K m s$^{-1}$. (b) the intensity-weighted mean velocity map from 30 km s$^{-1}$ to 90 km s$^{-1}$ (color) overlaid on the intensity contours shown in (a). The color bar denotes the unit of m s$^{-1}$. The spectra of the DR, CR, and base of the PA are shown from Figure~\ref{fig-cs43spec1} to Figure~\ref{fig-cs43spec3}, where the positions of the spectra are marked with the 9 points yellow crosses for each region. The regions of DR, CR, and the base of the PA are the top left, bottom right, and top right, respectively.}
\label{fig-cs43mom1}
\end{center}
\end{figure}

\begin{figure}
\begin{center}
\epsscale{0.5}
\includegraphics[angle=0,scale=0.5]{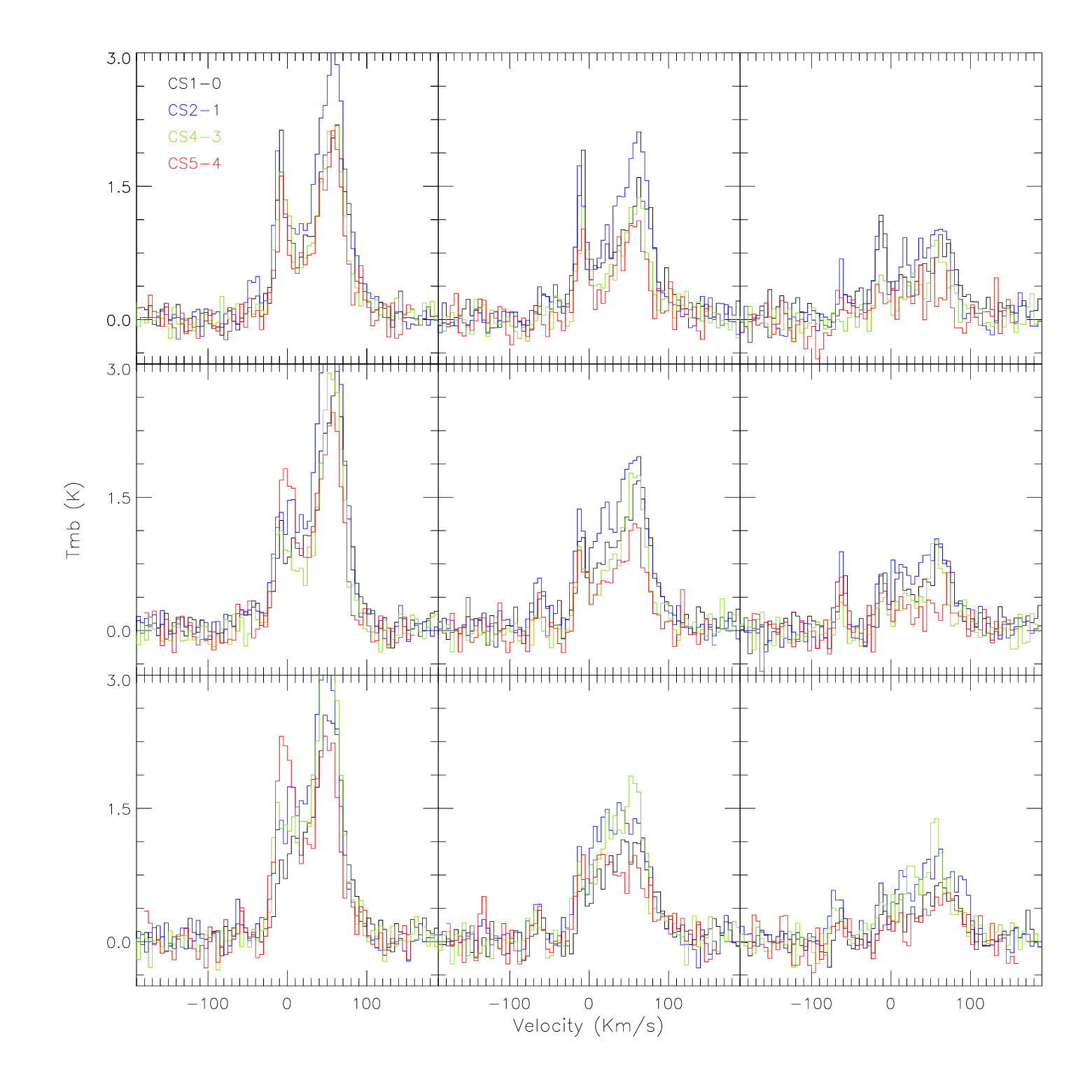}
\caption[]{Spectra of the DR. The CS(J = 5--4), CS(J = 4--3),
CS(J = 2--1), and CS(J = 1--0) lines are shown for the 9 points (1 point is averaged with 40$\arcsec\times40\arcsec$ to measured the fluxes) shown in Figure~\ref{fig-cs43mom1}. The positions of the 3 by 3 grid spectra are shown in Figure~\ref{fig-cs43mom1} (top left box).
The DR is the positive velocity source with intensity peak at $\sim$55 km s$^{-1}$.}
\label{fig-cs43spec1}
\end{center}
\end{figure}

\begin{figure}
\begin{center}
\epsscale{0.5}
\includegraphics[angle=0,scale=0.5]{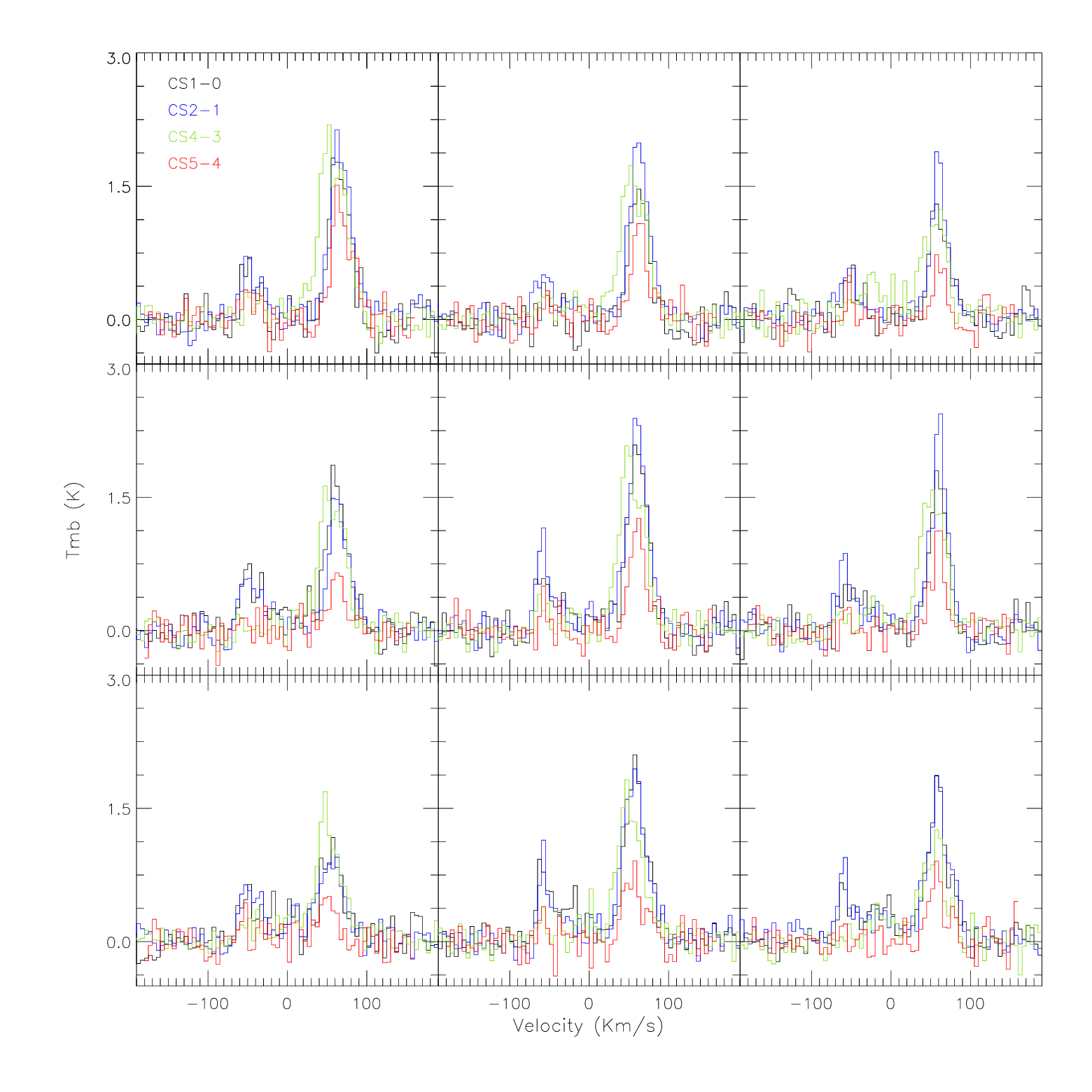}
\caption[]{Spectra of the CR. The CS(J = 5--4), CS(J = 4--3),
CS(J = 2--1), and CS(J = 1--0) lines are shown for the 9 points (1 point is averaged with 40$\arcsec\times40\arcsec$ to measured the fluxes) shown in Figure~\ref{fig-cs43mom1}. The positions of the 3 by 3 grid spectra are shown in Figure~\ref{fig-cs43mom1} (bottom right box).
The CR is the positive velocity source appears from 30 km s$^{-1}$ to 90 km s$^{-1}$ at FWZI. The emission shown at $\sim50$ km s$^{-1}$ is also shown in Figure~\ref{fig-pv-c1}.}
\label{fig-cs43spec2}
\end{center}
\end{figure}

\begin{figure}
\begin{center}
\epsscale{0.5}
\includegraphics[angle=0,scale=0.5]{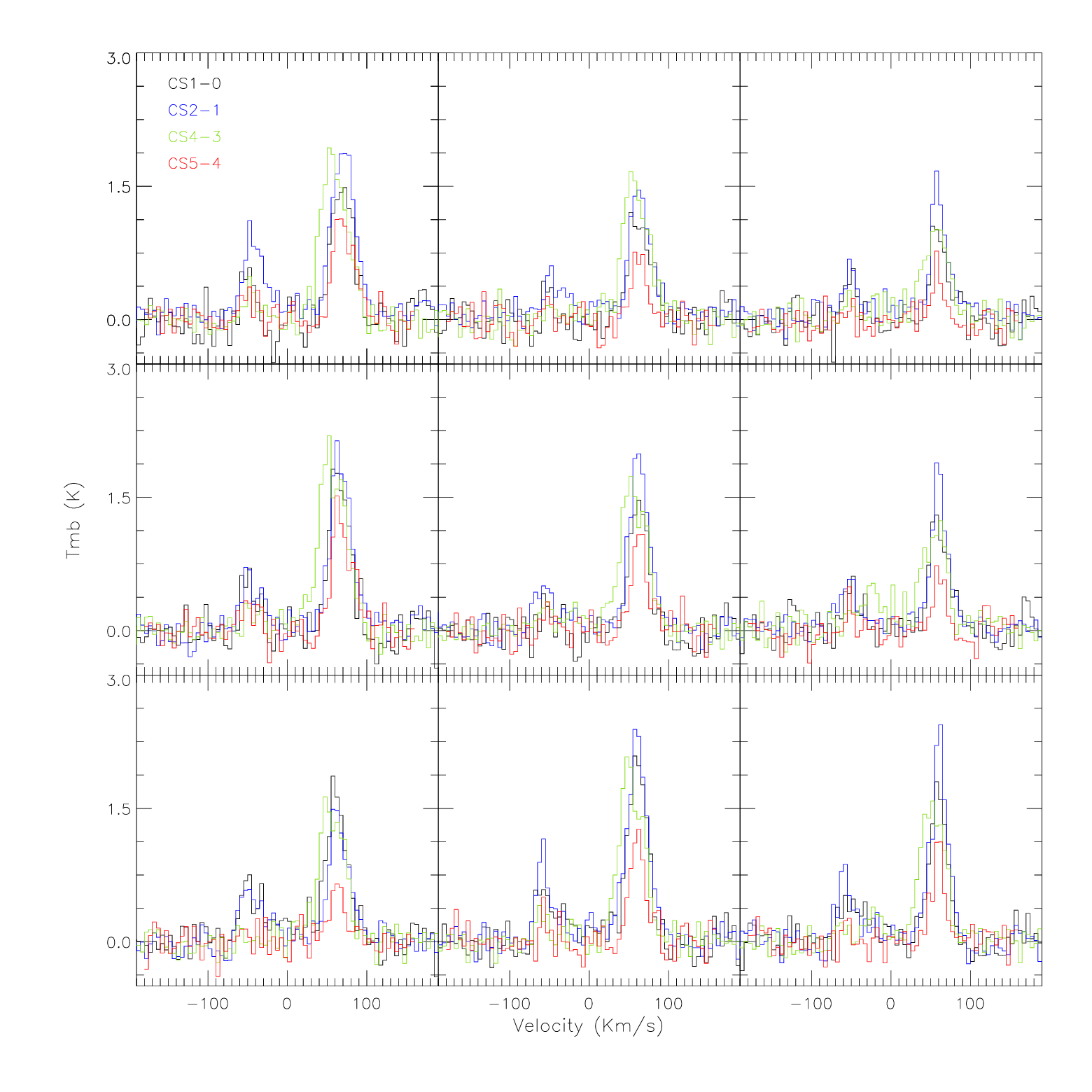}
\caption[]{Spectra of the base of the PA. The CS(J = 5--4), CS(J = 4--3),
CS(J = 2--1), and CS(J = 1--0) lines are shown for the 9 points (1 point is averaged with 40$\arcsec\times40\arcsec$ to measured the fluxes) shown in Figure~\ref{fig-cs43mom1}. The positions of the 3 by 3 grid spectra are shown in Figure~\ref{fig-cs43mom1} (top right box).
The PA is the positive velocity source with intensity peak at $\sim$62 km s$^{-1}$. The emission shown at $\sim50$ km s$^{-1}$ is also shown in Figure~\ref{fig-pv-c1}.}
\label{fig-cs43spec3}
\end{center}
\end{figure}

\begin{figure}
\begin{center}
\includegraphics[angle=0,scale=0.4]{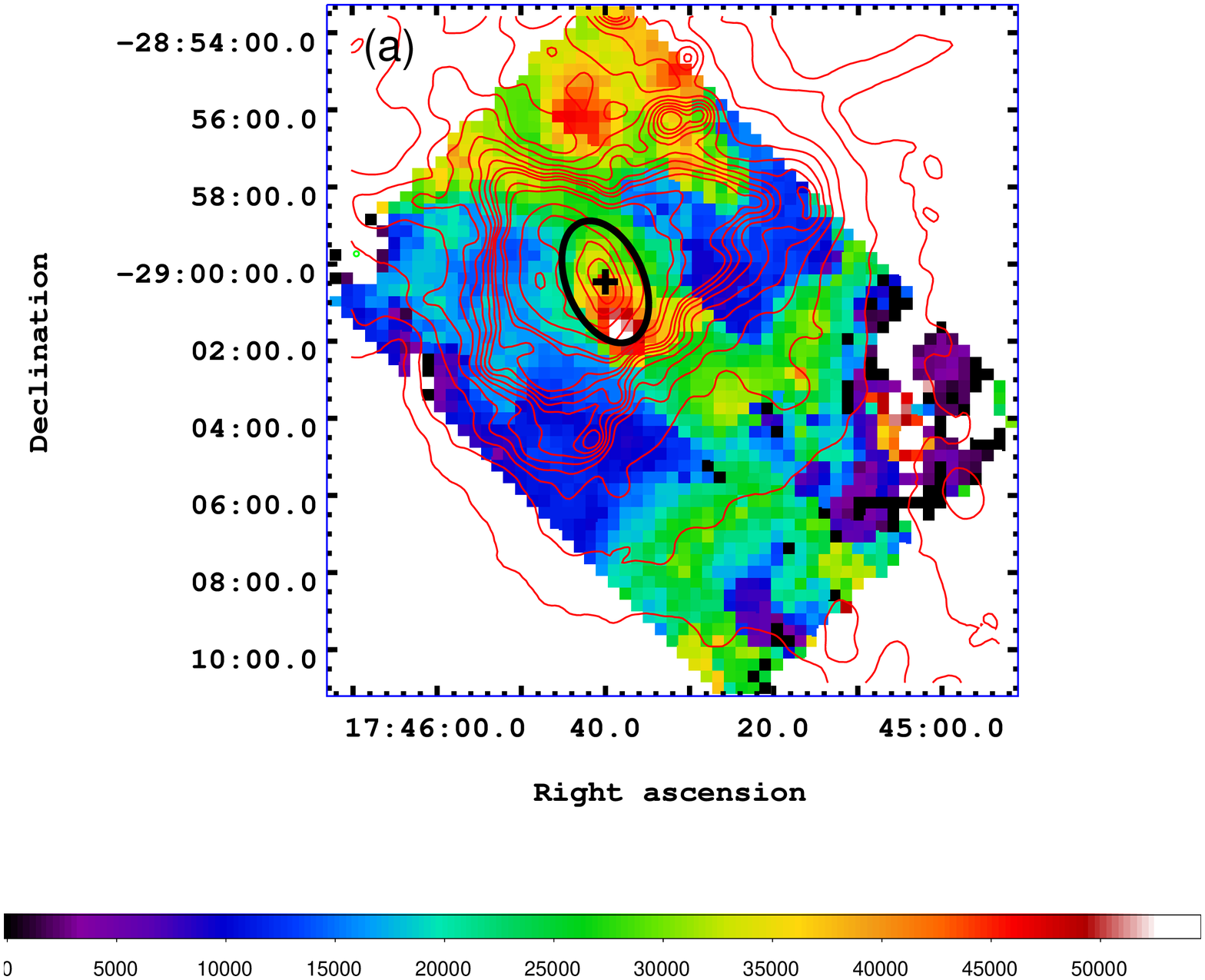}
\includegraphics[angle=0,scale=0.4]{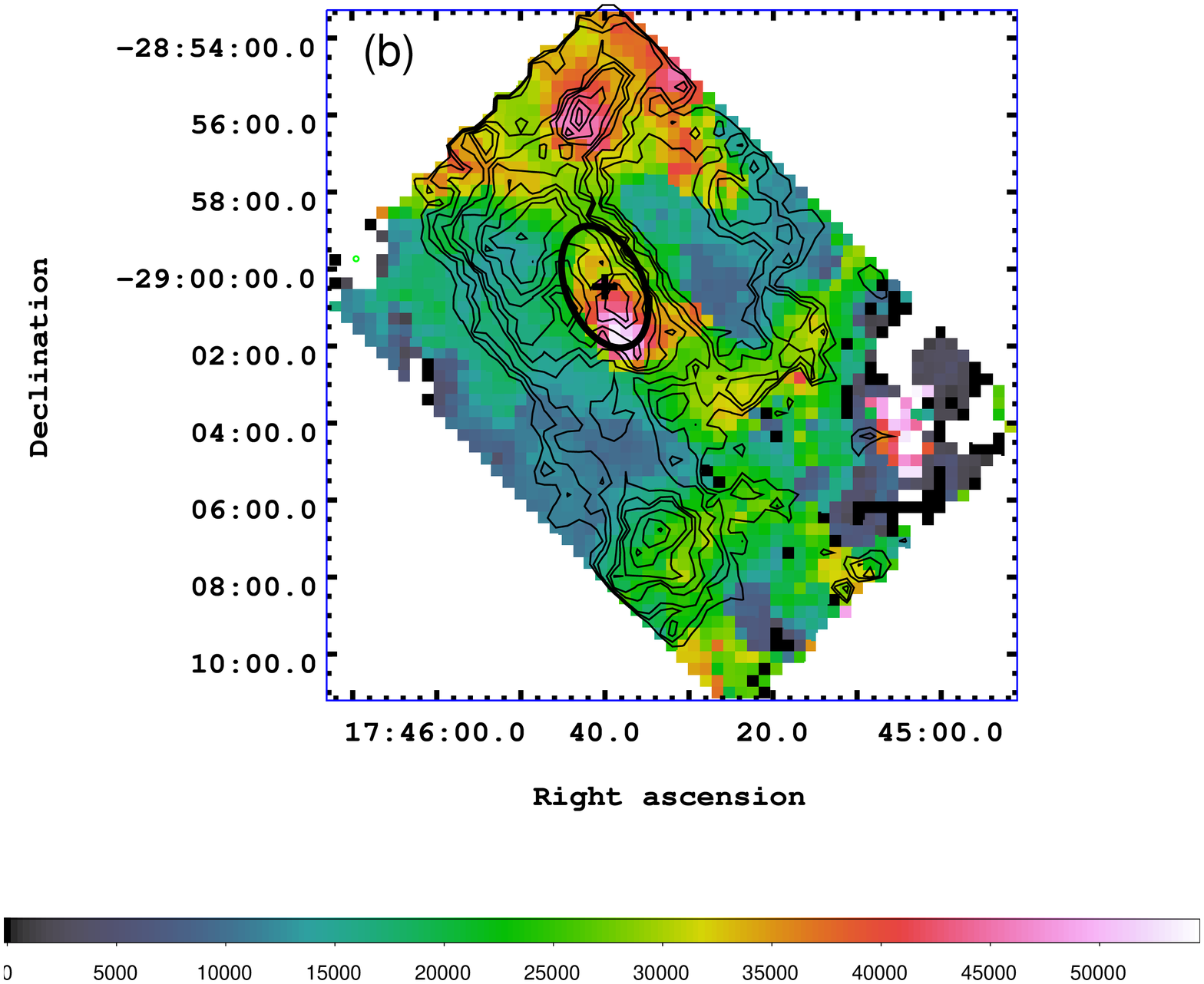}
\caption[]{(a) Velocity dispersion map (color) of the CS(J = 4--3) line made with $\pm133$ km s$^{-1}$. The contours of the 20 cm continuum emission are overlaid. The unit of the color bar is m s$^{-1}$. (b) The color map is the same with (a), and the contours of CS(J = 4--3) integrated intensity map ($\pm133$ km s$^{-1}$) are overlaid. The contour levels are 50, 60, 70, 90, 100, 130, 150, 170, 190 K km s$^{-1}$.}
\label{fig-cs43mom2}
\end{center}
\end{figure}

\clearpage

\begin{figure}
\begin{center}
\epsscale{0.5}
\includegraphics[angle=0,scale=0.25]{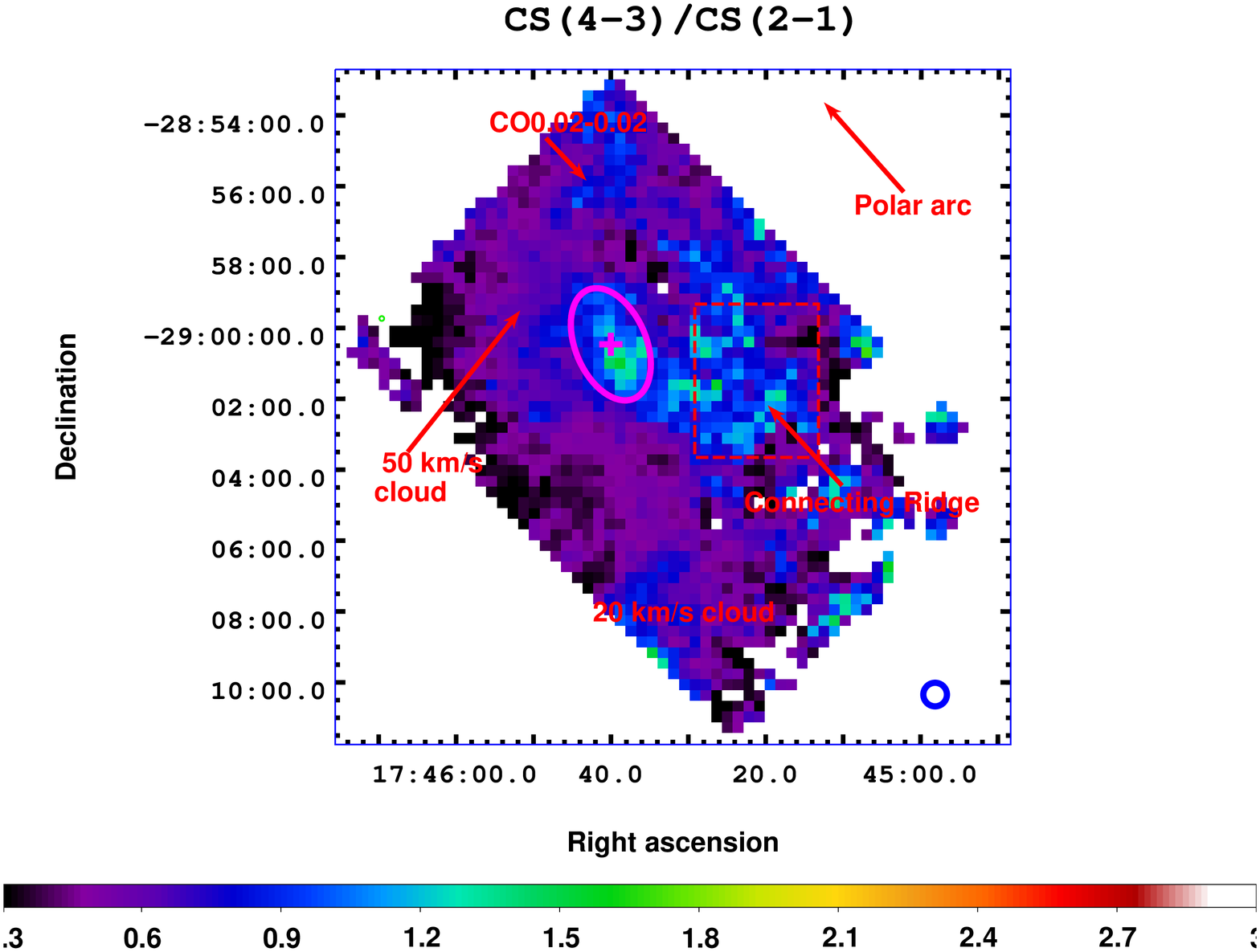}
\includegraphics[angle=0,scale=0.25]{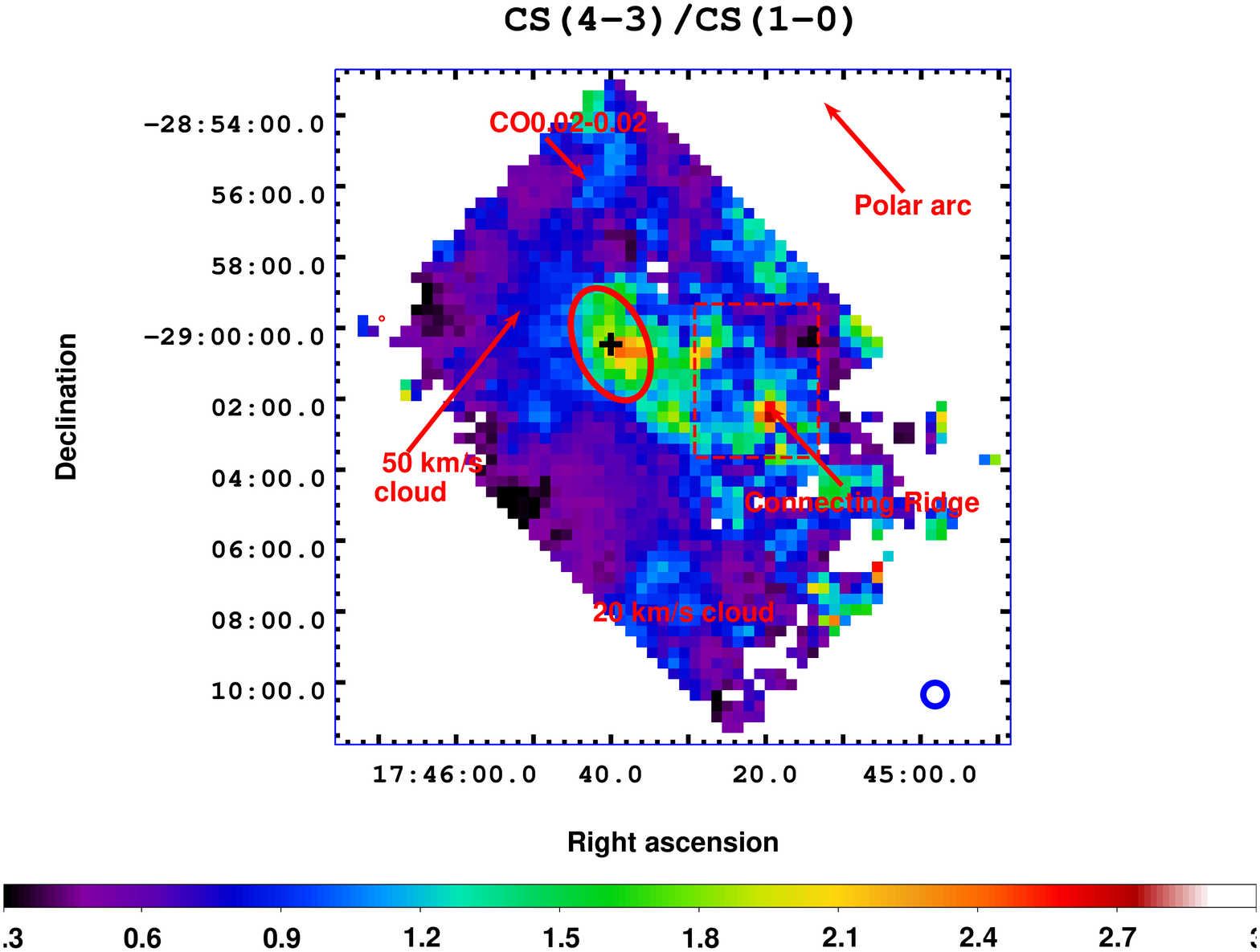}
\includegraphics[angle=0,scale=0.25]{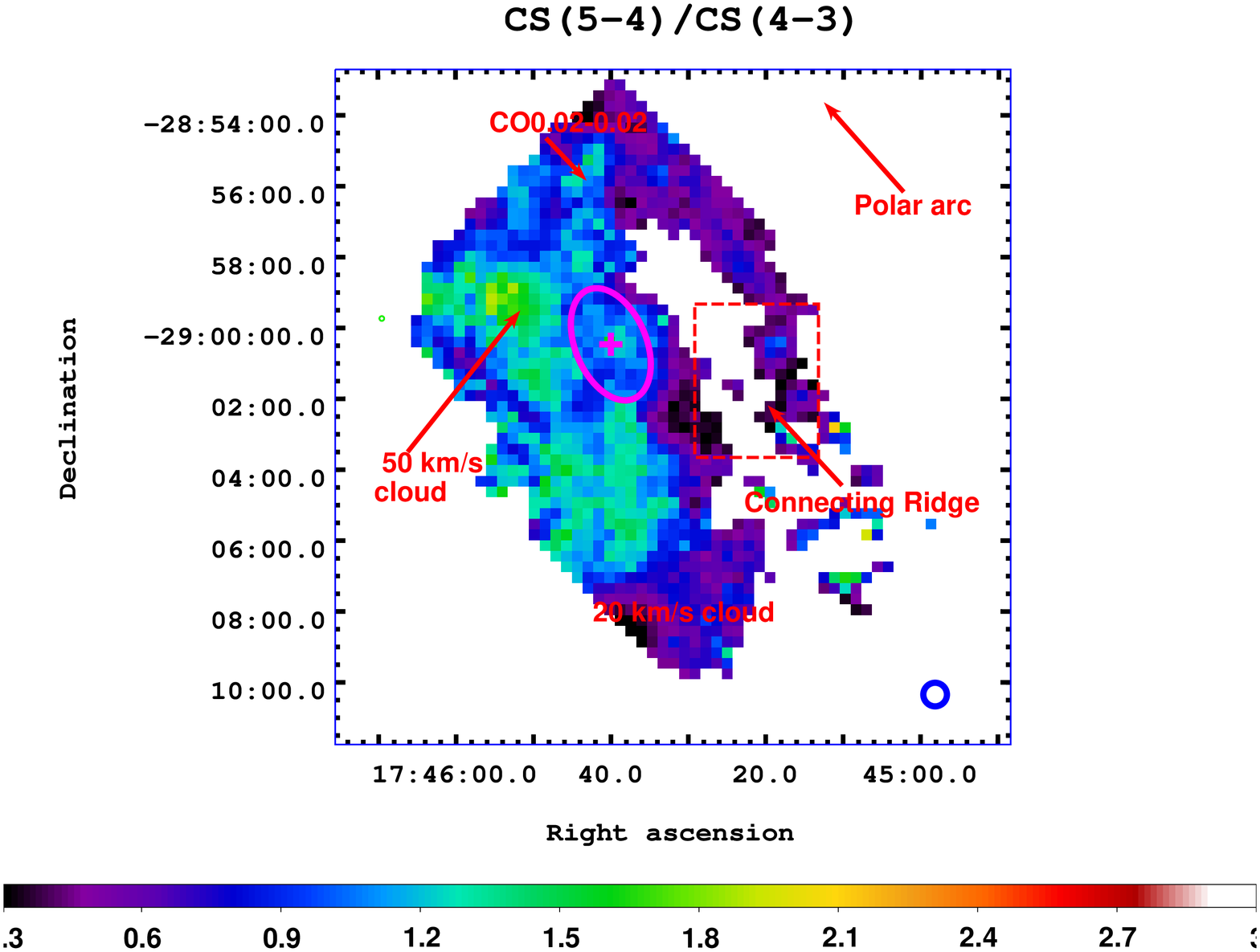}
\includegraphics[angle=0,scale=0.25]{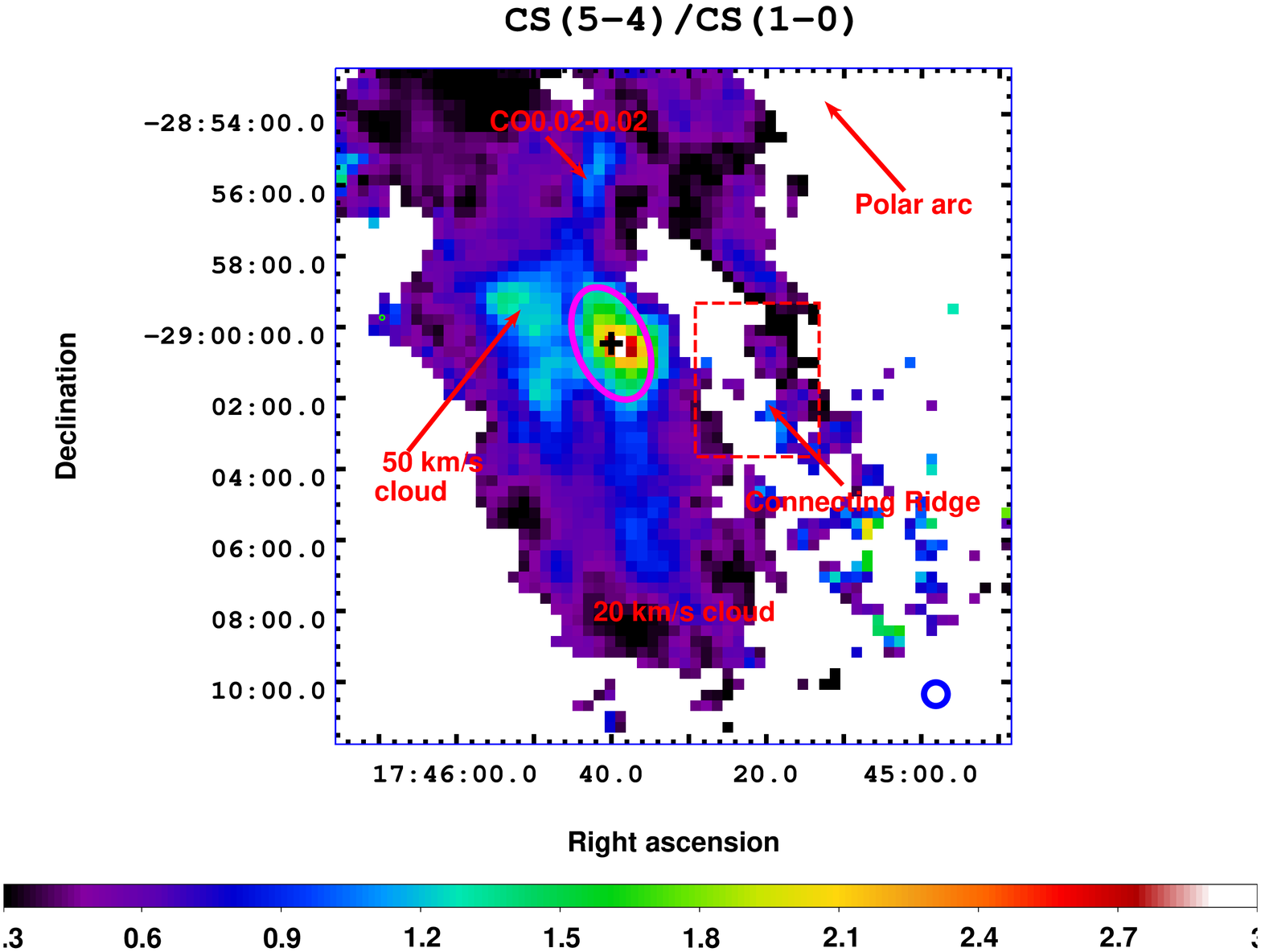}
\caption[]{Intensity ratio maps of the four CS lines. The color bar denotes the intensity ratios. We integrated the data from $-160$ km s$^{-1}$ to $+160$ km s$^{-1}$, and divided two maps to obtain the ratio maps. The ratios of CS(J = 4--3)/CS(J = 1--0), CS(J = 4--3)/CS(J = 1--0), CS(J = 5--4)/CS(J = 4--3), and CS(J = 5--4)/CS(J = 1--0) are presented. We smoothed the data to beam size of 40$\arcsec$ to make the ratio maps. The central ellipse and cross mark the location of the CND and SgrA*, respectively. The color scale was matched from 0.3 to 3 for all images for comparison.}
\label{fig-cs-ratio}
\end{center}
\end{figure}

\begin{figure}
\begin{center}
\epsscale{0.5}
\includegraphics[angle=0,scale=0.5]{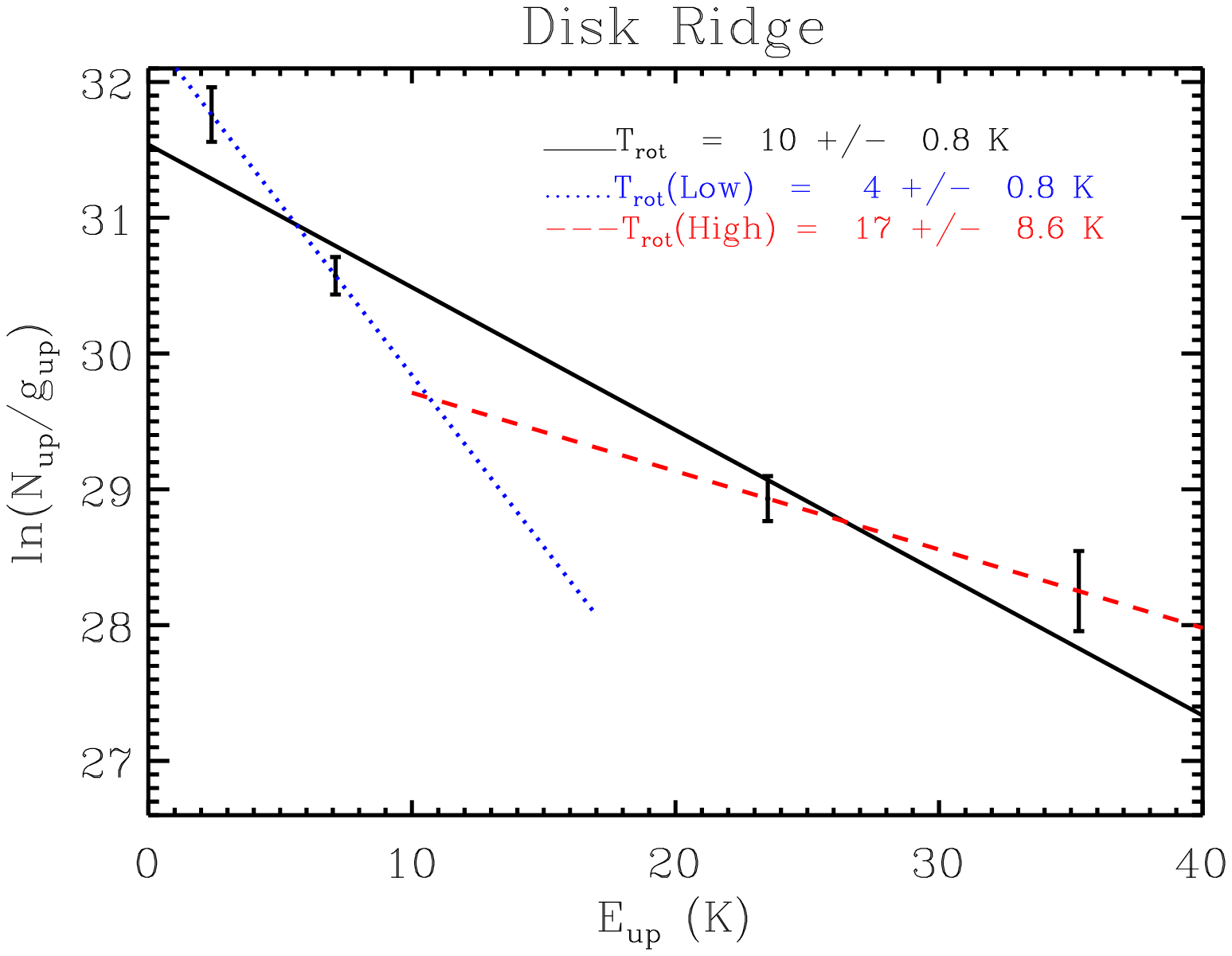}
\includegraphics[angle=0,scale=0.5]{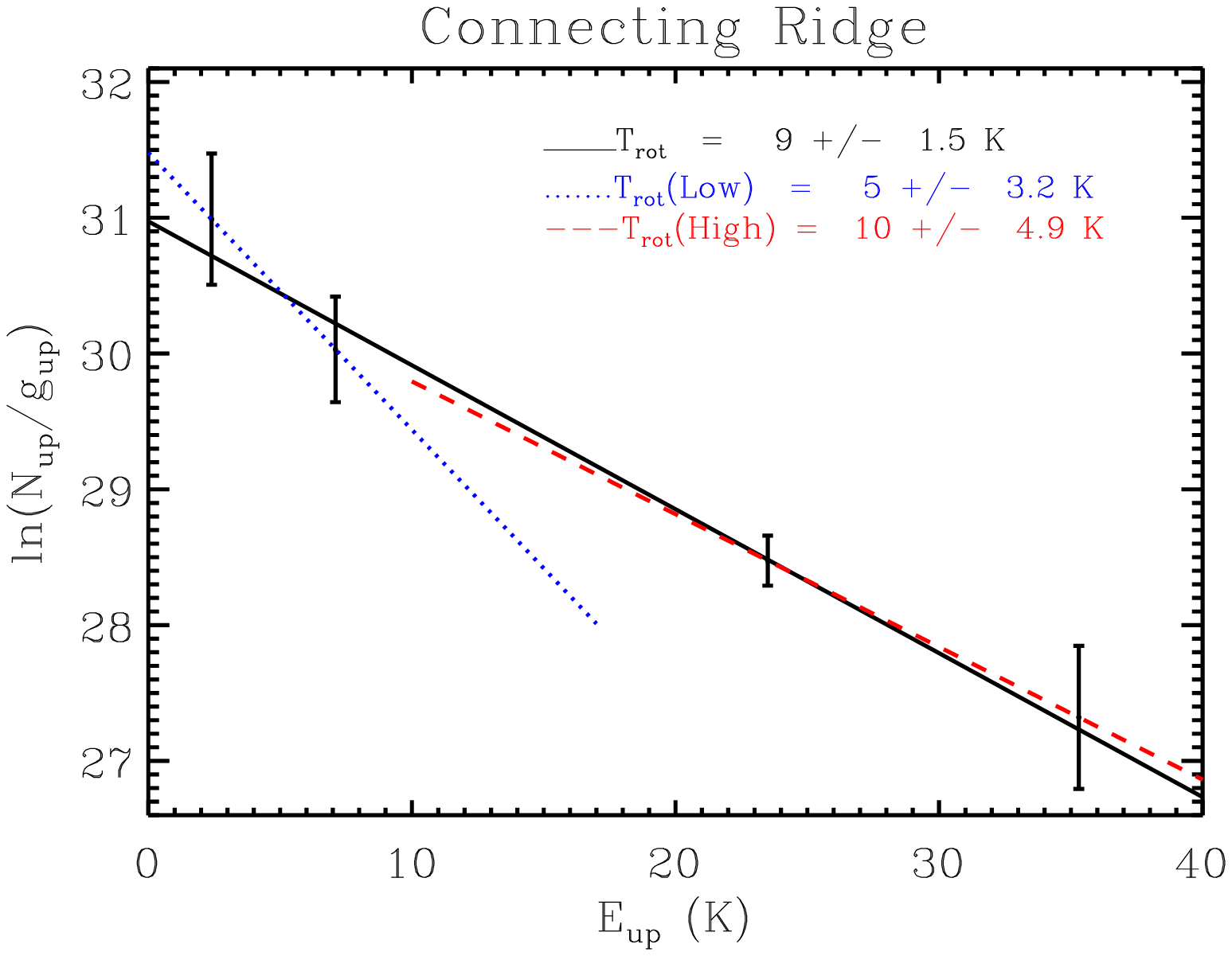}
\includegraphics[angle=0,scale=0.5]{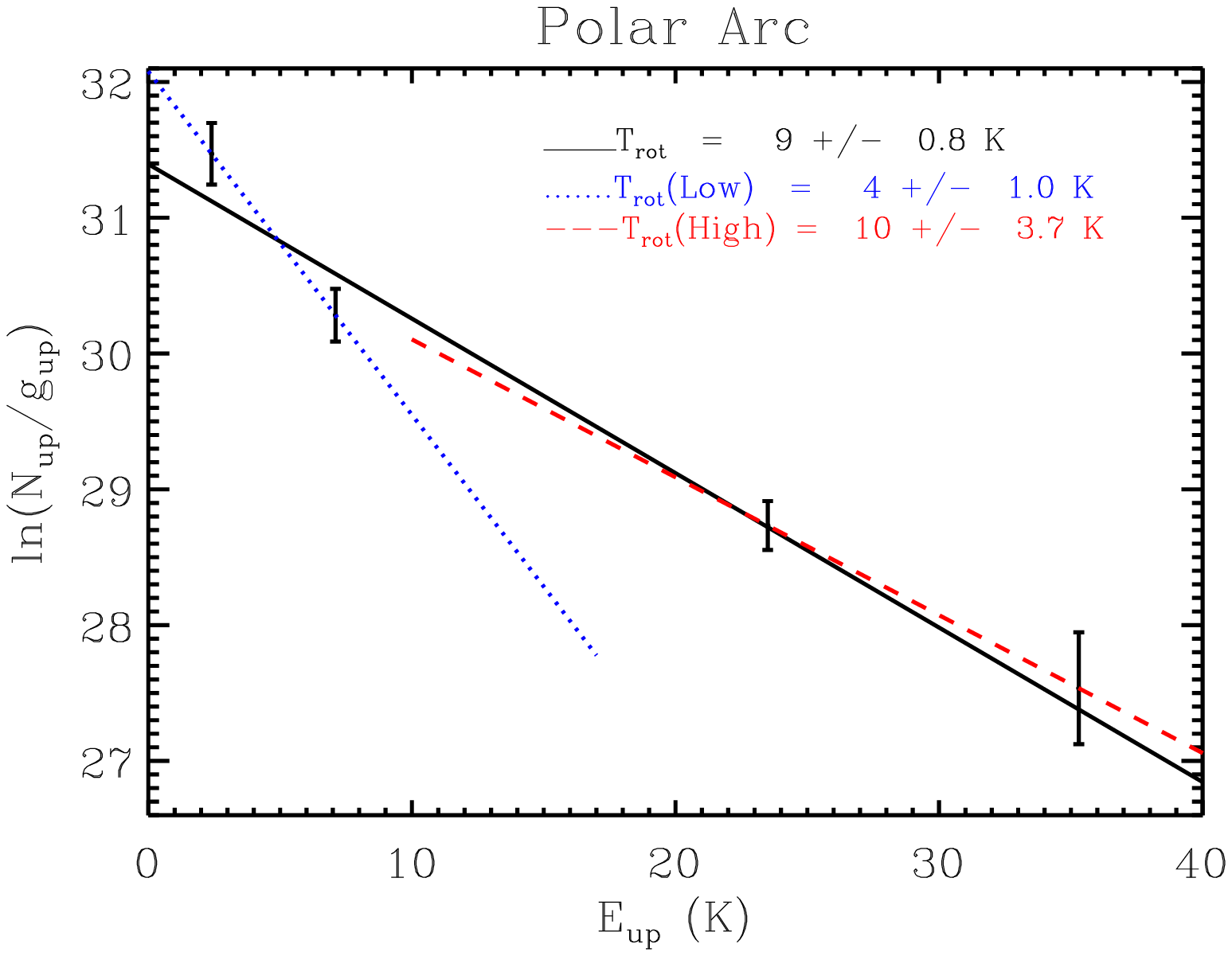}
\caption[]{We made the rotational diagrams with the CS(J = 1--0) \citep{tsuboi99}, CS(J = 2--1) \citep{hsieh15}, CS(J = 4--3), and CS(J = 5--4) lines. We show the rotation-diagrams of the DR, CR, and the base of the PA. The fluxes are taken from the average spectra presented in Figure~\ref{fig-cs43spec1}, Figure~\ref{fig-cs43spec2}, and Figure~\ref{fig-cs43spec3}. The derived rotational temperatures $T_{\rm rot}$, $T_{\rm rot}$(High), and $T_{\rm rot}$(Low) are labeled on the plots.
\label{fig-rot-diag}
}
\end{center}
\end{figure}

\clearpage

\begin{figure}
\begin{center}
\epsscale{0.5}
\includegraphics[angle=0,scale=0.6]{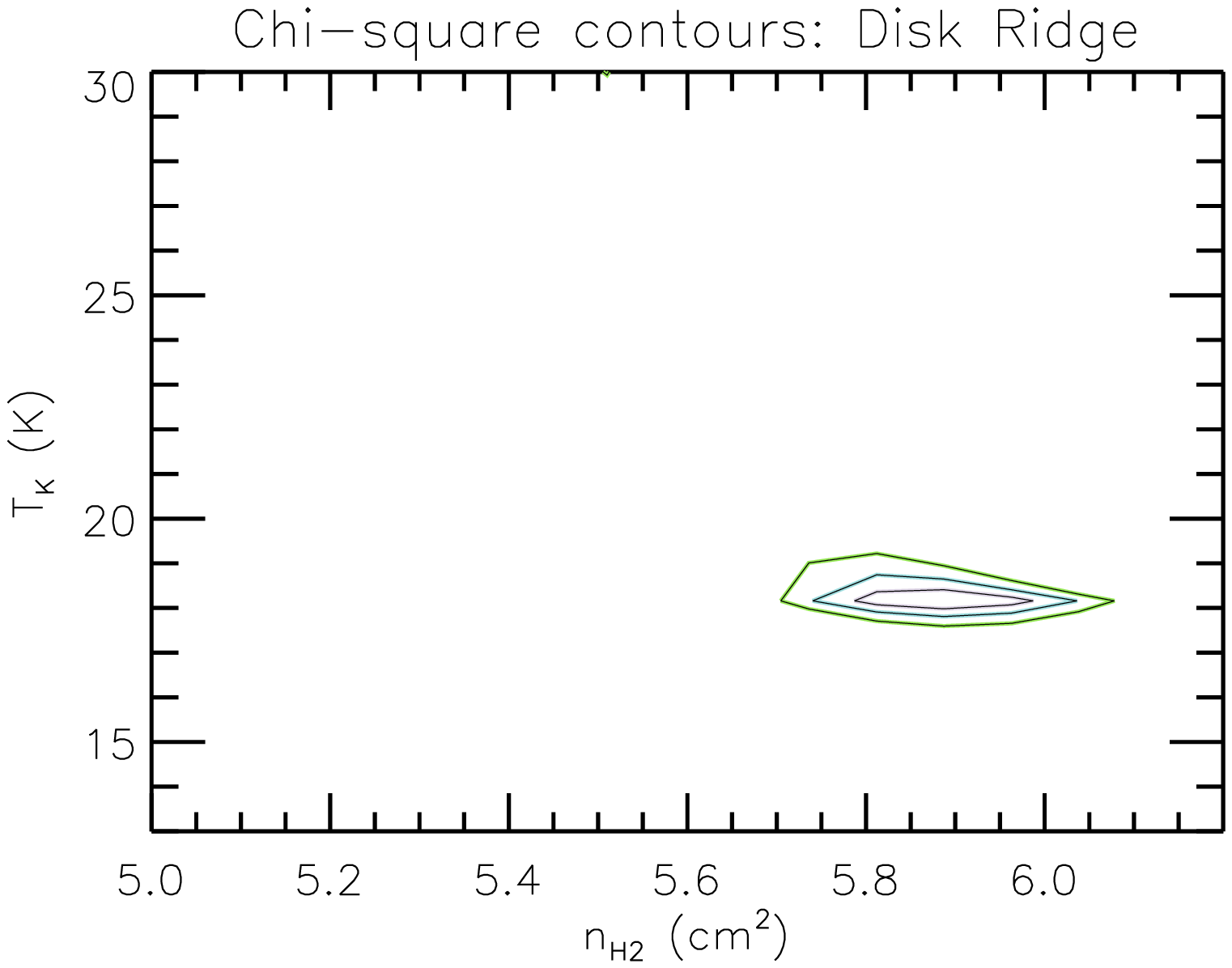}
\includegraphics[angle=0,scale=0.6]{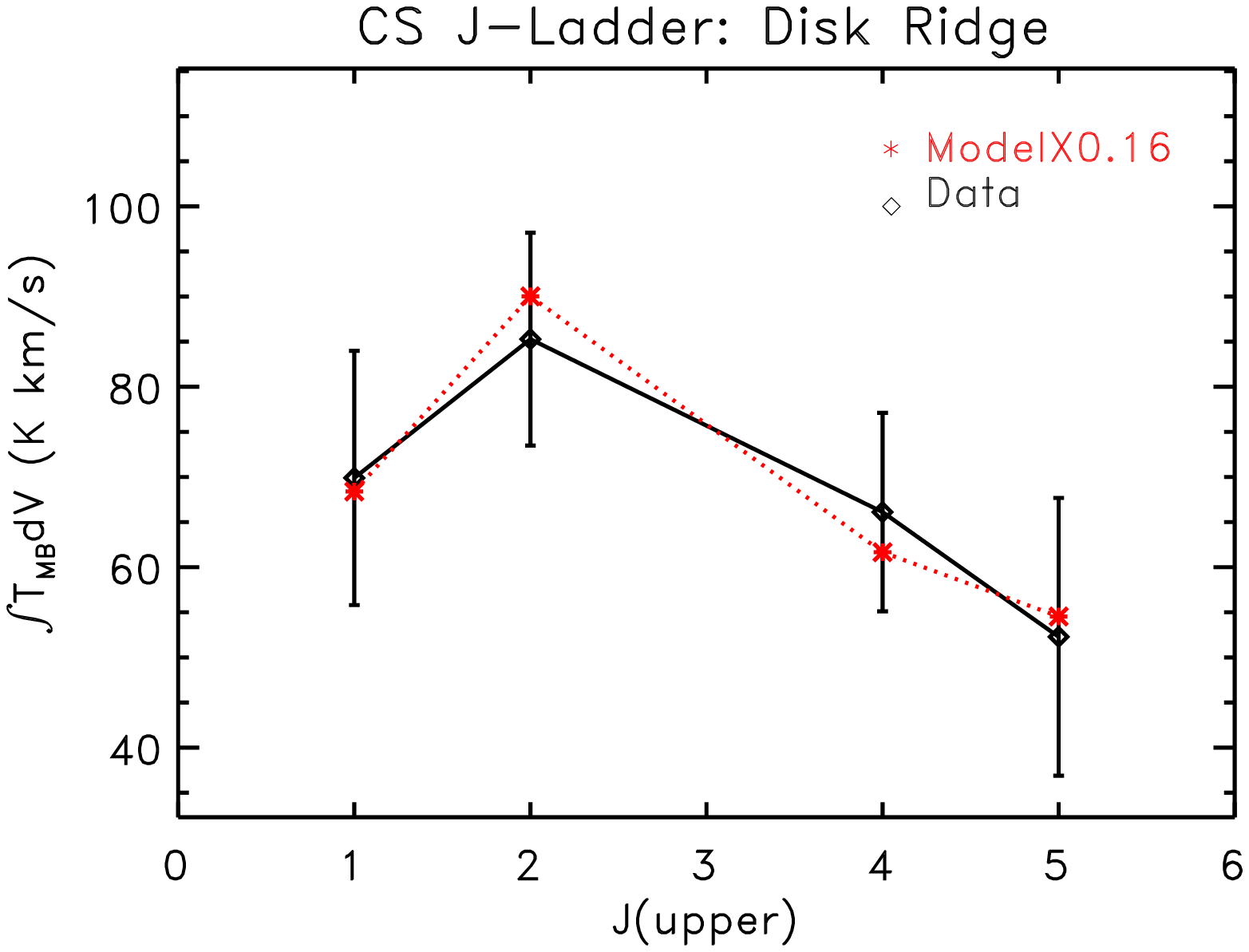}
\caption[]{CS molecule J-ladder plot of the DR. The x-axis denotes the upper J level. The y-axis denotes the integrated main beam brightness temperature. The fluxes measured from our data is shown in black line with diamond symbol. The error bar is the rms of the fluxes, and used for the $\chi^2$ fitting. The contours are the corresponding probabilities of 90\% (minimum $\chi^2=0.6$), 80\%, 70\%. The best fitted model (constrained by the intensity ratios) is shown in red asterisk. The model was multiplied by the average beam filling factor of 0.16. See Table~\ref{t.obspar1} for the derived parameters.
}
\label{fig-radex1}
\end{center}
\end{figure}
    
\begin{figure}
\begin{center}
\epsscale{0.5}
\includegraphics[angle=0,scale=0.6]{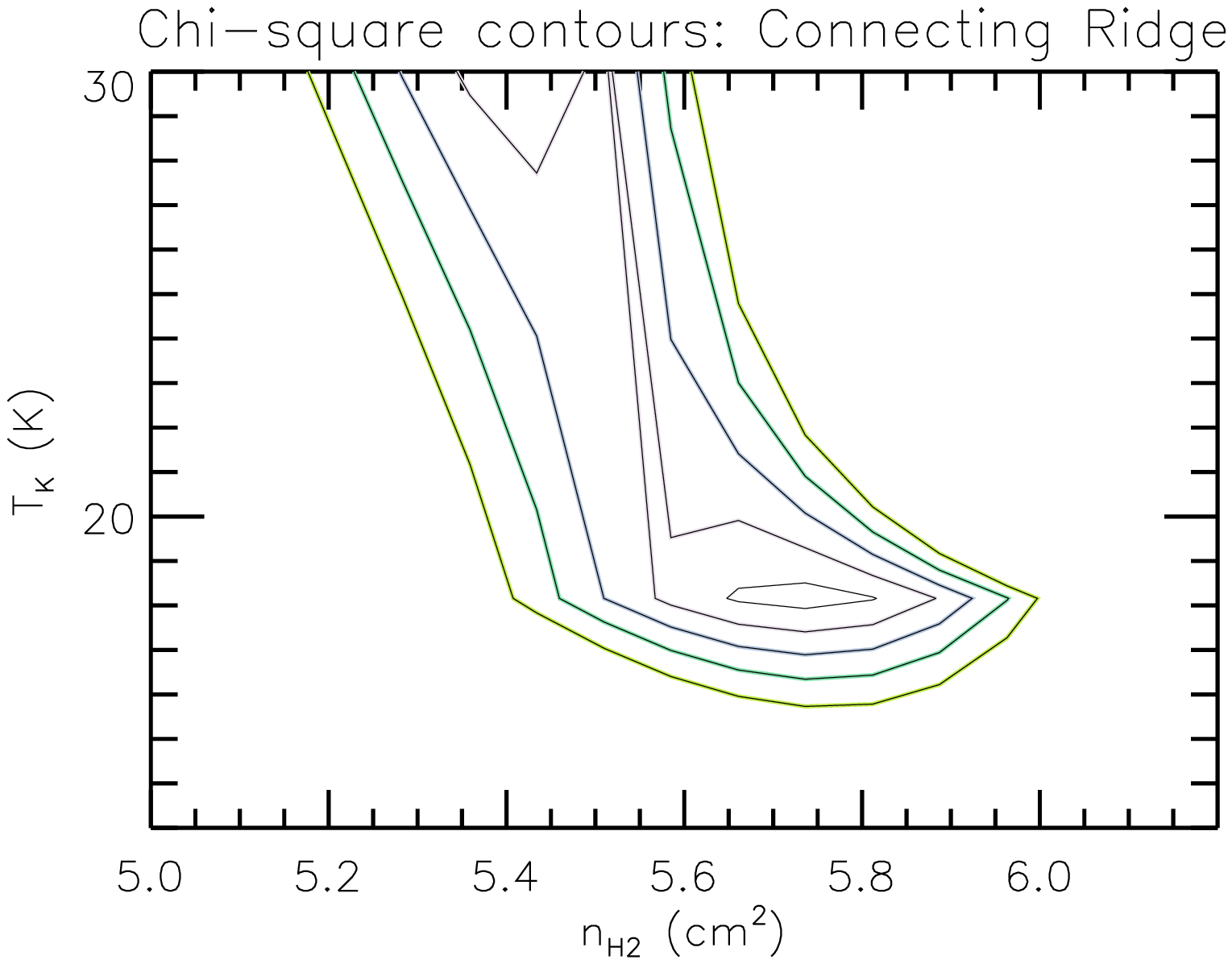}
\includegraphics[angle=0,scale=0.6]{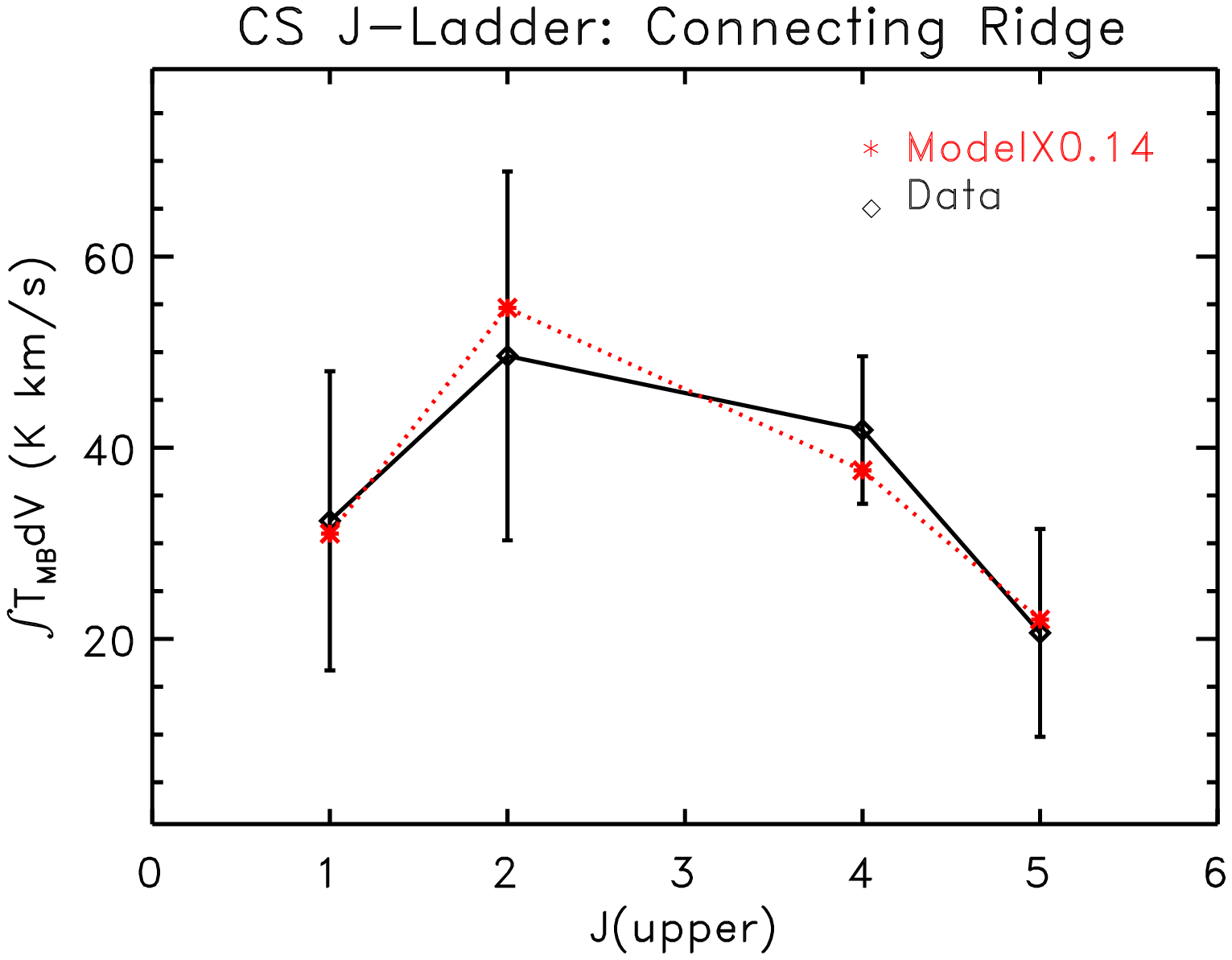}
\caption[]{CS molecule J-ladder plot of the CR. The x-axis denotes the upper J level. The y-axis denotes the integrated main beam brightness temperature. The fluxes measured from our data is shown in black line with diamond symbol. The error bar is the rms of the fluxes, and used for the $\chi^2$ fitting. The contours are the corresponding probabilities of 90\% (minimum $\chi^2=0.39$) to 50\%, 10\% in step. The best fitted model (constrained by the intensity ratios) is shown in red asterisk. The model was multiplied by the average beam filling factor of 0.14. See Table~\ref{t.obspar2} for the derived parameters.
}
\label{fig-radex2}
\end{center}
\end{figure}

\begin{figure}
\begin{center}
\epsscale{0.5}
\includegraphics[angle=0,scale=0.6]{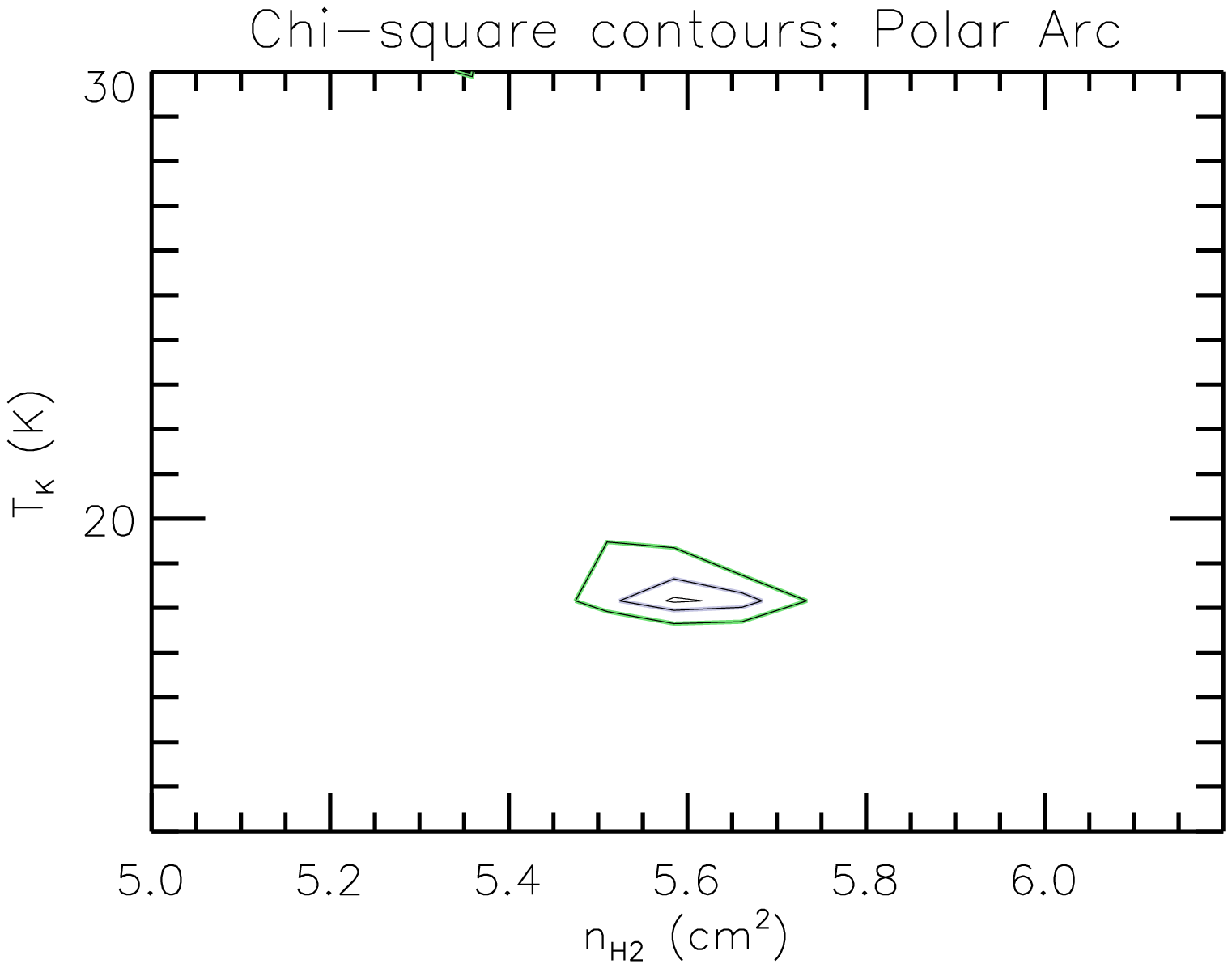}
\includegraphics[angle=0,scale=0.6]{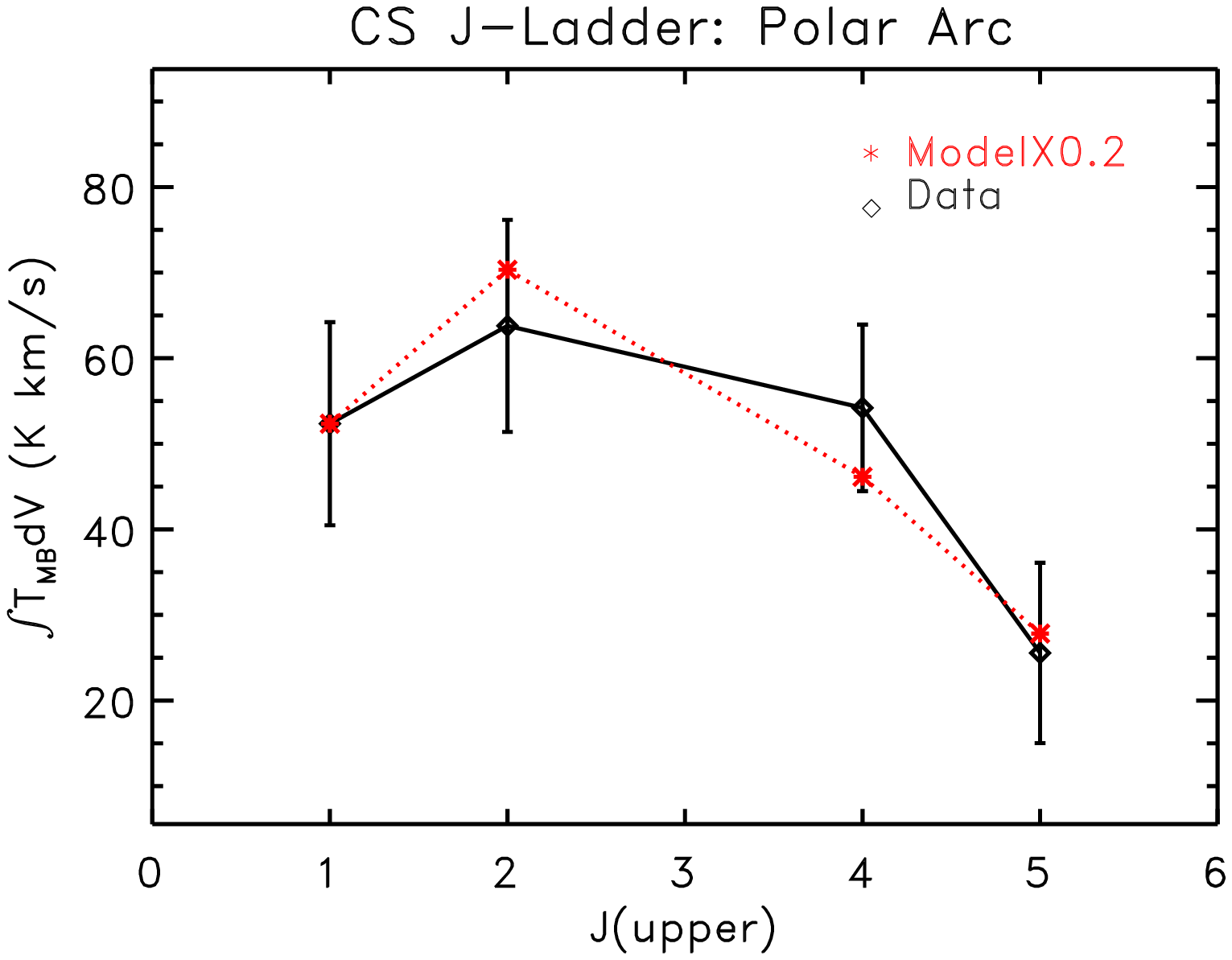}
\caption[]{CS molecule J-ladder plot of the base of the PA. The x-axis denotes the upper J level. The y-axis denotes the integrated main beam brightness temperature. The fluxes measured from our data is shown in black line with diamond symbol. The error bar is the rms of the fluxes, and used for the $\chi^2$ fitting. The contours are the corresponding probabilities of 70\% (minimum $\chi^2=1.5$), 60\%, and 50\%. The best fitted model (constrained by the intensity ratios) is shown in red asterisk. The model was multiplied by the average beam filling factor of 0.2. See Table~\ref{t.obspar3} for the derived parameters.
}
\label{fig-radex3}
\end{center}
\end{figure}

\begin{figure}
\begin{center}
\epsscale{0.5}
\includegraphics[angle=0,scale=0.4]{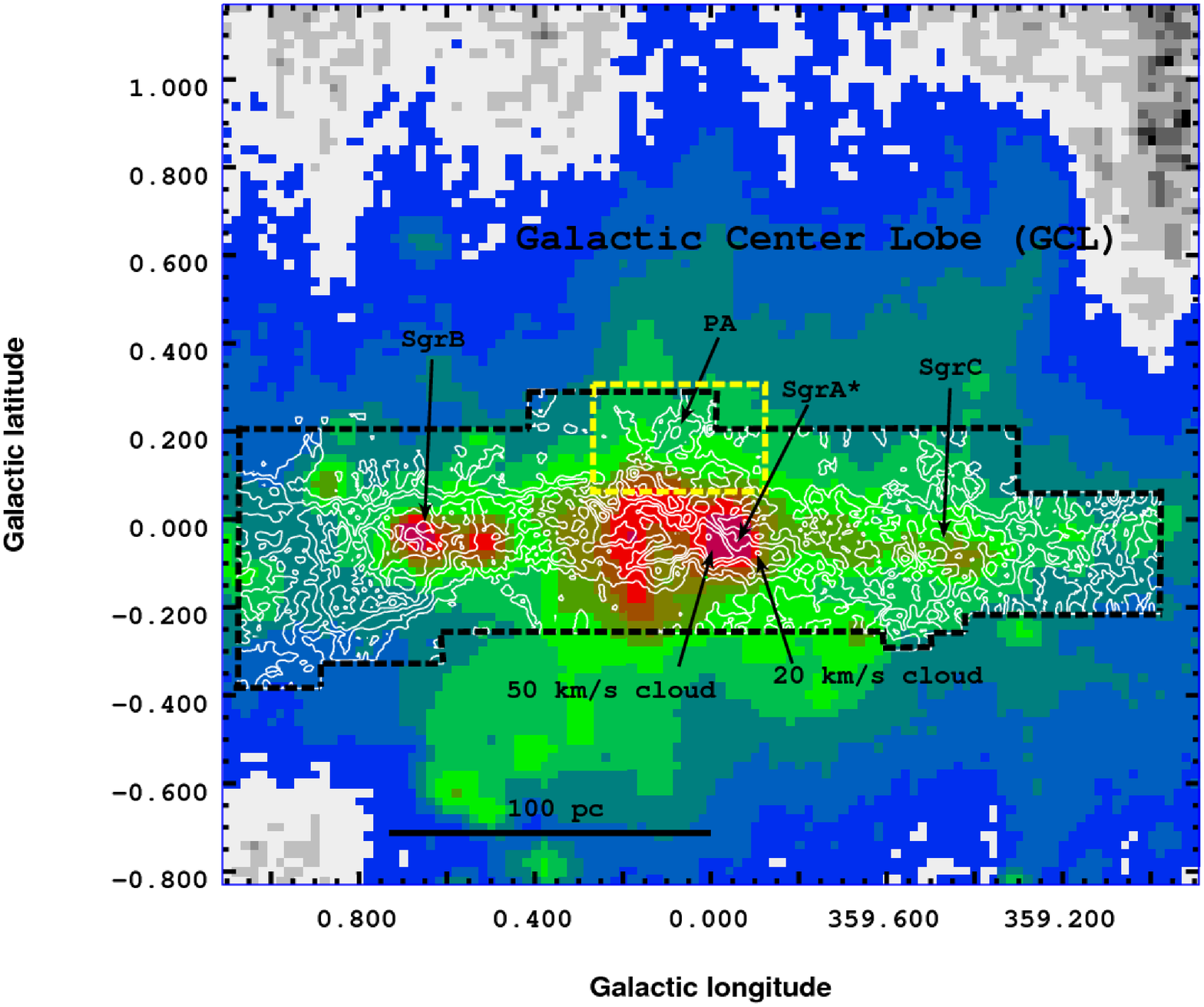}
\caption[]{10 GHz continuum \citep{handa87} (color image) overlaid on the contours of CS(J = 1--0) line \citep{tsuboi99} (60$\arcsec$ beam) integrated over $\pm217.5$ km s$^{-1}$. The contours are 15, 30, 50, 70, and 90 K km s$^{-1}$. The map area of the CS(J = 1--0) line is denoted by the black dashed polygon. The yellow dashed line box denotes the PA.}
\label{fig-10ghz}
\end{center}
\end{figure}

\end{document}